\documentclass[a4paper,11pt] {article}
\usepackage[left=2.5cm,right=2.5cm,top=1cm,bottom=4cm]{geometry}
\usepackage{color}
\usepackage{colortbl}
\usepackage{fancyhdr}
\usepackage{graphicx}
\usepackage{eurosym}
\usepackage{enumitem}
\usepackage{lastpage}
\usepackage{booktabs}
\usepackage[yyyymmdd]{datetime}
\usepackage{multirow}
\usepackage{pifont}
\usepackage[colorlinks=true, urlcolor=blue, linkcolor=black, citecolor=red]{hyperref}
\usepackage{amssymb} %maths
\usepackage[color]{changebar}
\usepackage{xcolor}
\usepackage{listings}
\usepackage{threeparttable}
\usepackage{flafter}
\usepackage{longtable}
\usepackage{hyperref}
\usepackage{eurosym}
\usepackage{multicol}
\usepackage{tabularx}
\usepackage{wrapfig}
\usepackage{cuted}
\usepackage{caption}
\usepackage{colortbl}
\usepackage{color}
\usepackage{placeins}
\usepackage{aas_macros}

\usepackage{outlines}
\usepackage{siunitx}

\usepackage{enumitem}
\setitemize{noitemsep,topsep=0pt,parsep=0pt,partopsep=0pt}

\usepackage{libertine}
\usepackage[libertine]{newtxmath}

\usepackage[printonlyused]{acronym}

\usepackage{tikz}
\usetikzlibrary{fit,chains,shapes}
\usetikzlibrary{shapes.geometric,shapes.arrows,decorations.pathmorphing}
\usetikzlibrary{matrix,chains,scopes,positioning,arrows,fit}
\usetikzlibrary{arrows,calc,shapes,decorations.pathreplacing,decorations.markings}

\DeclareSIUnit{\rthz}{\ensuremath{\sqrt{\text{\hertz}}}}
\begin{document}

 %%%%%%%%%%%%%%% Set the global variables values  %%%%%%%%%%%%%%%
\newcommand{\ReportTitle}{LISA Sensitivity and SNR Calculations}           % Report Title
\newcommand{\ReportShortTitle}{LISA Sensitvity \& SNR}     % Report Title
\newcommand{\ReportIssue}{1}                                                                  % Issue number
\newcommand{\ReportRevision}{1}                                                             % Revision number
\newcommand{\ReportRef}{LISA-LCST-SGS-TN-001}                                           % Report reference number
\newcommand{\ReportCurrentAuthor}{A. Petiteau} %  Current author
\newcommand{\ReportCurrentAuthorInstitute}{APC}           % Institute of the current author

 %%%%%%%%%%%%%%% %%%%%%%%%%%%%%% %%%%%%%%%%%%%%%

%%%%%%%%%%%%%%%
%%% New commands, environment, etc
\definecolor{pink}{rgb}{0.55,0,0.52}
\definecolor{mygreen}{rgb}{0.19,0.55,0.11}
\definecolor{dkgreen}{rgb}{0,0.6,0}
\definecolor{gray}{rgb}{0.5,0.5,0.5}
\definecolor{mauve}{rgb}{0.58,0,0.82}
\definecolor{verbgray}{gray}{0.9}
\definecolor{lightblue}{rgb}{0.85,0.9,1}
\definecolor{lightgreen}{rgb}{0.85,1,0.85}
\definecolor{lightorange}{rgb}{1,0.94,0.8}
\definecolor{forestgreen}{rgb}{0.1,0.49,0.07}

\newcommand{\todo}[1]{\textcolor{red}{[TODO: #1]}}
\newcommand{\TBC}{\textcolor{red}{(to be checked)}}
\newcommand{\TBD}[1]{\textcolor{red}{\textit{TBD} #1}}
\newcommand{\tbc}[0]{\textcolor{red}{TBC}}

\newcommand{\PriorityCaptain}[2]{ \textcolor{red}{[priority: #1, bookcaptain: #2]} }

%%%% Comments :
\newcommand{\cAP}[1]{\textcolor{red}{\textit{[Antoine] #1}}}
\newcommand{\sAP}[2]{\textcolor{red}{#1 \textit{[Antoine] #2}}}
\newcommand{\cSB}[1]{\textcolor{forestgreen}{\textit{[Stas] #1}}}

%%%%%%%%%%%%%%%%%%%%%%%%%%%%%%%%%%%%%%%%%%%%%%%%%%%%%%%%%%%%%%%%%%%%%%
% Some convenience commands
%%%%%%%%%%%%%%%%%%%%%%%%%%%%%%%%%%%%%%%%%%%%%%%%%%%%%%%%%%%%%%%%%%%%%%
\newcommand{\PSD}[2][h]{\ensuremath{S_{#1}(#2)}}
\newcommand{\lPSD}[2][h]{\ensuremath{\sqrt{S_{#1}(#2)}}}
\newcommand{\PSDhat}[2][h]{\ensuremath{\hat S_{#1}(#2)}}

\def\be{\begin{equation}}
\def\en{\end{equation}}
\def\bea{\begin{eqnarray}}
\def\ena{\end{eqnarray}}

\setlength{\parskip}{\baselineskip}%
\setlength{\parindent}{0pt}%

\newcommand{\cQ}[1]{\textcolor{red}{\textit{[?] #1}}}

\setcounter{secnumdepth}{4}
\setcounter{tocdepth}{4}

\lstnewenvironment{rawtxt}{%
  \lstset{backgroundcolor=\color{verbgray},
  frame=single,
  framerule=0pt,
  basicstyle=\ttfamily,
  columns=fullflexible}}{}

\newcommand{\delay}[1]{\mathcal{D}_{#1}}
\renewcommand{\Re}{\operatorname{Re}}
\newtheorem{app}{Approximation}[section]
%%%%%%%%%%%%%%%

\pagenumbering{arabic}

\begin{titlepage}
\thispagestyle{fancy}

%\lhead{\begin{tabular}{c m{0.4\textwidth}}  \multirow{2}{*}{\includegraphics[scale=0.3]{Figures/LISAConstellation.png}} & \Large{\textbf{Laser Interferometer}} \\   &  \Large{\textbf{Space Antenna}} \end{tabular} }
%\chead{}
%\rhead{\begin{tabular}{|m{0.2\textwidth}|m{0.15\textwidth}|} \hline \multicolumn{2}{|l|}{Ref : \ReportRef} \\ \hline Issue :  \ReportIssue &  Revision : \ReportRevision \\ \hline Date :  \today  & Page : \thepage / \lastpageref{LastPage}  \\ \hline \end{tabular}}

%\lhead{\begin{tabular}{c m{0.4\textwidth}}  \multirow{2}{*}{\includegraphics[scale=0.3,trim={0 0 2.5cm 0},clip]{Figures/Logo_LISA_ESA_1711.pdf}} & \Large{\textbf{Laser Interferometer}} \\   &  \Large{\textbf{Space Antenna}} \end{tabular} }
\lhead{\begin{tabular}{c m{0.4\textwidth}}  \multirow{2}{*}{
\includegraphics[scale=0.22]{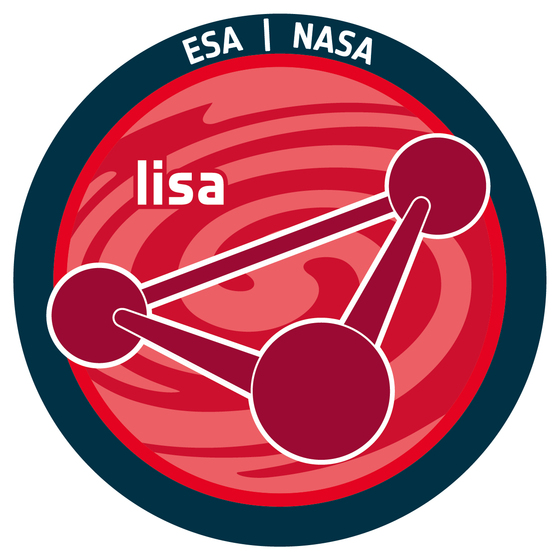}
\includegraphics[scale=0.07]{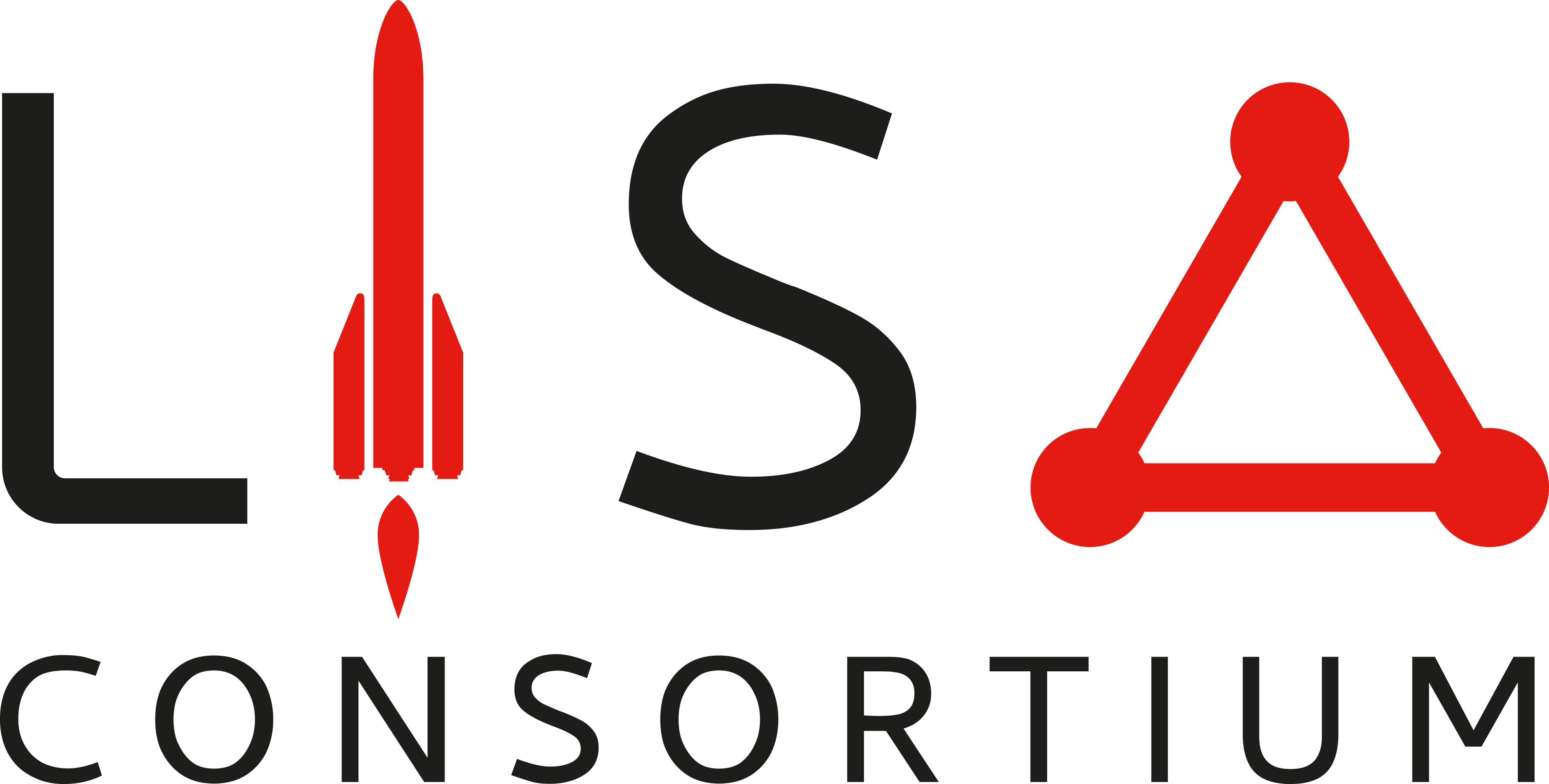}
} & \Large{\textbf{Laser Interferometer}} \\   &  \Large{\textbf{Space Antenna}} \end{tabular} }
\chead{}
\rhead{\begin{tabular}{|m{0.2\textwidth}|m{0.15\textwidth}|} \hline \multicolumn{2}{|l|}{Ref : \ReportRef} \\ \hline Issue :  \ReportIssue &  Revision : \ReportRevision \\ \hline Date :  2021/12/16  & Page : \thepage / \pageref{LastPage}  \\ \hline \end{tabular}}

\cfoot{{\tiny This document is the property of the \ac{LISA} Consortium and cannot be reproduced or distributed without its authorization.}}

\begin{center}

{\Huge \textbf{ \ReportTitle}}\\
\vspace{1cm}

\begin{tabular}{@{}|p{0.12\textwidth} | p{0.8\textwidth} | @{}}
\hline
 & \\
{\bf N/Ref}  : &  \ReportRef  \\
 & \\
\hline
 & \\
 {\bf Title} & \textbf{\ReportTitle} \\
 & \\
\hline
 & \\
 %%%%%%%%%%%%%%% ABSTRACT HERE  %%%%%%%%%%%%%%%
{\bf Abstract} & 
This Technical Note (LISA reference \ReportRef) describes the computation of the noise power spectral density, the  sensitivity curve and the signal-to-noise ratio for LISA (Laser Interferometer Antenna). It is an applicable document for ESA (European Space Agency) and the reference for the LISA Science Requirement Document.
%\cSB{I am not sure the title reflects the content, but let you decide.}
 %%%%%%%%%%%%%%% %%%%%%%%%%%%%%% %%%%%%%%%%%%%%%
\\
&
\\
\hline
\end{tabular}

\vfill

%\begin{tabular}{| m{0.18\textwidth} | m{0.28\textwidth} | m{0.12\textwidth} | m{0.22\textwidth} |}
%\hline
% & {\bf Name} & {\bf Date} & {\bf Signature} \\
%\hline
%Prepared by & S. Babak, M. Hewitson and A. Petiteau & \today &  \\[20pt]
%\hline
%Checked by & LISA Science Study Team & 2021/06/30 &  \\[20pt]
%\hline
%%Checked by (QA)  & &  &  \\[20pt]
%%\hline
%Approved by  & LISA Science Study Team & 2021/06/30  &  \\[20pt]
%\hline
%\end{tabular}

\begin{tabular}{| m{0.18\textwidth} | m{0.38\textwidth} | m{0.15\textwidth} |}
\hline
 & {\bf Name} & {\bf Date} \\
\hline
Prepared by & S. Babak (APC), M. Hewitson (AEI) and A. Petiteau$^1$ (CEA/IRFU/DPhP \& APC) & \today  \\[20pt]
\hline
Checked by & LISA Science Study Team & 2021/06/30  \\[20pt]
\hline
%Checked by (QA)  & &  &  \\[20pt]
%\hline
Approved by  & LISA Science Study Team & 2021/06/30  \\[20pt]
\hline
\end{tabular}

\end{center}
$^1$ contact: antoine.petiteau@cea.fr

%%%%%%%%%%%%%%%%%%%%%%%%%%%%%%%%%
\newpage

\lhead{\begin{tabular}{c m{0.4\textwidth}}  \multirow{2}{*}{
\includegraphics[scale=0.25]{Figures/Small_Lisa_logo.jpg}
\includegraphics[scale=0.07]{Figures/Standard_LISA_Consortium_RGB.png}
} & \large{\textbf{LISA}} \\   &  \large{\ReportShortTitle} \end{tabular} }

\begin{center}

 %%%%%%%%%%%%%%% CONTRIBUTOR  LIST  %%%%%%%%%%%%%%%
%\vspace{1.0cm}

{\Large \textbf{Contributor List}}\\
\vspace{0.5cm}

\begin{tabular}{|m{0.25\textwidth}| >{\centering\arraybackslash}m{0.3\textwidth}| >{\centering\arraybackslash} m{0.3\textwidth} |}
\hline
{\bf Author's name} & {\bf Institute} & {\bf Location } \\
\hline
Babak Stas     & APC & Paris-France \\
Petiteau Antoine     & CEA/IRFU/DPhP \& APC & Gif-sur-Yvette \\
Hewitson Martin     & AEI & Hannover-Germany \\
\hline

\hline
\end{tabular}
 %%%%%%%%%%%%%%% %%%%%%%%%%%%%%% %%%%%%%%%%%%%%%

\vspace{1.0cm}

{\Large \textbf{Document Change Record}}\\
\vspace{0.5cm}

 %%%%%%%%%%%%%%% TABLE FOR CHANGE RECORD  %%%%%%%%%%%%%%%
\begin{tabular}{|m{0.05\textwidth}|m{0.12\textwidth}|m{0.24\textwidth}|m{0.35\textwidth}|m{0.1\textwidth}|}
\hline
{\bf Ver}. & {\bf Date} & {\bf Author} & {\bf Description} & {\bf Pages} \\
\hline
\hline
0.0 & 2018/05/22 &  S. Babak (APC), M. Hewitson(AEI), A. Petiteau (APC)\ & First Version & all\\
\hline
0.1 & 2018/05/22 & M. Hewitson (AEI)\ & Criculate in SST + reviewers for comments & all\\
\hline
0.2 & 2018/08/13 & A. Petiteau (APC) & Implementing comments/corrections from D. Shoemaker, P. Jetzer and M. Colpi & all \\
\hline
0.3 & ... & S. Babak (APC) & Some small corrections and comments implemented & all \\
\hline
0.4 & 2020/01/08 & A. Petiteau (APC) & Update all results using new LDC convetions; update TDI part & all \\
\hline
0.6 & 2020/05/20 & A. Petiteau (APC),  S. Babak (APC), M. Hewitson (AEI) & Add executive summary, more justification on sensitivity & all \\
\hline
0.7 & 2020/07/06 & S. Babak (APC) & Reorganisation & all \\
\hline
0.8 & 2021/01/19 & A. Petiteau (APC)  & Consolidation, Change of conventions & all \\
\hline
0.9 & 25/02/2021 & S. Babak (APC) & Rewriting semi-analytic response, adjusting Galactic foreground & all \\
\hline
%\ReportIssue.\ReportRevision & \today & \ReportCurrentAuthor\ (\ReportCurrentAuthorInstitute) & Consolidation (in progress) & all \\
0.10 & 19/03/2021 &  A. Petiteau (APC) & Rewriting semi-analytic response, adjusting Galactic foreground  & all \\
\hline
0.11 & 19/03/2021 &  A. Petiteau (APC) & Add small annexe deriving noise PSD from SciRD sensitivity  & appendix B \\
\hline
0.12 & 23/05/2021 &  A. Petiteau (APC) & Fix after review from SST & all \\
\hline
1.0 & 31/07/2021 &  A. Petiteau (APC) & First official release & all \\
\hline
1.1 & 16/12/2021 &  A. Petiteau (APC) & 2 small corrections on SGWB & 23-24, appendix D \\
\hline
\end{tabular}
 %%%%%%%%%%%%%%% %%%%%%%%%%%%%%% %%%%%%%%%%%%%%%

\vspace{1.0cm}

{\Large \textbf{Distribution list}}\\
\vspace{0.5cm}

 %%%%%%%%%%%%%%% DISTRIBUTION LIST  %%%%%%%%%%%%%%%
\begin{tabular}{|m{0.52\textwidth}| >{\centering\arraybackslash}m{0.2\textwidth}| >{\centering\arraybackslash} m{0.2\textwidth} |}
\hline
{\bf Recipient} & {\bf Restricted} & {\bf Not restricted} \\
\hline
LISA Consortium &  &  \ding{55} \\
\hline
\end{tabular}
 %%%%%%%%%%%%%%% %%%%%%%%%%%%%%% %%%%%%%%%%%%%%%

\end{center}

\end{titlepage}

% Set page counter to 3 (no reset after the title pages)
\setcounter{page}{3}

\tableofcontents  
\clearpage

%------------------------------------------------------------------------------

%\section*{Acronyms}
%
%\begin{acronym}
%\input{LISA_Acronyms.tex}
%\end{acronym}

\section{Executive Summary}

This note describes the calculations made to compute science performance for
 \ac{LISA} , in particular it presents a sensitivity curve that is in-line with the
formula given in the \ac{SciRD},\cite{SciRD}. 

\begin{figure}[htbp]
\centering
\includegraphics[width=0.8\textwidth]{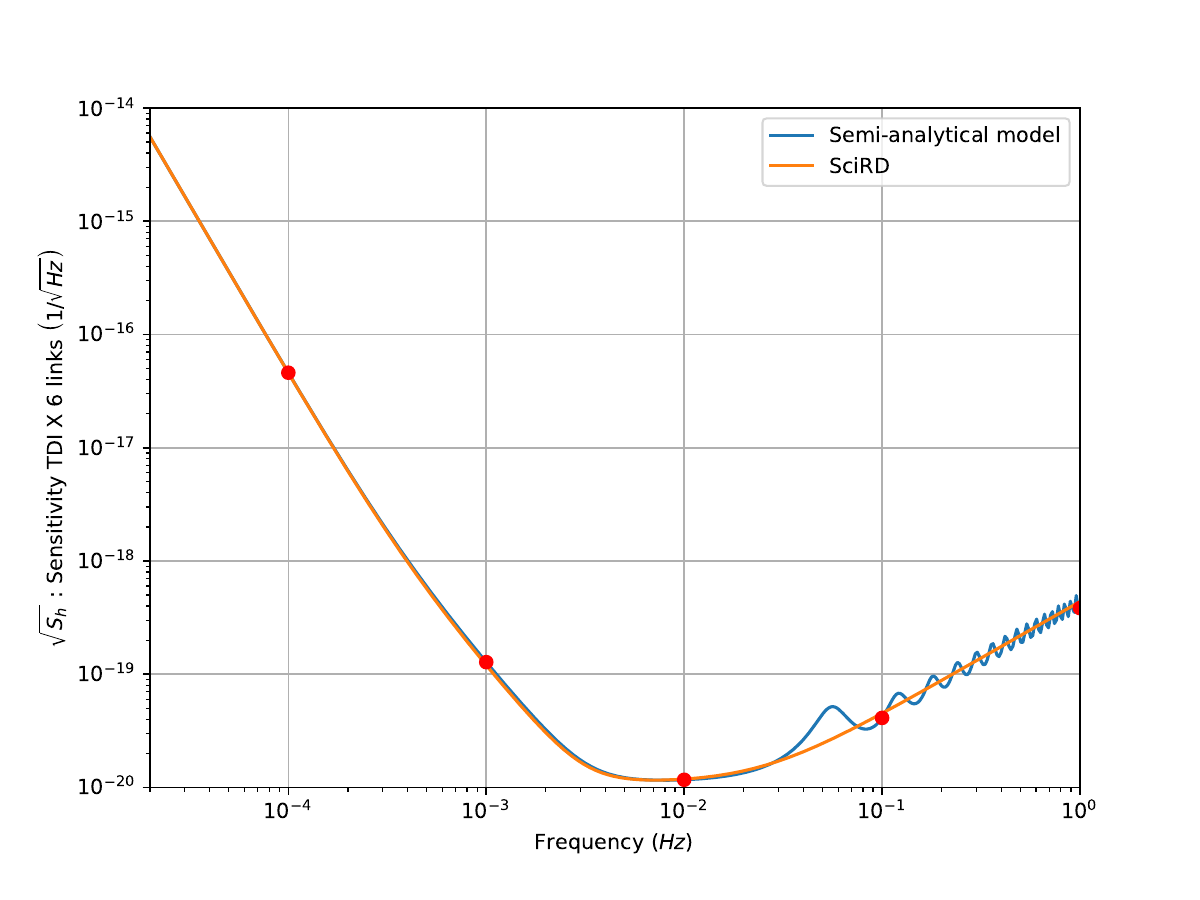}
\caption{A comparison of the \ac{SciRD}~\cite{SciRD} sensitivity formula (see appendix~\ref{app:SciRDNoise}) to the sensitivity model given in section~\ref{sec:Sensitivity}
for the 6-links configuration, i.e. X,Y and Z \ac{TDI}.
The plot also includes the points tabulated in Table~\ref{tab:StrainValues}.}
\label{fig:LISA_Sensitivity_Curves_Model}
\end{figure}

The formula for the required strain sensitivity curve is given in a way that is
implementation agnostic, i.e., it doesn't assume a particular observatory
configuration (interferometery, arm-length, number of links, etc). An
appropriate sky-averaged formulation of a sensitivity curve for the current set
of observatory parameters (6 interferometric links, each 2.5\,Gm long) is given
in Section~\ref{sec:Sensitivity}, and is shown graphically compared to the
\ac{SciRD} in Figure~\ref{fig:LISA_Sensitivity_Curves_Model}. This model assumes an
interferometer noise of 15\,pm$/\sqrt{ \rm Hz}$ and a test-mass acceleration
noise of 3\,$\textrm{fm}/\textrm{s}^2/\sqrt{\rm Hz}$ with the noise-shape
functions as described in that section. 

\textit{Note: the slight differences between the sensitivity model and the formulaic
curve from the \ac{SciRD} arise from the difference in the arm-response models.
}

Most of the time we work in the geometrical units $G=c=1$, however we restore $G$, $c$ when their presence is not obvious.

All computations and figures can be reproduced from the associated notebook available at \\ 
\href{https://gitlab.in2p3.fr/LISA/lisa_sensitivity_snr}{https://gitlab.in2p3.fr/LISA/lisa\_sensitivity\_snr/}~\cite{gitlabLISASensitvitySNR}.

%------------------------------------------------------------------------------
%\section{Introduction}
%
%
%This note describes the calculations made to compute science performance for
%LISA, in particular it presents a sensitivity curve that is in-line with the
%formula given in the Science Requirements Document and looks at computation of
%the SNR of different sources. It details the calculations that lead to `the
%usual' averaging factors used in various codes, and goes on to verify those with
%numerical simulations. Various different codes are compared, including different
%waveform models.

\section{Define the sensitivity}

The definition of the sensitivity is closely related to the signal-to-noise ratio \ac{SNR}
which for the deterministic source we define as 
 
\be
SNR^2  = 4 \Re \left( \int_{0}^{f_\textrm{max}}  df \frac{\tilde{X}\tilde{X}^*}{S_n(f)} \right),
\label{eq:SNR2}
\en
where we have introduced the one-sided {\bf noise} power spectral density  $S_n(f)$ and the signal 
$\tilde{X}(f)$ (in Fourier domain) as it appears at the detector's output plus post-processing (\ac{TDI} for example).
The noise \ac{PSD} is defined as 
 \be
E[\tilde{n}(f)\tilde{n}(f')] = \frac1{2} \delta(f-f') S_n(f)
 \en
 where $E[...]$ is the expectation value,
 $\tilde{n}(f)$ is the noise in Fourier domain
  and $\delta$ is the Dirac delta distribution.  
 For completeness we'll write it in the time domain as well:
 \be
E[n(t)n(t')] = \int_{-\infty}^{+\infty}df \  S_n(|f|) e^{-i2\pi f(t-t')}.
 \en  
  
The measurement  $\tilde{X}(f)$ (signal)  contains the detector's response and traces of any data post-processing (filters, \ac{TDI}, ...),  
we can write it as:
\be
\tilde{X}(f) = R_{+}(f, t, \theta_s, \phi_s, \psi) \tilde{h}_{+}(f) + R_{\times}(f, t, \theta_s, \phi_s, \psi) \tilde{h}_{\times}(f)
\en
where $h_{+, \times}$ is a GW signal in the source/radiation frame  and $R_{+,\times}$ is the detector's (LISA)
response to each polarization which depends on the sky location of the source $\theta_s, \phi_s$ and on polarization angle  $\psi$.
In general, response is a function of instantaneous frequency and the corresponding time (defining \ac{LISA}'s position on the orbit).  
We define the sensitivity using the ${\rm SNR}^2$ averaged over the sky and polarization: 

\be
<SNR^2> = 4 \Re \left( \int_{0}^{f_{\rm max}} df  \frac{\tilde{h}_{+}^2 +  \tilde{h}^2_{\times}}{S_n(f)/<|R_L|^2>} \right)
\label{eq:AvSNR2}
\en
where $<...>$ we use for the polarization and sky averaging: 
\be
<...> = \frac1{2\pi}\int_0^{2\pi}d\psi \frac1{4\pi} \int d^2\Omega ...
\en
and we have introduced $<|R_L|^2> = <|R_{+}|^2> = < |R_{\times}|^2>$ (see section~\ref{sec:AverageRespGWSemAn1} for detailed on the averaged antenna response functions). We {\bf define the sensitivity} as 

\be
S_{h}(f) = \frac{S_n(f)}{<|R_L|^2>}
\label{eq:SensDef}
\en
this definition could be extended to the isotropic stochastic signal with \ac{PSD} in each polarization $P_{+}, \; 
P_{\times}$:

\be
SNR^2 = T\int\,df \frac{P_{+}(f) + P_{\times}(f)}{S_n(f)/<|R_L|^2>}
\label{Eq:stochSens}
\en

\section{Instrumental noise}
\label{sec:InstrumentNoise}

The computation of the sensitivity curve presented above (see figure~\ref{fig:LISA_Sensitivity_Curves_Model}) is detailed in section~\ref{sec:SensitivityDefRef}. It should be essentially what we have in the
\ac{SciRD}~\cite{SciRD}, though the high-frequency wiggles are deliberately taken out in the SciRD. 
It corresponds to a simplified observatory with a noise described by only two 
components (see appendix~\ref{app:SciRDNoise} for correspondance between sensitivity and noise \ac{PSD}).

The first one is the high frequency noise component, given as a displacement noise 
of ${\rm 15 \ pm} / \sqrt{\rm Hz} $ rising with $f^2$, and described as 
\begin{eqnarray}
\sqrt{S_{OMS}}(f) & = & 15\,\left[\frac{\textrm{pm}}{\sqrt{\textrm{Hz}}}\right] 
\sqrt{1 + \left(\frac{2\times10^{-3}}{f}\right)^4}, \label{eq:SomsDisp}\\
\sqrt{S^{OMS}_{d\nu/\nu}}(f) & = & 15\times 10^{-12}\, \frac{2 \pi f}{c} \left[\frac{1}{\sqrt{\textrm{Hz}}}\right]
\sqrt{1 + \left(\frac{2\times10^{-3}}{f}\right)^4} \label{eq:SomsRelFreq}
\end{eqnarray}
The second one is the low frequency noise component, 
given as an acceleration noise of 3\,\si{fm.s^{-2}/\rthz}, rising with $f^2$ below \SI{0.4}{\milli\hertz},
and described as
\begin{eqnarray} 
\sqrt{S_{acc}}(f) & = &
3\,\left[\frac{\textrm{fm}.\textrm{s}^{-2}}{\sqrt{\textrm{Hz}}}\right]
\sqrt{1 + \left(\frac{0.4\times10^{-3}}{f}\right)^2}
\sqrt{1 + \left(\frac{f}{8\times10^{-3}}\right)^4} \label{eq:SaccAcc} \\
\sqrt{S^{acc}_{dL}}(f) & = &
{3 \over \left( 2 \pi f \right)^2 }\,\left[\frac{\textrm{fm}}{\sqrt{\textrm{Hz}}}\right]
\sqrt{1 + \left(\frac{0.4\times10^{-3}}{f}\right)^2}
\sqrt{1 + \left(\frac{f}{8\times10^{-3}}\right)^4}, \label{eq:SaccDisp} \\
\sqrt{S^{acc}_{d\nu/\nu}}(f) & = &
{3\times 10^{-15} \over  2 \pi f c}\,\left[\frac{1}{\sqrt{\textrm{Hz}}}\right]
\sqrt{1 + \left(\frac{0.4\times10^{-3}}{f}\right)^2}
\sqrt{1 + \left(\frac{f}{8\times10^{-3}}\right)^4}. \label{eq:SaccRelFreq}
\end{eqnarray}

The formulation \eqref{eq:SaccAcc} is in unit of acceleration, 
\eqref{eq:SomsDisp} and \eqref{eq:SaccDisp} in unit of displacement
and \eqref{eq:SomsRelFreq} and \eqref{eq:SaccRelFreq} in unit of relative frequency.

%The quantities we have outlined correspond to SciRD, for sake of comparison the MRD (mission requirement 
%document) adobes  $\sqrt{S_{OMS}} = 10\, \textrm{pm}/\sqrt{\textrm{Hz}}$,  
%$\sqrt{S_{acc}} = 2.4\, \textrm{fm s}^{-2}/\sqrt{\textrm{Hz}}$.

\subsection{Noise \ac{PSD}}
In the following we are describing the analytic model of the one-sided noise \ac{PSD}, $S_n(f)$,
and how it is derived from the noise components in each  interferometric measurements.
We will explicitly show $S_n(f)$  for a Michelson-type \ac{TDI} generator.
The one-sided noise \ac{PSD} is used in the matched filtering and, as a result, enters the likelihood function and  \ac{SNR}. 

%In this TN, it is used in the \ac{SNR} evaluation with Monte-Carlo Numerical
%computation ('PhD Num') and in \ac{TDI} average estimation ('PhD AvTDI')

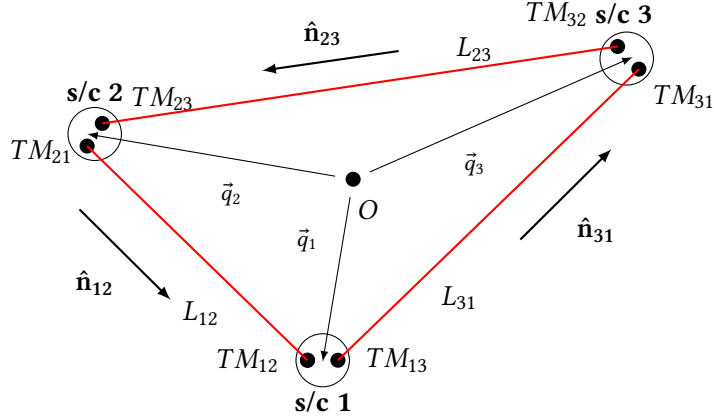
\begin{figure}[h]
	\label{F:LISAconstel} 
	\centering
	\begin{tikzpicture}
		\begin{scope}[scale=2]
			\tikzset{
				vertex/.style = {
					circle,
					fill      = black,
					outer sep = 4pt,
					inner sep = 2pt,
				}
			}
			\tikzset{arrow data/.style 2 args={%
					decoration={%
						markings,
						mark=at position #1 with \arrow{#2}},
					postaction=decorate}
			}%
			% We need to adjust the bounding box manually
			% as the control points enlarge it.
			%\path[use as bounding box] (-1,-0.5) rectangle (1.5,5.5);
			
			\coordinate (o) at (0,0);
			% \node[below] at (o) {root};
			
			%			\draw node[vertex, label=below:{1}] (1) at (0,0) {};
			%			\draw node[vertex, label=above:{3}] (3) at (2,2) {};
			
			\draw (2,2) circle[radius=5pt];
			\node[label=above:{\bf s/c 3}] (3) at (1.98,2.1){};
			\draw node[vertex, label={[xshift=-0.8cm, yshift=-0.1cm]$TM_{32}$}](32) at (1.94,2.08) {};
			\draw node[vertex, label={[xshift=0.6cm, yshift=-0.9cm]$TM_{31}$}](31) at (2.08,1.93) {};

			%			\draw node[vertex, label=above:{2}] (2) at (-1.5,1.5) {};			
			\draw (-1.5,1.5) circle[radius=5pt];
			\node[label=above:{\bf s/c 2}] (2) at (-1.5,1.58){};
			\draw node[vertex, label={[xshift=-0.6cm, yshift=-0.6cm]$TM_{21}$}](21) at (-1.55,1.42) {};
			\draw node[vertex, label={[xshift=0.8cm, yshift=-0.2cm] $TM_{23}$}](23) at (-1.45,1.57) {};
			
			%			\draw node[vertex, label=below:{1}] ($TM_{13}$) at (0.1,0){};
			
			\draw (0,0) circle[radius=5pt] ;
			\draw node[vertex] ($TM_{12}$) at (-0.1,0) {};
			\draw node[vertex, label=left:{$TM_{12}$}](12) at (-0.1,0) {};
			\draw node[vertex, label=right:{$TM_{13}$}](13) at (0.1,0) {};
			\node[label=below:{\bf s/c 1}] (1) at (0,-0.07){};
			
			\draw node[vertex] (0) at (0.2, 1.2) {};
			label=[xshift=1.0cm, yshift=0.3cm]
			
			\node[] (A2) at (-1.42, 1.8) {};
			\node[] (B2) at (-1.8, 1.4){};
			
			\node[] (A3) at (1.82, 2.2) {};
			\node[] (B3) at (2.2, 1.8) {};
			%\draw [fill=black] (A) circle (1pt);
			% \draw [fill=black] (B) circle (1pt);

			%\draw[-latex, black, thick]    (-1.8, 1.0) -- (-1.1, 0.3);  %n3
			%\draw[-latex, black, thick]    (1.4, 0.6) --(2.3, 1.5);   % n2
			%\draw[-latex, black, thick]     (0.4, 2.2) -- (-1., 2.0); % n1
			\draw[-latex, black, thick]    (-1.6, 1.0) -- (-1., 0.4);  %n3
			\draw[-latex, black, thick]    (1.3, 0.8) --(1.9, 1.4);   % n2
			%			\draw[-latex, black, thick]     (0., 2.05) -- (-0.9, 1.9); % n1
			\draw[-latex, black, thick]     (0.5, 2.05) -- (-0.4, 1.92); % n1

			%			\node[label=$\bf{\hat{n}}_{23}$] (n1) at (-0.5, 1.95) {};  
			\node[label=$\bf{\hat{n}}_{23}$] (n1) at (-0.0, 1.95) {};  
			\node[label=$\bf{\hat{n}}_{31}$] (n2) at (1.8, 0.65) {}; 
			\node[label=$\bf{\hat{n}}_{12}$] (n3) at (-1.5, 0.3) {}; 
			
			%			\draw[red, thick] (0,0) -- (2,2);   % 1->3
			\draw[red, thick] (0.1,0) -- (2.08,1.93); % 1-3
			\draw[red, thick] (-0.1,0) -- (-1.55,1.42);   % 1->2
			\draw[red, thick] (1.94,2.08) -- (-1.45,1.57);   % 3->2
			
			\draw[-latex, black]  (0) -- (0.0, -0.01); % q1
			\draw[-latex, black]  (0) -- (-1.55, 1.5); % q2
			\draw[-latex, black]  (0) -- (2.04, 2.01); % q3
			%			\draw[-latex, black]  (0) -- (2); % q2
			%			\draw[-latex, black]  (0) -- (3); % q3
			
			\node[label=$L_{12}$] (L1) at (-0.8, 0.1) {};
			\node[label=$L_{31}$] (L2) at (0.9, 0.2) {};
			\node[label=$L_{23}$] (L3) at (1.0, 1.85) {};
			\node[label=$O$] (O) at (0.3, 0.8) {};
			
			%\node[scale=0.1,label=$\bf{\hat{q}}_1 $] (q1) at (-0.1, 0.1) {}; 
			\node[scale=0.8] (q1) at (-0.1, 0.8) {$\vec{q}_1 $}; 
			\node[scale=0.8] (q2) at (-0.6, 1.1) {$\vec{q}_2 $}; 
			\node[scale=0.8] (q3) at (1.0, 1.3) {$\vec{q}_3 $};   
		\end{scope}  
	\end{tikzpicture}
	
	\caption{\label{F:LISAconstel} Schematic representation of the LISA constellation. The bullet points inside of each spacecraft  denote the test masses (TMs). The spacecrafts are numbered and the direction of propagation of the laser light is explicitly labeled as subscripts. We do not detail the connection between TMs inside the s/c. }
\end{figure}

We refer to Fig.~\ref{F:LISAconstel} for labeling of the spacecrafts (s/c) and test masses (\ac{TM}).
We just consider the top-level noise sources here, i.e. the two 
components described above, namely 
$S_{acc_{ij}}$ representing the low-frequency noise on the \ac{MOSA} of 
spacecraft $i$ and facing spacecraft $j$ 
and $S_{OMS_{ij}}$ describing the high-frequency noise of the same \ac{MOSA}.
We refer to the Performance TN~\cite{PerformanceTN} for more details on the noise budget.

The reference measurement is the \ac{TDI} generator X given, for generation 1.5 and 2.0, by:
\begin{eqnarray}
X_{1.5} & = & 
\eta_{13} + D_{13}  \eta_{31} + D_{13}  D_{31} \eta_{12} + D_{13}  D_{31} D_{12} \eta_{21}
- \eta_{12} - D_{12} \eta_{21} - D_{12} D_{21} \eta_{13} - D_{12} D_{21} D_{13}  \eta_{31} 
\label{eq:TDI15}\\
X_{2.0} & = &
 \eta_{13} %
+ D_{13}  \eta_{31} %
+ D_{13}  D_{31} \eta_{12} %
+ D_{13}  D_{31} D_{12} \eta_{21} %
+ D_{13}  D_{31} D_{12} D_{21} \eta_{12} \nonumber \\ %
& & + D_{13}  D_{31} D_{12} D_{21} D_{12} \eta_{21} %
+ D_{13}  D_{31} D_{12} D_{21} D_{12} D_{21} \eta_{13} %
+ D_{13}  D_{31} D_{12} D_{21} D_{12} D_{21} D_{13}  \eta_{31} \\
& & - \eta_{12} 
- D_{12} \eta_{21} 
- D_{12} D_{21} \eta_{13}
- D_{12} D_{21} D_{13}  \eta_{31}
- D_{12} D_{21} D_{13}  D_{31} \eta_{13} \nonumber \\ %
& & - D_{12} D_{21} D_{13}  D_{31} D_{13}  \eta_{31} %
- D_{12} D_{21} D_{13}  D_{31} D_{13}  D_{31} \eta_{12} %
- D_{12} D_{21} D_{13}  D_{31} D_{13}  D_{31} D_{12} \eta_{21}
\label{eq:TDI20}
\end{eqnarray}
with $\eta_{ij}$ the cleaned science measurement \ac{MOSA} on spacecraft $i$ and receiving the signal from the s/c $j$ and $D_{ij} x(t) = x(t - L_{ij}/c)$ the delay 
operator along the link from s/c $j$ to s/c $i$.

Here we give the final expression for the noise \ac{PSD} with different level of approximation.
The detailed derivation is given in the Appendix~\ref{app:NoisePSDDerivation}.

\textbf{Approximation TDI-1.}: Armlength are constant, i.e. the delay operators are
commuting and we can use \ac{TDI} generation 1.5, 

\begin{eqnarray}
S_{n,X_{1.5}} = PSD_{X_{1.5}}  =   4 \sin^2 \left( \omega { L_{31} + L_{13} \over 2} \right) 
\left( S_{OMS_{12}} + S_{OMS_{21}} + 4 S_{acc_{21}} + 4 \cos^2 \left( \omega { L_{12} + L_{21} \over 2} \right) S_{acc_{12}} \right) \nonumber \\
+ 4 \sin^2 \left( \omega {L_{21} + L_{12} \over 2} \right) \left( S_{OMS_{13}} + S_{OMS_{31}} + 4 S_{acc_{31}} + 4 \cos^2 \left( \omega { L_{31} + L_{13} \over 2} \right) S_{acc_{13}} \right)
\label{eq:PSDnoiseX15}
\end{eqnarray}

\textbf{Approximation TDI-2.}: All armlength equal : $L_{ij} = L$.
With this approximation, the \ac{PSD} is:
\begin{eqnarray}
S_{n,X_{1.5}} = PSD_{X_{1.5}}  =   4 \sin^2 \left( \omega L \right)
\left( S_{OMS_{12}} + S_{OMS_{21}}  + S_{OMS_{13}} + S_{OMS_{31}}
+ 4 \left( S_{acc_{21}} + S_{acc_{31}}
+ \cos^2 \left( \omega L \right)  \left( S_{acc_{12}} + S_{acc_{13}}  \right) \right) 
\right) \nonumber \\
\label{eq:PSDnoiseX15_EqArm}
\end{eqnarray}

\textbf{Approximation TDI-3.}: All noises of the same type have the same \ac{PSD} :
$S_{OMS_{ij}}= S_{OMS}$ and $S_{acc_{ij}} = S_{acc}$
With this approximation, the \ac{PSD} is:
\begin{eqnarray}
S_{n,X_{1.5}} = PSD_{X_{1.5}}  =   16 \sin^2 \left( \omega L \right)
\left( S_{OMS} + \left( 3 + \cos \left( 2 \omega L \right)  \right)  S_{acc} \right)
\label{eq:PSDX15EqLSameLevel}
\end{eqnarray}

\begin{figure}[htbp]
	\centering
	\includegraphics[width=0.8\textwidth]{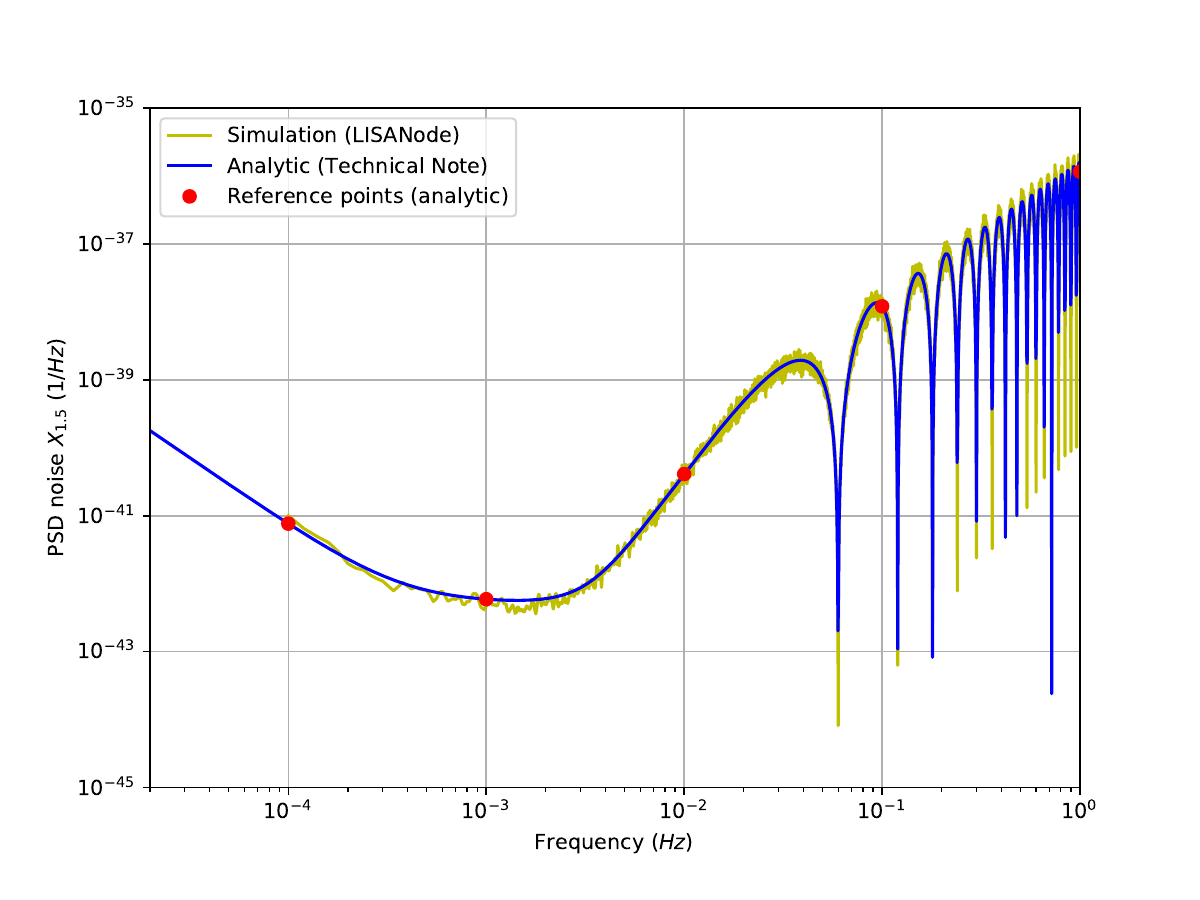}
	\caption{Noise of \ac{TDI} $X_{1.5}$: comparison between numerical simulation with the simulator
		LISANode~\cite{2019PhRvD..99h4023B,bayle:tel-03120731} (version 1.2; varying armlength 
		according to the keplerian orbit),
		LDC and analytical formulation~\eqref{eq:PSDX15EqLSameLevel}.
		The plot also includes the points tabulated in Table~\ref{tab:TDINoiseValues}.
		The discrepancy of the simulation at high frequency is due to the low sampling rate and simplified anti-aliasing filtering used for this particular simulation.}
	\label{fig:PSDX15_Num_An_Ref}
\end{figure}

The figure~\ref{fig:PSDX15_Num_An_Ref} shows the comparison for TDI 1.5 
between numerical simulation by the simulator
LISACode~\cite{2006AIPC..873..633P,petiteau:tel-00383222} and the analytical formulation~\eqref{eq:PSDX15EqLSameLevel}.

In reality, in order to suppress the laser noise we need to consider the flexing of the arms, and, therefore,  use \ac{TDI} $X_{2.0}$: 

\begin{eqnarray}
S_{n,X_{2.0}} = PSD_{X_{2.0}}  =   64 \sin^2 \left( \omega L \right) \sin^2 \left(2 \omega L \right)
\left( S_{OMS} + \left( 3 + \cos \left( 2 \omega L \right)  \right)  S_{acc} \right)
\label{eq:PSDX20EqLSameLevel}
\end{eqnarray}

\begin{figure}[htbp]
	\centering
	\includegraphics[width=0.8\textwidth]{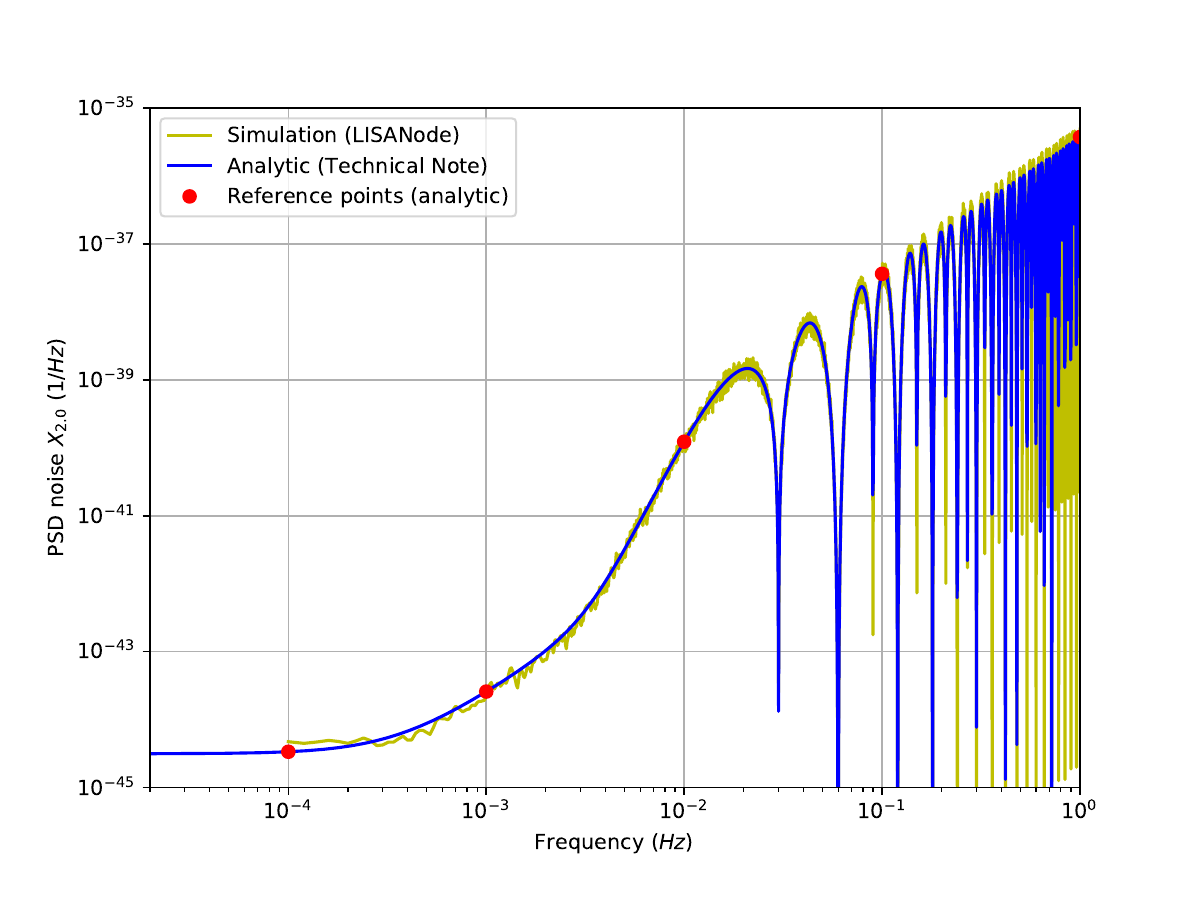}
	\caption{Noise of \ac{TDI} $X_{2.0}$: comparison between numerical simulation with
		LISANode,
		LDC 
		and analytical formulation~\eqref{eq:PSDX20EqLSameLevel}.
		The plot also includes the points tabulated in Table~\ref{tab:TDINoiseValues}.}
	\label{fig:PSDX20_Num_An_Ref}
\end{figure}

The figure~\ref{fig:PSDX20_Num_An_Ref} shows the comparison for TDI 2.0 between numerical simulation by
LISANode~\cite{2019PhRvD..99h4023B,bayle:tel-03120731}
%(see appendix~\ref{app:ImplementationLISANodeNoise} for implementation details) 
and the analytical formulation~\eqref{eq:PSDX20EqLSameLevel}.

The table~\ref{tab:TDINoiseValues} is providing the value for the reference 
frequency points.

\begin{table}[ht]
	\begin{center}
		\begin{tabular}{|c|c|c|} \hline
			Frequency (Hz) &  $PSD_{X_{1.5}, \rm SciRD}$ & $PSD_{X_{2.0}, \rm SciRD}$ \\ \hline \hline
			%0.00002 & 1.787856e-40 \\
			0.00010 & 7.645865e-42 & 3.358340e-45 \\
                         0.00100 & 5.901951e-43 & 2.580670e-44 \\
                         0.01000 & 4.099978e-41 & 1.227169e-40 \\
                         0.10000 & 1.206944e-38 & 3.628698e-38 \\
                         1.00000 & 1.152922e-36 & 3.737383e-36 \\
			\hline
		\end{tabular}
	\end{center}
	\caption{The numerical values of the \ac{PSD} of noises in \ac{TDI} $X_{1.5}$ and  $X_{2.0}$ for some fixed frequencies. 
	For reference we also give the \ac{PSD} for the curve given in the SciRD~\cite{SciRD}.}
	\label{tab:TDINoiseValues}
\end{table}

\section {GW response for X-Michelson \ac{TDI}}
\label{sec:RespGWX}

Here we will introduce the notation for GW signal as it is seen at the output of applying  \ac{TDI} procedure.  Most of our notations are presented schematically in Fig.~\ref{F:LISAconstel}, note that for the GW response 
we will always assume that $L_{sr} = L_{rs}=L$. 
The single link response (the laser light emitted by ``s"ender to the ``r"eceiver, 
traveling along link ``rs")
%~\footnote{there are several notations 
%floating in the literature: dropping the middle index ``{\it l}", or opposite dropping the first and last indices, the three index-notation contains redundancy but it is superseeds other notations and requires less of the brain work })
to GW is given in relative frequency as 
\be
y_{rs}^{GW} = \frac{\Phi_{rs}(t_s - \hat{k}.\vec{R}_s(t_s)) - \Phi_{rs}(t - \hat{k}.\vec{R}_r(t))}{2(1-\hat{k}.\hat{n}_{rs})}
\en
where $\hat{k}$ is direction of GW propagation, $\vec{R}_{s,r}$ vector position of a sender/receiver, $\hat{n}_{rs}$ a unit vector
connecting sender and receiver (link).
We have used the following notations
\be
\Phi_{rs} = \hat{n}_{rs} h_{ij} \hat{n}_{rs}
\en
which is projection of the GW strain on the  link, $t_s = t - |\vec{R}_r(t) - \vec{R}_s(t_s) | \approx t - L_{rs}$, and 
$L\hat{n}_{rs} \approx \vec{R}_r(t) - \vec{R}_s(t_s)$, using this approximation we have 
\be
y_{rs}^{GW} = \frac{\Phi_{rs}(t - \hat{k} . \vec{R}_s - L_{rs}) - \Phi_{rs}(t - \hat{k} . \vec{R}_r)}{2(1-\hat{k} . \hat{n}_{rs})}
\en
Substitute $h_{ij}$ decomposed in the polarization basis in the \ac{SSB} we get
\bea
\Phi_{rs} = \hat{n}_{rs} \hat{n}_{rs} \left[ 
h^{SSB}_{+}(\hat{u} \otimes \hat{u} - \hat{v} \otimes \hat{v}) + h^{SSB}_{\times}(\hat{u}\otimes \hat{v} + \hat{v}\otimes\hat{u}) 
\right]
\ena
The polarization basis is chosen as 
\bea
\hat{u} &=& -\hat{e}_{\phi} \propto \left(\frac{\partial \hat{k}}{\partial \lambda}\right) = \{ \sin{\lambda}, 
\cos{\lambda}, 0 \} \\
\hat{v} &=& -\hat{e}_{\theta} \propto \left( \frac{\partial  \hat{k}}{\partial \beta} \right) = \{-\sin{\beta}\cos{\lambda},  -\sin{\beta}\sin{\lambda}, \cos{\beta}\}
\ena
where $\beta$ is ecliptic latitude and $\lambda$ ecliptic longitude of the source. 
\be
\hat{k} = -\hat{e}_r = -\{ \cos{\beta}\cos{\lambda}, \cos{\beta}\sin{\lambda}, \sin{\beta}\}.
\en
 The polarizations in the \ac{SSB} are related to the strain 
in the source (radiation frame) via polarization angle $\psi$: 
\bea
h_{ij}^{SSB} &=& (h_+ \cos{2\psi} - h_{\times}\sin{2\psi})\epsilon^+_{ij}  + 
 (h_+ \sin{2\psi} + h_{\times}\cos{2\psi})\epsilon^+_{ij} \\
 \epsilon^+_{ij} &=& (\hat{u} \otimes \hat{u} - \hat{v} \otimes \hat{v})_{ij}\\
 \epsilon^{\times}_{ij} &=& (\hat{u} \otimes \hat{v} + \hat{v} \otimes \hat{u})_{ij}
\ena
Finally, to get the \ac{GW} response to TDI X for greneration 1.5 and generation 2.0,
we use the eqns~(\ref{eq:TDI15}) and~(\ref{eq:TDI20}) respectively 
without the noise in the measurement, i.e. $\eta_{ij} = y_{ij}$. 
We assume that the GW signal in the source frame can be presented as 
\bea
h_{+} = A_{+} e^{i\Phi(t)},\;\;\;\; h_{\times}=  -i A_{\times}
e^{i\Phi(t)}
\ena
assuming that we add complex conjugate to make it real. Then we can write
\bea
X^{GW}_{1.5} &=& (\omega L) \sin{(\omega L)} e^{-i[\Phi(t - \hat{k}\vec{R}_1) - \omega L]}\{ 
A_{+} [F_{13}^{+}\Upsilon_{13} - F_{12}^{+}\Upsilon_{12} ] + 
A_{\times} [F_{13}^{\times}\Upsilon_{13} - F_{12}^{\times}\Upsilon_{12} ]
\} \label{E:XGW1.5}\\
X^{GW}_{2.0} &=&  2i \sin{(2\omega L)} e^{-2 i \omega L} X^{GW}_{1.5}. \label{E:XGW2.0}\
\ena
We have used the following notations:
\bea
F_{rs}^{+} = \hat{n}_{rs}^i \hat{n}_{rs}^j \left[
\epsilon^{+}_{ij} \cos{2 \psi}  + \epsilon^{\times}_{ij}\sin{2\psi} 
\right],\,\,
F_{rs}^{\times} = \hat{n}_{rs}^i \hat{n}_{rs}^j \left[
- \epsilon^{+}_{ij} \sin{2 \psi} +  \epsilon^{\times}_{ij} \cos{2\psi}
\right]\\
\Upsilon_{rs} = Sinc\left[  \frac{\omega L}{2} 
(1 - \hat{k}.\hat{n}_{rs})\right]
e^{-i  \frac{\omega L}{2} 
	(1 - \hat{k}.\hat{n}_{rs})} + Sinc\left[  \frac{\omega L}{2} 
(1 - \hat{k}.\hat{n}_{sr})\right]
e^{-i  \frac{\omega L}{2} 
(3 + \hat{k}.\hat{n}_{sr})}
\ena
This form is convenient when computing the average response as we show in the next section.

%The $X_{1.5}$ \ac{TDI} is given as 
%\bea
%X^{GW}_{1.5} &=& D_{12}D_{31}D_{13} y_{13'2} + D_{13} D_{31}y_{231} + D_{13} y_{123} + y_{32'1} - \nonumber \\
%& &\left[  D_{12}D_{21}D_{13} y_{123} + D_{12}D_{21}y_{32'1} + D_{12}y_{13'2} + y_{231}
%\right],
%\ena
%where we have used again the delay operator $D_{13} y = y(t-L_{13})$
%of the link $\hat{n}_2$. 

%\section{Sensitivity: derivation}
%\label{sec:Sensitivity}

\section{Average GW response}
\label{sec:AverageRespGW}

By convention the sensitivity of a \ac{GW} instrument is computed with the
 response averaged over source polarization and sky position. 
 It is a useful form for calculating the observational capability.
 It can be computed using numerical simulation or semi-analytical formulation.

\subsection{Using numerical simulations}
\label{sec:AverageRespGWSimu}

\begin{figure}[htbp]
	\centering
	\includegraphics[width=0.8\textwidth]{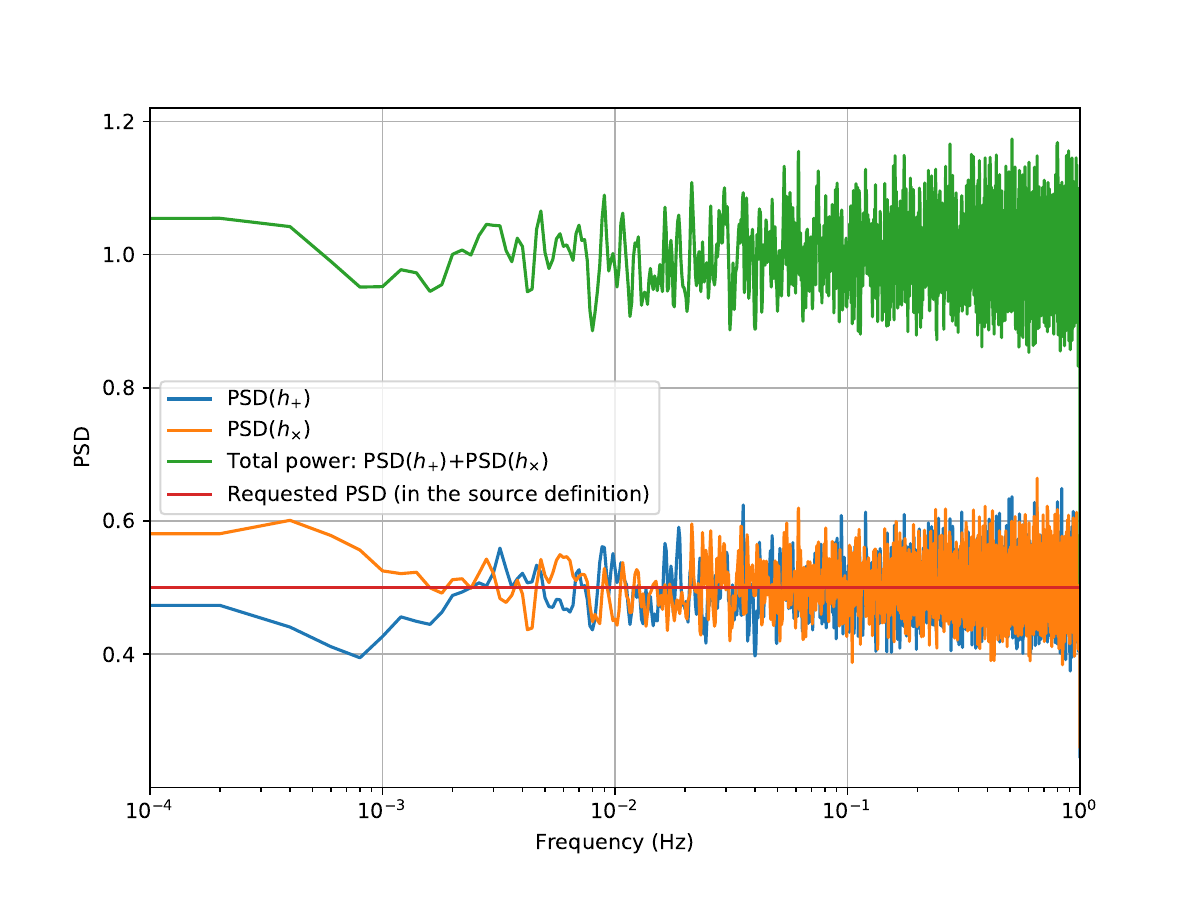}
	\caption{\ac{PSD}s of simulation for one \ac{SGWB} source with LISACode and configured
		as a white noise with at \ac{PSD}=0.5 .}
	\label{fig:PSDcheck_SGWB_1src}
\end{figure}

The easiest way to numerically compute the response of the instrument  averaged over polarization and sky
is to use~\eqref{Eq:stochSens}. We generate the point-like independent noise sources uniformly distributed on the sky and we randomly assign polarization angle to each source. If we use the noise level for each polarization component 
$1/2$, then the average \ac{PSD} of a generate signal 
$$
<\rm{P_{GW}}> = <|R_L|>(P_{+}(f) + P_{\times}(f) ) =   <|R_L|>
$$ 
so we directly obtain the average response which can be used in computing the sensitivity.
The advantage of using the white noise is that we cover all frequencies at once. 
\footnote{It's equivalent to computing the response to an isotropic flat Stochastic GW Background (SGWB)}.
The alternative derivation based on the monochromatic sources will be presented in the next section. 
The simulated \ac{PSD} for each source is shown in  figure~\ref{fig:PSDcheck_SGWB_1src}. 
We are numericlally averaging over polarization angle (performing monte carlo simulation) and we are simulating \ac{LISA} 
observation (\ac{TDI} computation) using LISACode and 192 white noise sources uniformly distributed over the sky .  
The \ac{PSD} of each source requested by the simulator is $PSD_{src} = \frac{PSD_{total}}{2 N_{src}}$. 

The figure~\ref{fig:PSDX15_RespGW_Num_An_Ref}  shows the numerical response for 
TDI $X_{1.5}$  according to the procedure described above.

\subsection{Semi-analytical computation}
\label{sec:AverageRespGWSemAn1}

A semi-analytical treatment for computing the average response to \ac{GW} is 
described in \cite{LarsonHellings2000}. 

We can reproduce it using the eqns.~(\ref{E:XGW1.5}),~(\ref{E:XGW2.0}) and using the definition~(\ref{eq:SensDef}). In other words we need to compute $<|X_{1.5}^{GW}|^2>$ .  The averaging over polarization and sky position can be done analytically for the long-wavelength approximation and will be discussed in section~\ref{sec:SensLW}. We introduce yet another notation:
\bea
F^{+}_{X} \equiv  \frac1{4} \left[ F_{13}^{+}\Upsilon_{13} - F_{12}^{+}\Upsilon_{12}\right], \;\; 
F^{\times}_X \equiv \frac1{4} \left[ F_{13}^{\times}\Upsilon_{13} - F_{12}^{\times}\Upsilon_{12} \right]
\ena
in order to get the compact form:
\bea
<|X_{1.5}^{GW}|^2> = (4\omega L)^2 \sin^2{(\omega L)}
	\left[ A_{+}^2 <(F^{+}_{X})^2> + 
	A_{\times}^2 <(F^{\times}_{X})^2> + 
	2A_{+} A_{\times}<F^{+}_{X}F^{\times *}_{X} +
	F^{+*}_{X}F^{\times}_{X}>  
	\right]
\ena
and note that $F^{+, \times}_{X} = F^{+, \times}_{X}(\psi, 
\hat{k}; t)$. Unfortunately we have to perform averaging numerically. It can be shown (i.e.~\cite{LarsonHellings2000})
that the averaged antenna response functions $<(F^{+, \times}_{X})^2>$ do not depend on time, moreover 
$<(F^{+}_{X})^2> = <(F^{\times}_{X})^2>$,
 and the cross-term averages to zero 
 $<F^{+}_{X}F^{\times *}_{X} +
 F^{+*}_{X}F^{\times}_{X}> = 0$. This implies:
 \bea
 <|X_{1.5}^{GW}|^2> = (4 \omega L)^2 \sin^2{(\omega L)}
 <(F^{+}_{X})^2> [ A_{+}^2  +  A_{\times}^2].
 \ena
Using definition ~(\ref{eq:SensDef}) we get for the response 
($X_{1.5}$) function to GW:

\be
<R_{L, X_{1.5}}(f)> =  (4\omega L)^2 \sin^2{(\omega L)}
<(F^{+}_{X})^2>.
\label{eq:RGWanalyticX15}
\en
Using the relation between 1.5 and 2.0 generation of TDI, given by eqn.~(\ref{E:XGW2.0}) we get the average response $X_{2.0}$ to \ac{GW}:
\be
<R_{L, X_{2.0}}(f)> =  (4\omega L)^2 \sin^2{(\omega L)}
(2 \sin{(2\omega L)})^2  <(F^{+}_{X})^2>.
\label{eq:RGWanalyticX20}
\en

%\be
%<R_{L,X_{2.0}}(f)> = \left( 4 \sin \left( {2 \pi f L \over c} \right) \right)^2 
%\left( 2 \sin \left( {4 \pi f L \over c} \right) \right)^2
%\left( {L \over c} \right)^2
%\left(2 \pi f\right)^2
%R_{\Sigma}(f, L) 
%\label{eq:RGWanalyticX20}
%\en

%The equation (45) of~\cite{LarsonHellings2000} gives the transfer function\footnote{The results 
%has to be slightly renormalised based the expected value ${ \sin^2{60^\circ} \over 5 } $.}, 
%$R_{\Sigma}$,
%for a standard Michelson for a measurement expressed in
%phase divided by the laser frequency and the light travel time along one arm $L/c$. 
%To convert it in unit of relative frequency fluctuation it has to be multiply by $(2 \pi f)^2$
%and $(L/c)^2$.
%To convert the measurement for a standard Michelson to a TDI  $X_{1.5}$ measurement, it has 
%to be multiply by $\left( 4 \sin \left( {2 \pi f L \over c} \right) \right)^2$. 
%Then, the average response $X_{1.5}$ to \ac{GW} is:
%\be
%<R_{L,X_{1.5}}(f)> = \left( 4 \sin \left( {2 \pi f L \over c} \right) \right)^2 
%\left( {L \over c} \right)^2
%\left(2 \pi f\right)^2
%R_{\Sigma}(f, L) 
%\label{eq:RGWanalyticX15}
%\en

\begin{figure}[htbp]
	\centering
	\includegraphics[width=0.8\textwidth]{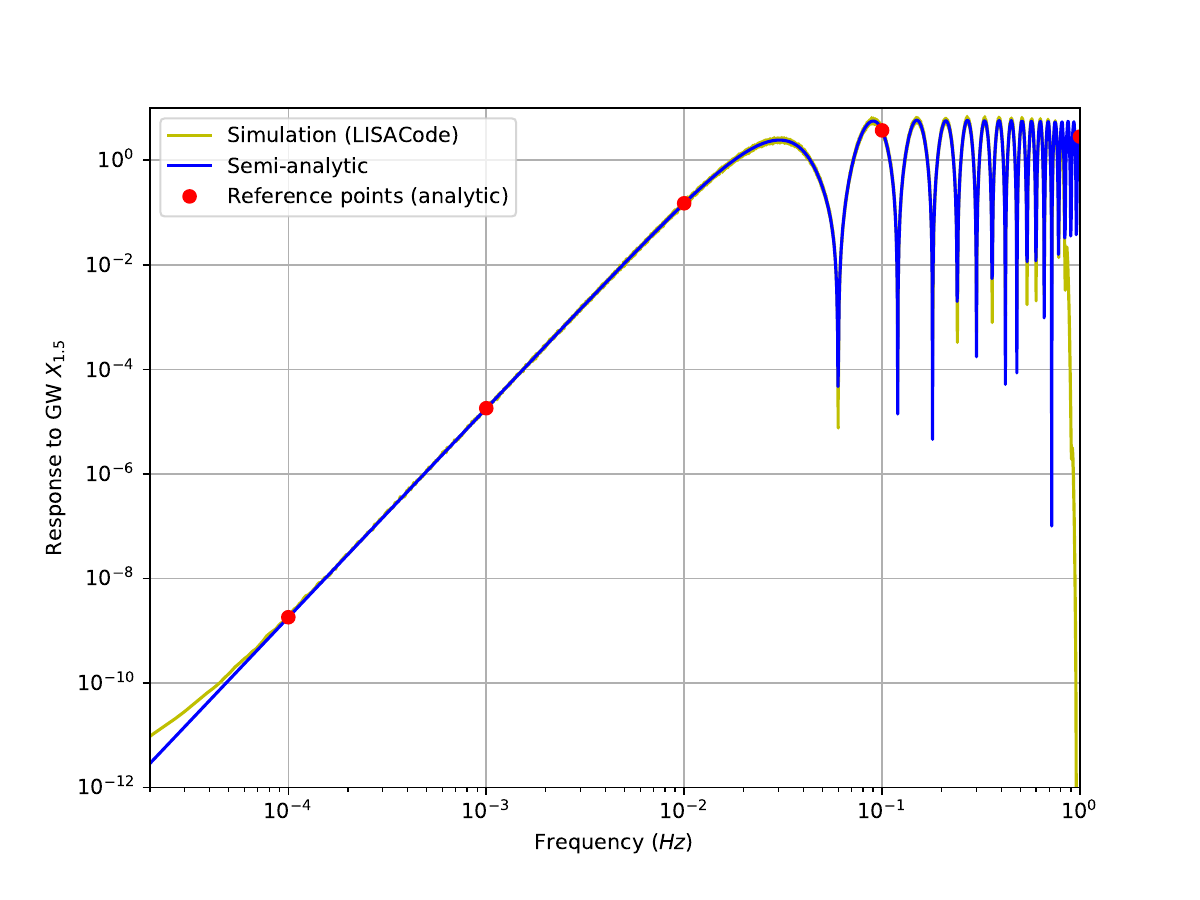}
	\caption{The numerical values of the response to GW for \ac{TDI} $X_{1.5}$ for
		some fixed frequencies. The yellow curve corresponds to the reference numerical result
		(valid for $ 0.2 mHz < f < 0.5 Hz$)  
		and the blue to the analytical version from equation~\eqref{eq:RGWanalyticX15}.
		The plot also includes in red the points tabulated in Table~\ref{tab:RespGWValues}.}
	\label{fig:PSDX15_RespGW_Num_An_Ref}
\end{figure}

The figure~\ref{fig:PSDX15_RespGW_Num_An_Ref} shows the comparison between numerical response
obtained for TDI $X_{1.5}$ and the semi-analytic treatment.
%\be
%<R^2_{L}(f)> = \left( 4 \sin \left( {2 \pi f L \over c} \right) \right)^2 { \sin^2{60^\circ} \over 5 } 
%L^2 \ T_{\rm arm}^2(f, L) 
%\left( {2 \pi f \over c} \right)^2
%\label{eq:RGWanalytic}
%\en
%where the $T_{\rm arm}(f)$ is obtained numerically as described in \cite{LarsonHellings2000}.
 Reference points are given in the Table~\ref{tab:RespGWValues}.

\begin{table}[ht]
	\begin{center}
		\begin{tabular}{|c|c|c|c|c|} \hline
			Frequency (Hz) &  Analytic approx. $X_{1.5}$ & $R_{\rm LISACode,X_{1.5}}$ &  Analytic approx. $X_{2.0}$ & $R_{\rm LISACode,X_{2.0}}$ \\ \hline \hline
0.00010 & 1.808897e-09 & 1.906063e-09 & 7.945155e-13 & 7.372150e-12 \\
0.00100 & 1.805606e-05 & 1.750219e-05 & 7.902030e-07 & 8.349357e-07 \\
0.01000 & 1.503154e-01 & 1.593552e-01 & 4.513257e-01 & 4.427565e-01 \\
0.10000 & 3.726972e+00 & 3.825690e+00 & 1.127471e+01 & 1.104438e+01 \\
1.00000 & 2.836169e+00 & - & 6.804830e+00 & - \\
			\hline
		\end{tabular}
	\end{center}
	\caption{The numerical values of the response of \ac{TDI} $X_{1.5}$ to GW for some fixed frequencies. }
	\label{tab:RespGWValues}
\end{table}

%To convert the measurement for a standard Michelson to a TDI  $X_{2.0}$ measurement, it has 
%to be multiply by $\left( 4 \sin \left( {2 \pi f L \over c} \right) \right)^2 \left( 2 \sin \left( {4 \pi f L \over c} \right) \right)^2$. 
%Then, the average response $X_{2.0}$ to \ac{GW} is:
%\be
%<R_{L,X_{2.0}}(f)> = \left( 4 \sin \left( {2 \pi f L \over c} \right) \right)^2 
%\left( 2 \sin \left( {4 \pi f L \over c} \right) \right)^2
%\left( {L \over c} \right)^2
%\left(2 \pi f\right)^2
%R_{\Sigma}(f, L) 
%\label{eq:RGWanalyticX20}
%\en

\begin{figure}[htbp]
	\centering
	\includegraphics[width=0.8\textwidth]{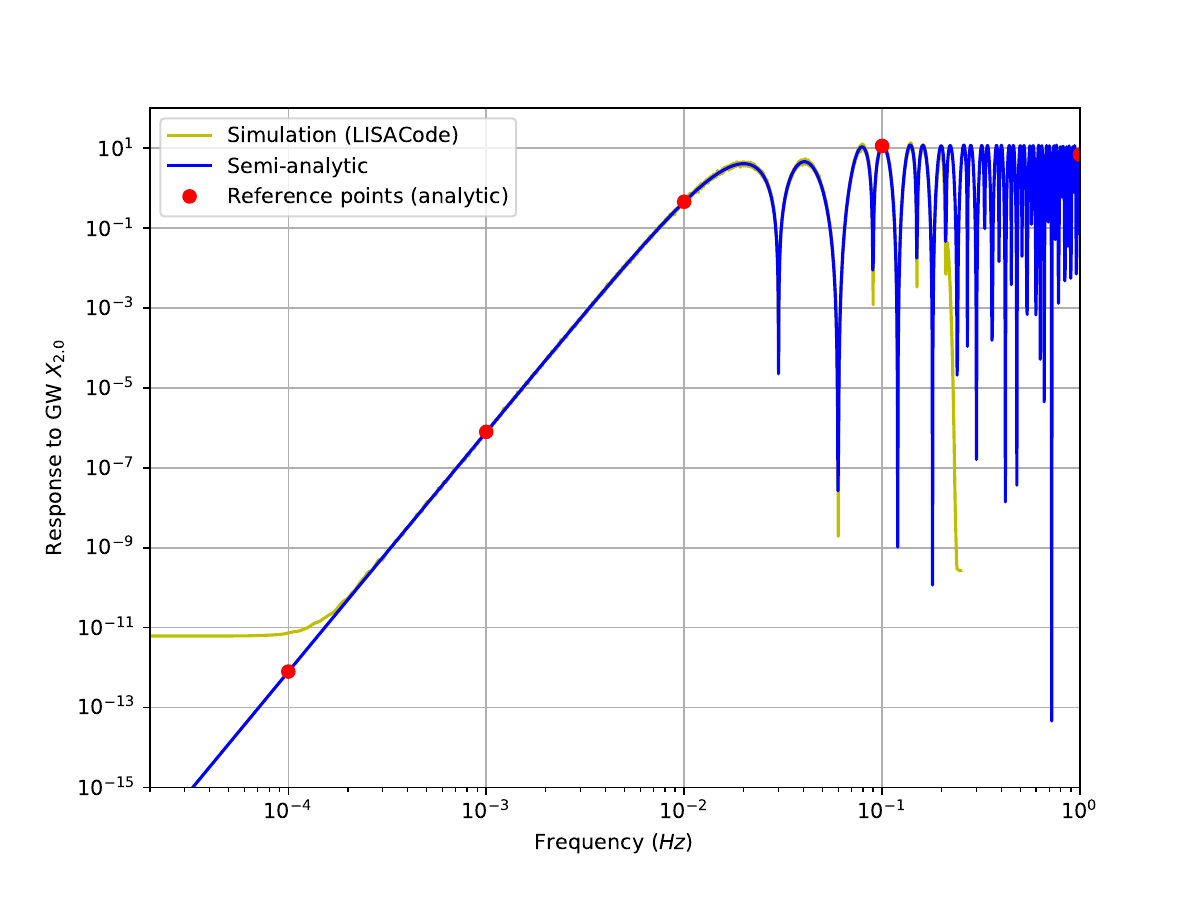}
	\caption{The numerical values of the response to GW for \ac{TDI} $X_{2.0}$ for
		some fixed frequencies. The yellow curve corresponds to the reference numerical result 
		(valid for $ 0.2 mHz < f < 0.1 Hz$) 
		and the blue to the analytical version from equation~\eqref{eq:RGWanalyticX20}.
		The plot also includes in red the points tabulated in Table~\ref{tab:RespGWValues}.}
	\label{fig:PSDX20_RespGW_Num_An_Ref}
\end{figure}

The comparison between the numerical and the semi-analytic response for TDI $X_{2.0}$ and the simulated data are shown on figure~\ref{fig:PSDX20_RespGW_Num_An_Ref}.
%\be
%<R^2_{L}(f)> = \left( 4 \sin \left( {2 \pi f L \over c} \right) \right)^2 { \sin^2{60^\circ} \over 5 } 
%L^2 \ T_{\rm arm}^2(f, L) 
%\left( {2 \pi f \over c} \right)^2
%\label{eq:RGWanalytic}
%\en
%where the $T_{\rm arm}(f)$ is obtained numerically as described in \cite{LarsonHellings2000}.
 Reference points are given in the Table~\ref{tab:RespGWValues}

%%%%%%%%%%%%%%%%%%%%%%%%%%%%%%%%%%%%%%%%%%%%
\section{Sensitivity}
\label{sec:Sensitivity}

\subsection{Definition and reference}
\label{sec:SensitivityDefRef}

The sensitivity is obtained using the averaged response for $X$ described in section~\ref{sec:AverageRespGW}. It is defined as (similar to~\eqref{eq:SensDef}): 
\bea
S_{h, X} = \frac{PSD_{X}}{<|R_{L,X}|^2>}
\label{eq:SensDef2}
\ena

We can also substitute explicitly the \ac{PSD} and the averaged response function from previous subsections (formulas
\eqref{eq:RGWanalyticX15}, \eqref{eq:PSDX15EqLSameLevel},\eqref{eq:RGWanalyticX20}, \eqref{eq:PSDX20EqLSameLevel}):
%\bea
%S_{h, X}  =  \frac{ S_{OMS} + \left( 3 + \cos \left( \frac{2 \omega L}{c} \right)  \right)  S_{acc} }
%{ \left( {\omega L \over c} \right)^2 \ R_{\Sigma}(f, L) }
%\ena
\bea
S_{h, X}  =  \frac{ S_{OMS} + \left( 3 + \cos \left( 2 \omega L \right)  \right)  S_{acc} }
{ \left( {\omega L} \right)^2 \ <(F^{+}_{X})^2> }
\ena
NOTE: The expression of the  sensitivity is the same for $X_{1.5}$ and for $X_{2.0}$ since the ratio of noise PSD and response to GW are identical, i.e 
\eqref{eq:PSDX15EqLSameLevel} over \eqref{eq:RGWanalyticX15} is equivalent to \eqref{eq:PSDX20EqLSameLevel} over \eqref{eq:RGWanalyticX20}.

\begin{figure}[htbp]
	\centering
	\includegraphics[width=0.8\textwidth]{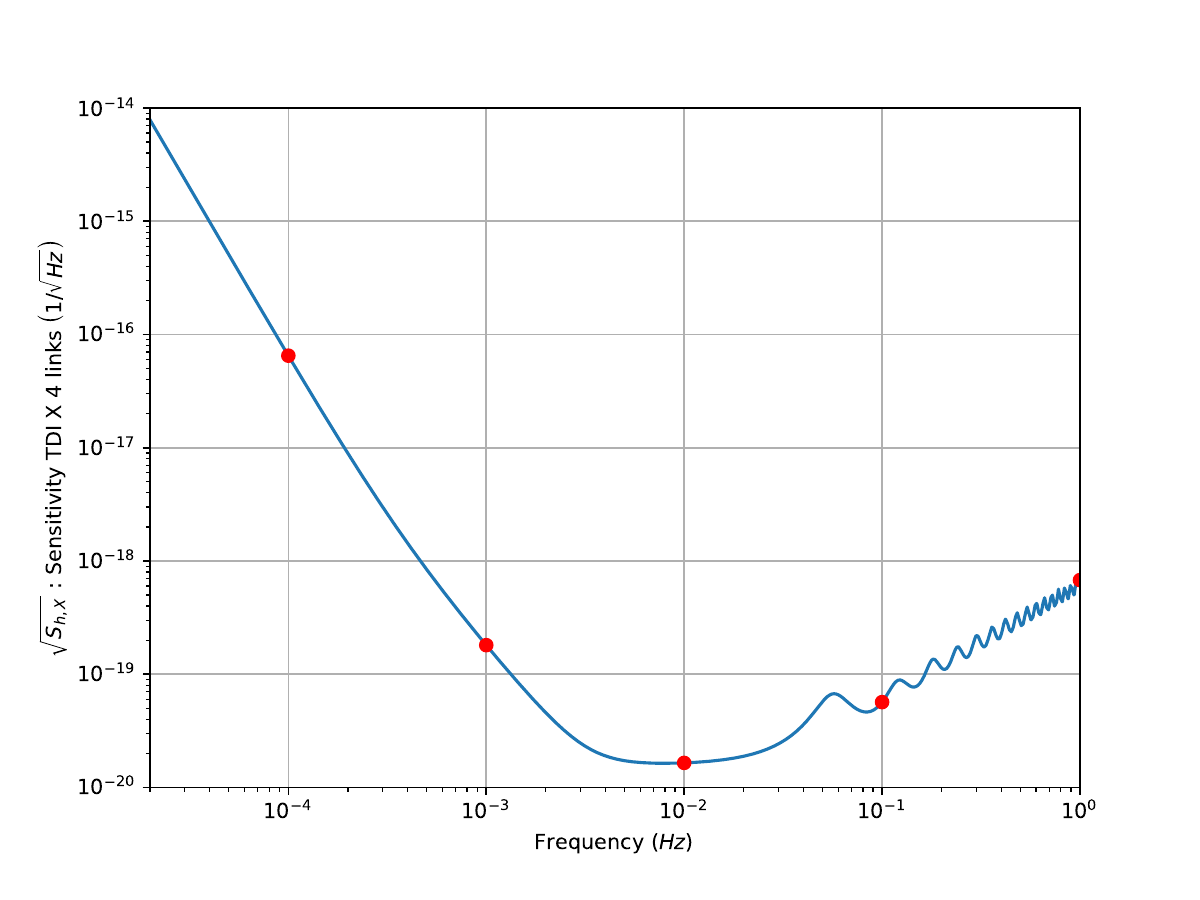}
	\caption{Sensitivity curve for a single TDI-X observable. The plot also includes the points tabulated in Table~\ref{tab:StrainValues}.}
	\label{fig:Sensitivty_X15_An_Ref}
\end{figure}

We on purpose kept the subscript $X$ to emphasize that this is a sensitivity for 4-links measurement which consists
 of only one Michelson \ac{TDI} combination. 
 For the 6 links (current \ac{LISA} configuration), we can form 3 Michelson \ac{TDI} combinations ($X, Y, Z$) 
 which are not independent as they share one arm. We can form another \ac{TDI} combination (referred as $A, E, T$)
 \cite{Prince} which have (under simplified assumptions adopted here) independent noise and roughly maximise to ``$+$", ``$\times$" polarizations and ``null" stream (free of GW signal). 
This interpretation is especially good for long-wavelength approximation $\omega L \ll 1$. At high frequencies all three \ac{TDI} combinations (X, Y, Z) have a similar response to a GW signal. The key point is that if we use combined \ac{SNR} (see appendix~\ref{app:CombinedSNR}) to define the sensitivity then 
\be
S_h = S_{h, X}/2. 
\label{eq:Sens6L}
\en
The strain sensitivity to only $X$, $\sqrt{S_{h, X}}$, is plotted in fig.~\ref{fig:Sensitivty_X15_An_Ref} and some points are tabulated in 
table~\ref{tab:StrainValues}. 

\begin{table}[ht]
	\begin{center}
		\begin{tabular}{|c|c|c|} \hline
			Frequency (Hz) & $S_{h, X}$ & $S_{h}$ \\ \hline
			0.00010 & 4.227356e-33 & 2.113678e-33 \\
			0.00100 & 3.266037e-38 & 1.633019e-38 \\
			0.01000 & 2.719112e-40 & 1.359556e-40 \\
			0.10000 & 3.218770e-39 & 1.609385e-39 \\
			1.00000 & 4.598240e-37 & 2.299120e-37 \\
			\hline
		\end{tabular}
	\end{center}
	\caption{The numerical values of the strain sensitivity curve for some fixed frequencies.}
	\label{tab:StrainValues}
\end{table}

The strain sensitivity to 6 links, $\sqrt{S_{h}}$, is plotted in fig.~\ref{fig:LISA_Sensitivity_Curves_Model} and some points are tabulated in 
table~\ref{tab:StrainValues}. 

\subsection{Sensitivity of ground-based detectors}

We start with a slight detour to show the sensitivity (as we have defined it) for LIGO-like detectors.
The LIGO response is usually defined in the frame associated with the beam splitter and the characteristic wavelength 
of the \ac{GW}s observed on the ground is usually larger than the size of the interferometer (for LIGO and other 3-4km detectors). So we can choose approximately an inertial frame covering the whole measuring device which simplifies the calculations. 
However, for the sake of comparison to \ac{LISA}, 
we will use the ``TT" (transverse-traceless) frame which can be seen as ``co-moving" with the \ac{GW}. 
In this frame the coordinate, distance between the beam splitter and the end mirrors does not change 
(but the proper distance does!): instead we can attribute the interaction with the \ac{GW} to the blue/red shift 
of the laser frequency. To compute this effect one needs to integrate the propagation of laser's photon 
in a field of (weak) \ac{GW}: $g^{\alpha\beta}{\phi^{l}}_{,\alpha}{\phi^{l}}_{,\beta}=0$, where $g^{\alpha\beta}$ is a metric  (for example $ds^2 = -dt^2 + [1+h_+(t-z)]dx^2 + [1-h_+(t-z)]dy^2 + dz^2$ is the metric of a GW propagating in $z$-direction and which has only $+$-polarization) and $\phi^l$ is the phase of the laser light.  We can compute the phase shift after the light has traveled  from the beam splitter to the end-mirrors (located along $x$ and $y$ axis) and back to the beam splitter (round trip):
\be
\Delta \phi^l = \phi^l|_x - \phi^l|_y = -\nu_l 
\left[-2(L_x - L_y ) + \frac1{2}H(t-2L_x) + \frac1{2}H(t-2L_y) -H(t) \right]
\en 
where $\nu_l$ is the laser nominal frequency,  $H(t)= \int_0^t h(t') dt'$ and $L_x, L_y$ are coordinate distances from the beam splitter to the end mirrors, and in case of LIGO/Virgo $L_x=L_y=L$ (see details in~\cite{2017mcpo.book.....T}). Note that this expression does not require the long wavelength approximation $\omega L \ll 1$, but it is the case for the GW sources observed on the ground so we have 
\be
\Delta \phi^l \approx 2\nu_l L h_{+}(t) 
\en
which allows us to define the \ac{GW} strain (for GW propagating in arbitrary direction and not necessarily linearly polarized): 
\be
h(t) = \frac{\Delta \phi^l}{2\nu_l L} = \frac1{2} (n^{(1)}_{i}n^{(1)}_{j} - n^{(2)}_{i}n^{(2)}_{i}) h^{ij} \equiv F_{+}h_{+} + F_{\times}h_{\times},
\label{Eq:LIGOstrain}
\en
where the factor half accounts for the round trip. If we assume that $h_{+, \times}$ above is defined in the source frame then the antenna function takes the form: 
\bea
F_{+} = - \frac1{2} (1+\cos^2\theta)\cos{2\phi}\cos{2\psi} - \cos{\theta}\sin{2\phi}\sin{2\psi} \label{eq:Fp}\\
F_{\times} = \frac1{2} (1+\cos^2\theta)\cos{2\phi}\sin{2\psi} - \cos{\theta}\sin{2\phi}\cos{2\psi} \label{eq:Fc}
\ena
We use the convention of the \ac{LDC}. 
\footnote{The other convention used in the litterature corresponds to $2 \psi \rightarrow \pi - 2 \psi$.}

Now we compute the square of \ac{SNR} for the monochromatic source:
\bea
SNR^2 = 4 \Re  \left( \int_0^{f_{\rm max}}\, df \frac{|\tilde{h}(f)|^2}{S_n(f)} \right)
\ena
and average it over the sky (uniform in $\phi, \, \cos{\theta}$) and over the polarization (uniform), taking into account that $<F_+ F_{\times}>=0$ and $<F_+^2> = <F_{\times}^2> = 1/5$:

\bea
<SNR^2> = 4 \Re \left(  \int_0^{f_{\rm max}}\, df \frac{\frac1{5}[\tilde{h}_{+}^2(f) + \tilde{h}_{\times}^2(f)]}{S_n(f)} \right).
\ena
Interestingly, in LIGO/Virgo,  the numerical factor $1/5$ is usually absorbed into the signal keeping $S_h = S_n$.
However according to our definition~\eqref{eq:SensDef}, $S_h(f) = 5S_n(f)$, so we differ by a constant factor from the LIGO usual $S_h$.
One needs to be careful when putting LIGO and \ac{LISA} sensitivity on the sample plot! 
The reason for keeping this factor in the numerator is that we can also average over the inclination  and define the sensitivity as the characteristic amplitude of the sky, polarization, inclination averaged monochromatic source which produces \ac{SNR}=$1$ at a given frequency.

\subsection{\ac{LISA} sensitivity in the long wavelength limit}
\label{sec:SensLW}

Now we are ready to derive the sensitivity for \ac{LISA} at  low frequencies ($\omega L \ll 1$). We assume only 4 links and use only X-TDI (analogue of the Michelson interferometer). The single link response in the long wavelength limit is 
\be
y \approx -\frac{L}{2} n_{rs}^i n_{rs}^j  \dot{h}_{ij}(t-\hat{k}\vec{R}),
\en
where the dot denotes the time derivative. 
%and we have used notations introduced in the subsection \ref{sec:RespGWX}. 
For $X_{1.5}$ \ac{TDI} in the long wavelength limit we obtain 
\be
{X}_{1.5} \approx 2L^2 (n_{12}^in_{12}^j  - n_{13}^in_{13}^j  ) \ddot{h}_{ij}.
\en
It is convenient to switch to the \ac{LISA}-based frame (the source sky position and polarization vary in time) where we can write:  
\be
\tilde{X}_{1.5} \approx (4\omega L) \sin{\omega L} \frac{\sqrt{3}}{2} (F_{+} \tilde{h}_{+} + F_{\times} \tilde{h}_{\times}),
\en
where $\sqrt{3}/2$ comes from $\sin{(\pi/3)}$ - the angle between links and $F_{+,\times}$
are the antenna functions defined in the previous subsection.  Finally we get for the  average \ac{SNR}:

\be
<SNR^2> = 4 \Re \left( \int_{0}^{f_{\rm max}} (4\omega L)^2  \sin^2({\omega L})  \frac{3}{20}  \frac{\tilde{h}_{+}^2(f) + \tilde{h}_{\times}^2(f)}{S_{n,X_{1.5}}(f)}\, df \right).
\en 
This result of averaging is the same in moving or static LISA. 
The sensitivity at low frequencies ($\omega L \ll 1$):

\be
S_{h, X_{1.5}}^{LW}(f) = \frac{20}{3} \frac{S_{n,X_{1.5}}(f)}{(4\omega L)^2  \sin^2({\omega L})},
\en 
where $S_{n,X_{1.5}}$ is the noise \ac{PSD} given by expression (\ref{eq:PSDX15EqLSameLevel})
and this sensitivity corresponds to 4 links only (need additional factor 2 for 6-links \ac{SNR}, in other words we have $10/3$ factor for 6 links). One can get a very good approximation to the numerical sensitivity with the help of an additional factor:
%
%We have found that 
%we can further improve this long wavelength expression (push it to higher frequencies)  if we define: 
\be
S_{h, X_{1.5}}(f) \approx \frac{20}{3}
\left(1+0.6(\omega L)^2\right)
 \frac{S_{n,X_{1.5}}(f)}{(4\omega L)^2  \sin^2({\omega L})},
\label{Eq.LWsens}
\en  
and similarly for  TDI $X_{2.0}$
\be
S_{h, X_{2.0}}(f) \approx \frac{20}{3}
\left(1+0.6(\omega L)^2\right)
\frac{S_{n,X_{2.0}}(f)}{(4\omega L)^2  \sin^2({\omega L})
(2 \sin{(2\omega L)})^2}.
\label{eq:SensX2approx}
\en

%\be
%S_{h, X_{2.0}}(f) = \frac{20}{3} (2\omega L)^{-2} (2\sin(\omega L))^{-2} (2\sin(2 \omega L))^{-2} S_{n,X_{2.0}}(f).
%\label{Eq.LWsens}
%\en  

%Note: In the long wavelength approximation, the response to \ac{GW} for TDI $X_{1.5}$ and $X_{2.0}$ are simply $S_{h, X_{1.5}}/S_{n,X_{1.5}}$ and $S_{h, X_{2.0}}/S_{n,X_{2.0}}$.

%$S_h^{NSA}$ is defined in case of LIGO/VIRGO type response as $S_h$ (polrization, sky averaged) dicvided by an averaging factor
%(this is expression 30) 

\section{Computing the \ac{SNR} for black hole binaries}
\label{S:Theory}

We start with the definition of the waveform for inspiralling binaries in
the frequency domain. This covers  Black Hole Binaries (BHB) 
with sufficient frequency evolution. 
%We use geometrical units $G=c=1$.

%\subsection{Frequency domain signal, static \ac{LISA}}
%\label{sec:theorySPA}

\subsection{Stationary phase approximation  waveform}

In this subsection we will deal with the leading order \ac{SPA} waveform, defined as:
\bea
\tilde{h}_+(f) &=& A \frac{M_c^{5/6}}{D_L}\frac{(1+\cos^2{\iota})}{2} f^{-7/6} e^{i\Psi(f)},\nonumber \\
\tilde{h}_{\times}(f) &=& iA \frac{M_c^{5/6}}{D_L}\cos{\iota} f^{-7/6}
e^{i\Psi(f)},
\label{Eq:SPA}
\ena
where the chirp mass in the observer frame and amplitude are given by
$$
M_c = M\eta^{3/5}= \frac{(m_1 m_2)^{3/5}}{(m_1+m_2)^{1/5}},\;\;\;
A = \frac2{\pi^{2/3}}\sqrt{\frac5{96}}.
$$
In addition: $D_L$ is the luminosity distance, $\iota$ is
inclination of the orbital angular momentum of the source to the line of sight, and, $\Psi(f)$ is the phase in the
frequency domain which we will not need here. This corresponds to the
waveform in time domain
\bea
h_+(t) &=& \frac{2M_c^{5/3}}{D_L} (\pi f_{\rm GW})^{2/3}(1+\cos^2{\iota})
\cos{\Phi(t)},\\
h_{\times}(t) &=& \frac{2M_c^{5/3}}{D_L} (\pi f_{\rm GW})^{2/3}2\cos{\iota}
\sin{\Phi(t)}.
\ena

with $f_{\rm GW}$, the frequency of GW in the observer frame (twice the orbital  frequency).

NOTE: Very often the amplitude of cross-polarization is defined with opposite sign!

We have neglected higher order modes considering only the leading order ($2,\pm 2$), and we consider 
non-precessing binaries.  Moreover, the expressions \eqref{Eq:SPA} are valid only for early inspiral.

Another way of writing two polarizations of \ac{GW}s is via mode decomposion
(decomposition in spin-weighted spherical harmonics) defined as:
\bea
\tilde{h}_{+} &=& \frac1{2} \sum_{l\le 2}\sum_{m=-l}^{l}
Y^{(-2)}_{lm} \tilde{h}_{lm} + Y^{*(-2)}_{lm} \tilde{h}^*_{lm}(-f),\\
\tilde{h}_{\times} &=& \frac{i}{2} \sum_{l\le 2}\sum_{m=-l}^{l}
Y^{(-2)}_{lm} \tilde{h}_{lm} - Y^{*(-2)}_{lm} \tilde{h}^*_{lm}(-f).
\ena
The advantage of this form is that the angular dependence of gravitational radiation is absorbed in the spherical
harmonics. If there is no precession (the case which we consider here),
$h_{l,-m} = (-1)^l h^{*}_{lm}$. We will consider here only the dominant
harmonic ($l=2, m=\pm 2$), so that

$$
\tilde{h}_{22} = A_{22}e^{i\Psi},\;\;\;\;
\tilde{h}_{2-2} = A_{22} e^{-i\Psi}.
$$
The two polarizations in this case (for positive frequencies) are
\bea
\tilde{h}_+ &=& \frac1{2}\left( Y_{22} + Y^*_{2-2}\right) A_{22}e^{i\Psi},\\
\tilde{h}_{\times} &=& \frac{i}{2}\left( Y_{22} - Y^*_{2-2}\right)
A_{22}e^{i\Psi}.
\ena

Note that those expressions are more general (we did not specify $A_{22}(f)$, $\Psi(f)$) and 
applicable also to the leading harmonic of phenomenological IMR (Inspiral-Merger-Ringdown) waveforms.

\subsubsection{SNR  averaged over sky, polarization and  inclination
of a coalescing binary}

A quantity often used in LISA to assess the scientific performance is the\textbf{ \ac{SNR} of a binary system defined by intrinsic parameters (masses, spins)  and distance
but on average over sky, polarization and {\bf inclination}}.  

In case of the ground-based detectors the \ac{SNR} is computed as 

%We start with the static \ac{LISA}, this is a valid approximation for 
%the transient signals (merging massive black holes (MBHBs)) which duration  (detectable part of the signal) is shorter than the 
%characteristic time of \ac{LISA}'s motion (say, year). In this approximation \ac{LISA} is often considered as ``LIGO in space" and the \ac{SNR} 
%is computed as 
\be
SNR^2  = 4 \Re \left( \int_{0}^{f_\textrm{max}} \frac{\tilde{h}(f)\tilde{h}^*(f)}{S_n(f)} df \right),
\en
 where $\tilde{h}(f)$ is given by LIGO-type strain~(\ref{Eq:LIGOstrain}) and $S_n(f)$ is the noise (or sensitivity, 
 again, in LIGO-sense).   We can average $|\tilde{h}(f)|^2$ over inclination, polarization and sky analytically.  Introduce an angle between 
 two arms as $\gamma$ ($\gamma=\pi/2$ for LIGO-like detectors and $\gamma=\pi/3$ for \ac{LISA} and ET) and consider a single 
 Michelson measurement ($X$ only in case of \ac{LISA}). The averaging procedure implies 
 \bea
 <|\tilde{h}(f)|^2>_{\iota, \psi, \rm sky} = \frac1{4\pi}\int\, d^2\Omega \frac1{2\pi}\int_{0}^{2\pi}\, d\psi 
 \frac1{2}\int_{-1}^1\,d\cos(\iota) |\tilde{h}(f)|^2
 \ena
Introduce polarization-dependent amplitudes $a_{+}=(1+\cos^2{\iota})/2,\,\, a_{\times} =  \cos{\iota}$ then the averaging procedure applied to $\tilde{h} = F_{+}\tilde{h}_{+}(f) + F_{\times}\tilde{h}_{\times}(f) $ implies
\be
\left<(F_+a_+)^2 + (F_{\times}a_{\times})^2 \right>_{\iota, \psi, \rm sky} =  \frac{4}{25} \sin^2{\gamma}
\en
and for \ac{SPA} gives
\bea
\left<\tilde{h}(f)\right>_{\rm LIGO} &=& \frac1{\sqrt{30}}  \frac{M_c^{5/6}}{\pi^{2/3}D_L} f^{-7/6}
e^{i\Psi(f)}. \\
\left<\tilde{h}(f)\right>_\textrm{LISA} &=& \frac1{2\sqrt{10}}  \frac{M_c^{5/6}}{\pi^{2/3}D_L} f^{-7/6} e^{i\Psi(f)}.
\label{Eq:LIGO_likeH}
\ena
Similar averaging procedure applied to phenomenological waveform restricted to $(2,\pm2)$ mode only
\bea
h_{+} &=& \sqrt{\frac5{64\pi}} e^{2i\phi_0}(1+\cos^2{\iota}) A_{22}(f)e^{i\Psi(f)},\\
h_{\times} &=& i\sqrt{\frac5{64\pi}} e^{2i\phi_0}2\cos{\iota} A_{22}(f)e^{i\Psi(f)},
\ena
gives us 
\bea
\left<\tilde{h}(f)\right>_\textrm{LISA} = \frac1{2}\sqrt{\frac3{20\pi}}
A_{22}(f)e^{i\Psi(f)}.
\ena

%\subsubsection{SNR for optimally oriented source}

Another used quantity is the\textbf{ \ac{SNR} of the optimally oriented source}, implying that the inclination is $\cos{\iota}=1$
which maximizes $a_{+, \times}=1$. The averaging over the sky and polarization for the face-on source gives 
\be
\textrm{max}_{\iota} \left<(F_+a_+)^2 + (F_{\times}a_{\times})^2 \right>_{\psi, \rm sky} =  \frac{2}{5} \sin^2{\gamma}.
\en

Return back to \ac{SNR}. In case of LISA we have
\be
<SNR^2_X> = 4 \Re \left( \int_0^{f_\textrm{max}} (4\omega L)^2  \sin^2({\omega L}) <(F_X^{+})^2>  \frac{\tilde{h}_{+}^2(f) + \tilde{h}_{\times}^2(f)}{S_{n,X_{1.5}}(f)}\, df \right) = 
 4 \Re \left( \int_0^{f_\textrm{max}} \frac{\tilde{h}_{+}^2(f) + \tilde{h}_{\times}^2(f)}{S_{h,X}(f)}\, df \right).
\en
What is left is to average the expression above over inclination, 
for a simple case of a dominant harmonic $|\tilde{h}_{+,\times}| = 
a_{+,\times}A(f)$ we have:
\be
<SNR^2_X>_{\iota, \psi, \rm sky}  = 4\Re \left( \int_0^{f_\textrm{max}} 
<a_+^2 + a_{\times}^2>_{\iota} \frac{A^2(f)}{S_{h,X}(f)}\, df \right) = 
 4\Re \left( \int_0^{f_\textrm{max}} 
 \frac{\frac4{5}A^2(f)}{S_{h,X}(f)}\, df \right).
\en

In the long wavelength approximation and for SPA this translates into
\be
\left<SNR_{X}^2\right>_{\iota, \psi, \rm sky} = 4 \Re \left( \int_{0}^{f_{\rm{max}}} 
\frac{ \sin^2{\gamma} <F^2_{+}a^2_{+} + F^2_{\times}a^2_{\times}>_{\psi, \iota,\rm{sky}}} 
{\sin^2{\gamma} <F^2_{+}>_{\psi, \rm sky}S^{LW}_{h,X}(f)}\left(A \frac{M_c^{5/6}}{D_L}\right)^2 f^{-7/3}\,  df \right)
\label{Eq:SNR_staticLISA}
\en
where $S^{LW}_{h,X}$ is given by eqn.~(\ref{Eq.LWsens}). 
The term in the numerator is just $\left< \tilde{h}(f)\right>$ given above and 
the factor  $\sin^2{\gamma} <F^2_{+}>_{\psi, \rm sky}$ (which is 
	3/20 in case of \ac{LISA}) is often referred as ``de-averaging'' factor. 
We can get a similar expression for a single mode of PhenomIMR waveforms. Finally the combined \ac{SNR} for \ac{LISA} 
is obtained by $\left<SNR^2\right>_{\iota, \psi, \rm sky}= 2\left<SNR_{X}^2\right>_{\iota, \psi, \rm sky}$ or replacing 
$S_{h,X}(f)$  by $S_h(f)$ in denominator.

%\subsubsection{SNR for SPA an sensitivity}

%Return back to \ac{SNR} averaged over polarisation sky and inclination and note that we have use the noise \ac{PSD} in the denominator and this is also the sensitivity in LIGO-sense. 
%In case of \ac{LISA} we need to connect it to the sensitvity, but we cannot use expression \eqref{Eq.LWsens} even for LW approximation. We have used the LIGO-like strain in the numerator so we have to use the LIGO-like link between sensitivity 
%and \ac{PSD} which is $S_n(f) = S_{h,X}^{LISA} \left<|F_+|^2 + |F_{\times}|^2 \right>$ and we get for the \ac{SNR}
%for SPA:
%
%\be
%\left<SNR_{X}^2\right>_{\iota, \psi, \rm sky} = 4 \Re \int_{0}^{f_{\rm{max}}} 
%\frac{<F^2_{+}a^2_{+} + F^2_{\times}a^2_{\times}>_{\psi, \iota,\rm{sky}}} 
%{<F^2_{+} + F^2_{\times}>_{\psi, \rm sky}S_{h,X}(f)}\left(A \frac{M_c^{5/6}}{D_L}\right)^2 f^{-7/3}\,  df 
%\label{Eq:SNR_staticLISA}
%\en

In the  Appendix~\ref{app:SNR_SMBHB_TestCases} we compute \ac{SNR} for several reference MBHBs systems using eq.~(\ref{Eq:SNR_staticLISA}) and three phenomenological 
IMR models: PhenomA, PhenomC,  PhenomD, those models perform similarly for equal mass non-spinning  systems but could give very different results as we increase mass ratio and spins. This is expected:
 PhenomD~\cite{PhenomD} supersedes PhenomC and PhenomA as it  was built based on the extended 
set of numerical waveforms (up to mass ratio 16).

\section{Relation of the sensitivity to \ac{SNR}}

To answer if  and how the sensitivity $S_h(f)$ is connected to the \ac{SNR}, we consider a monochromatic GW source 
given in time domain  by two polarizations in the source frame: 

\bea 
	h_{+}(t) &=& h_0 \frac{1+\cos^2{\iota}}{2}\cos(\omega_0 t + \phi_0) \\
		h_{\times}(t) &=& h_0 \cos{\iota} \cos(\omega_0 t + \phi_0)
\ena
While transforming to the frequency domain we (i) consider only positive frequencies (ii) neglect harmonics which appear due 
to relative motion of the source and the detector (relevant for \ac{LISA}) which leads to :
\bea
\tilde{h}_{+}(f) &=& \frac1{2}h_0  \frac{1+\cos^2{\iota}}{2} e^{i\phi_0}\delta_T(f-f_0),\\
\tilde{h}_{\times}(f) &=& \frac{-i}{2}h_0  \cos{\iota} e^{i\phi_0}\delta_T(f-f_0),
\ena 
where the delta function should be understood in reality as a finite time approximation 
$\delta_T(f-f_0)  \approx T \textrm{Sinc}\left[ T(f-f_0)\right]$ and $T$ is the observation duration.
We will use again the \ac{SNR} in LIGO-sense as described by (\ref{Eq:SNR_staticLISA}). For LIGO-like 
detectors which use $S_h^{\rm LIGO}(f) = S_n(f)$ as the sensitivity we obtain 
\be
\left<SNR_{X}^2\right>_{\iota, \psi, \rm sky} = 
\frac{<F^2_{+}A^2_{+} + F^2_{\times}A^2_{\times}>_{\psi, \iota,\rm{sky}}} 
{S_h^{\rm LIGO}(f)} h_0^2 T  = \frac{\left(\frac2{5}h_0\right)^2 T}{S_h^{\rm LIGO}(f)},
\en
This says, that $\sqrt{S_h^{\rm LIGO}(f)}$ can be interpreted as the amplitude of polarization, inclination and sky
averaged monochromatic GW signal which gives $SNR=1$ after observing it for $T_{\rm obs} =T$ with a  single detector.  

Let us try to interprete \ac{LISA} sensitivity in a similar way. We consider static \ac{LISA} and use  (\ref{Eq:SNR_staticLISA}) 
taking into account that 
\bea
<F^2_{+}a^2_{+} + F^2_{\times}a^2_{\times}>_{\psi, \iota,\rm{sky}} = \frac4{25}\sin^2\gamma \\
<F^2_{+}>_{\psi, \rm sky}   =  <F^2_{\times}>_{\psi, \rm sky} =  \frac1{5}\sin^2\gamma
\ena
we obtain for sensitvity (combined out of two noise independent data streams):
\be
\left<SNR^2\right>_{\iota, \psi, \rm sky} = 
\frac{<F^2_{+}a^2_{+} + F^2_{\times}a^2_{\times}>_{\psi, \iota,\rm{sky}}} 
{<F^2_{+}>_{\psi, \rm sky}S_h(f)}\left(h_0^2 T\right)^2   =  
5  \frac{\left(\frac2{5}h_0\right)^2 T}{S_h(f)},
\en
Clearly it has different interpretations: (combined) \ac{LISA} sensitivity $\sqrt{S_h(f)}$ can be seen
as the  amplitude of polarization, inclination and sky
averaged monochromatic GW signal which gives a combined $SNR^2=5$ after observing it for $T_{\rm obs} =T$. We should however emphasize that we have defined ``polarization, inclination and sky
averaged monochromatic GW signal" excluding the factor $\sin^2{(\gamma)}$ for comparison with the ground-based GW detectors. 
If we include the angle between the arms into the definition of an average signal
$\left(\sin{(\gamma)}\frac2{5}h_0\right)$, then the combined 
\ac{SNR} should be $SNR^2=20/3$.

\section{Galactic confusion noise}
\label{S:GBs}

The numerical results for the Galactic confusion noise were first obtained
using in \cite{Confusion}  for several durations of
observation. % In this method  (Cormish \& Littenberg) 
The level of Galactic noise is assessed by generating the simulated data which contains  
the Galactic population of \ac{WD} binaries (based on a fiducial model \cite{Korol:2017qcx}). 
The bright sources are removed in a iterative manner using smooth fit in estimated \ac{PSD} (combination of the noise and 
GW signals).  The first  estimator of the \ac{PSD} will be severely biased by the loud GW signals, removal of the bright sources and 
re-evaluation of the \ac{PSD} is converging after about 8-10 iterations. As the result we have a number resolvable sources (assuming
 perfect removal of bright sources)  and the residual stochastic foreground.  NOTE that this procedure is sensitive 
 to (i) \ac{SNR} threshold assumed for identification and removal of the sources (ii) the method used to evaluate the smooth
 \ac{PSD} (running median or mean). The fit below was obtained assuming combined \ac{SNR} threshold $SNR>7$ and running mean 
 for smoothing \ac{PSD} estimator (see for details  Karnesis+ [2021], in preparation).

The analytic fit expression of the strain sensitivity curve is given as
	\bea
S_{\rm Gal} = A f^{-7/3} e^{-\left(\frac{f}{f_1}\right) ^{\alpha} }
\frac1{2} \left[1.0 + \tanh\left (-\frac{f-f_k}{f_2}\right) \right],
\label{eq:SGal}
\ena

where $A = 1.14\times 10^{-44}, \, \alpha=1.8, \,  f_2 = 0.31\, \rm{mHz}$ and $f_1, 
f_k$ depend on the observation time (we resolve and remove more signals for longer observation). 
\bea
\log_{10}(f_1) = a_1 \log_{10}(T_{obs}) + b_1,\,\,\,\,  
\log_{10}(f_k) = a_k \log_{10}(T_{obs}) + b_k
\ena
Here $T_{obs}$ is in years and $a_1,\, b_1$ are $-0.25,\,   -2.7$ and 
$a_k,\, b_k$ are $-0.27, -2.47$. 
 The first term in this expression corresponds to what we expect (\ac{PSD}) from the population of monochromatic GW sources. 
 The second term reflects fact that we have fewer sources at high frequencies and that the bright sources are removed. 
The last term is determined by the data analysis: $f_k$ is a characteristic frequency above which we should be able to resolve and remove all GB sources.

\begin{figure}[htbp]
	\centering
	\includegraphics[width=0.8\textwidth]{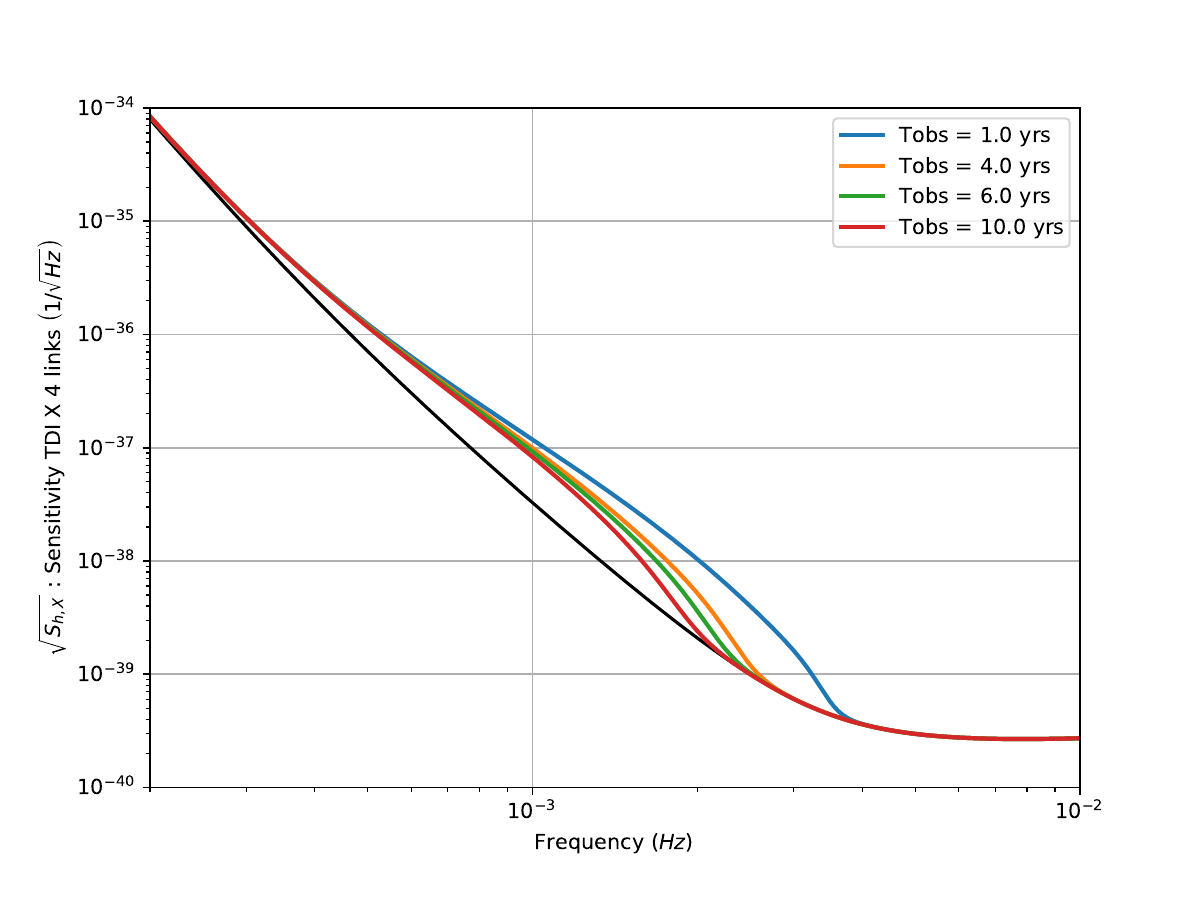}
	\caption{Analytic fits to the numerical results for estiamtion of the stochastic GW signal from the WD Galactic binaries.}
		\label{fig:GBconf}
\end{figure}

In the figure \ref{fig:GBconf} we show evolution of Galactic foreground  for $(1, 4, 6, 10)$ years of observation. 

\FloatBarrier

%------------------------------------------------------------------------------
\section{Cosmology}

This note uses the following cosmological parameters for conversion of
redshift to luminosity distance. The conversion is typically done according
to the description given in~\cite{Cosmology}.

The parameters used are Lambda CDM taken from~\cite{Planck} and are quoted
here:

\begin{eqnarray}
\Omega_M      &=& 0.3175~\textrm{(Matter)} \\
\Omega_k      &=& 0~\textrm{(Curvature)} \\
\Omega_\Lambda &=& 0.6825~\textrm{(Density)} \\
h   &=& 0.671 \\
H_0  &=& 67.1\,\textrm{km}\,\textrm{s}^{-1}\,\textrm{Mpc}^{-1}
\end{eqnarray}

With these parameters the luminosity distance corresponding to $z=1$ is $D_L = 6823 Mpc$.

\section{Stochastic GW backgrounds}

The quantity usually used to describe the sensitivity to \ac{SGWB} is the 
\textbf{energy density} sensitivity defined by : 
\begin{eqnarray}
\Omega_{Sens} (f) & = & { 4 \pi^2 \over 3 H_0^2 } f^3 S_{h}(f) \\
h^2 \Omega_{Sens} (f) & = & { 4 \pi^2 \over 3 h_0^2 } f^3 S_{h}(f)
\label{eq:OmegaGW_Sh}
\end{eqnarray}
with $H_0 = h \ h_0 $,  $ h_0 = 100 \  \rm{km}\, \rm{s}^{-1}\,\rm{Mpc}^{-1} = 3.24 \times 10^{-18} Hz$, $h$ the reduced Hubble constant and
$S_h(f)$ is the strain sensitvity in $Hz^{-1}$.

\begin{figure}[htbp]
	\centering
	\includegraphics[width=0.8\textwidth]{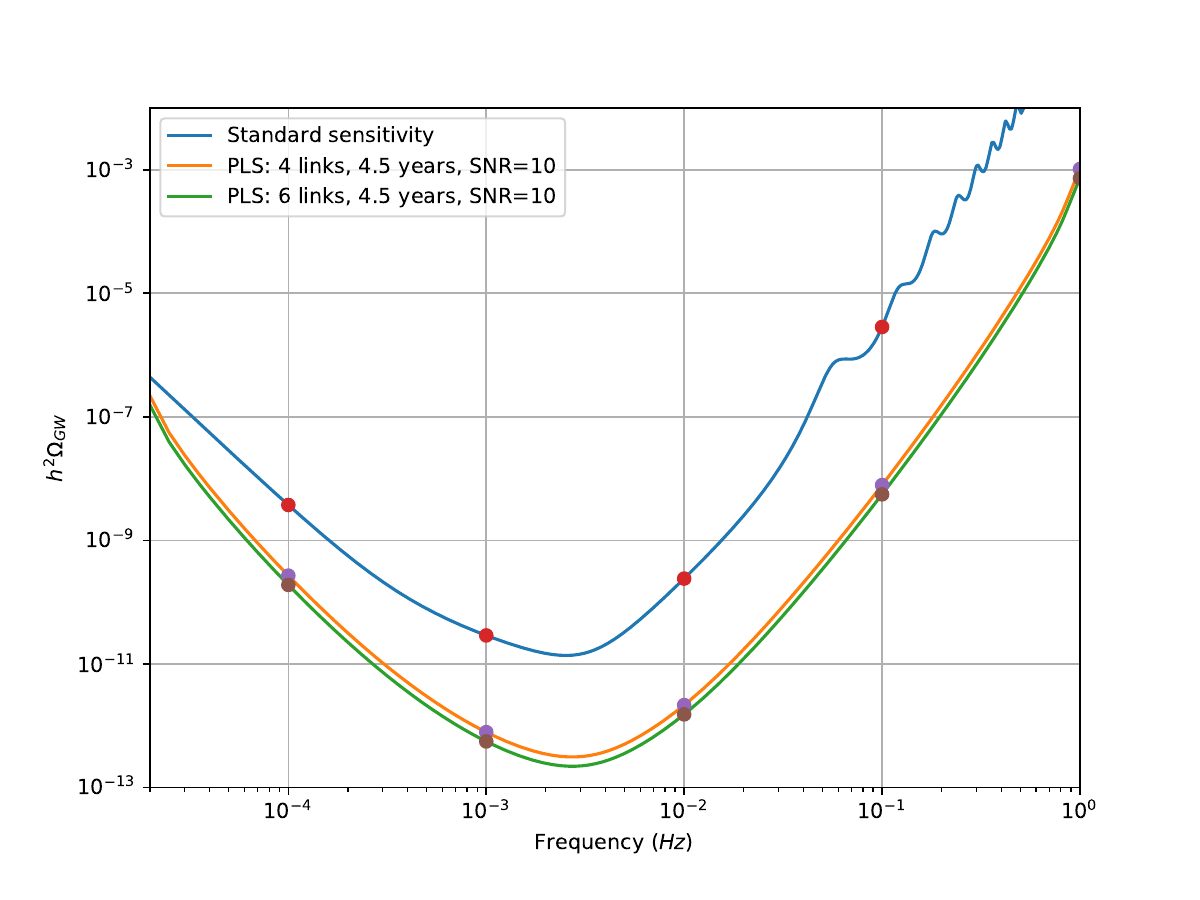}
	\caption{The sensitivity in $h^2 \Omega_{GW}$ and PLS for 4.5 years and SNR=10.
	The plot also includes the points tabulated in Table~\ref{tab:PLSh2OmGWValues}.}
	\label{fig:PLS_4ysnr20}
\end{figure}

The figure \ref{fig:PLS_4ysnr20} is showing the sensitivity in energy density.

%%%%%%%%%%%%%%%%%%%%%%%%%%
%\begin{figure}
%\centering
%\includegraphics[width=10cm,clip=true,angle=0]{PLS_1yr_LISA6_Adams.pdf}
%\caption{Various sensitivity in energy density for ``old \ac{LISA}" 6 links with one year observation : the green curve is the standard sensitivity, the red, blue, magenta and brown curves are the power-law sensitivity for respectively \ac{SNR} $\rho$=1, 5, 10 and 20. The black line is the stochastic background detected by Adams \& Cornish~\cite{2014PhRvD..89b2001A} }
%\label{fig:LISA6PLSAdams}
%\end{figure}
%%%%%%%%%%%%%%%%%%%%%%%%%%

The \ac{SNR} of a \ac{SGWB} with energy density $\Omega_{GW} (f)$ and observed for a duration $T_{obs},$ is computed by the following integration over the frequency :
\begin{equation}
SNR = \sqrt{T_{obs} \int_{f_{min}}^{f_{max}} df { (h^2 \Omega_{GW} (f))^2 \over (h^2 \Omega_{Sens} (f))^2 }}
\label{eq:SNR_SGWB}
\end{equation}

\begin{figure}[htbp]
	\centering
	\includegraphics[width=0.8\textwidth]{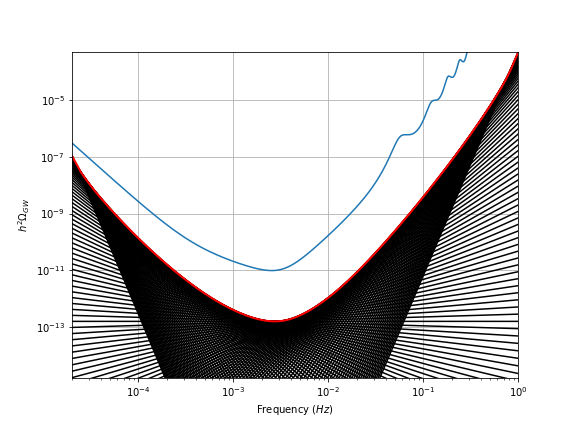}
	\caption{Computation of the \ac{PLS} for 4.5 years and SNR=10 (red).
	The black lines are all power laws of various index with SNR=10.
	The plot also includes the sensitivty as reference (blue).}
	\label{fig:PLS_4ysnr20_details}
\end{figure}

In order to quickly estimate the detectability of a power law \ac{SGWB}, i.e. in the form
\begin{equation}
h^2 \Omega_{GW} (f) = \Omega_{\beta} \left( {f \over f_{ref}} \right)^{\beta},
\end{equation}
within the observed frequency range, i.e.  $f\in[2 \times10^{-5}, 1] $Hz, the \ac{PLS} has been introduced~\cite{2013PhRvD..88l4032T}.
For an observation duration $T_{obs}$, it is computed as the enveloppe of all power law \ac{SGWB} with $SNR=SNR_{thr}$, scanning the slope $\beta$ (see figure~\ref{fig:PLS_4ysnr20_details}).

The use of the \ac{PLS} is straightforward: if one part of a power law representing the signal is higher than the \ac{PLS}, the signal is detectable. 

We usually choose $SNR_{thr} = 10$ (for example~\cite{2021JCAP...01..059F}). It is based on the preliminary results about \ac{SGWB} detectability~\cite{2014PhRvD..89b2001A}, adding some margin.
The nominal duration of the mission is 4.5 years. The corresponding \ac{PLS} is shown in the figure~\ref{fig:PLS_4ysnr20} and some points are tabulated in table~\ref{tab:PLSh2OmGWValues}.
See appendix~\ref{app:CombinedSNR} for details about the 6 links configuration.

\begin{table}[ht]
	\begin{center}
		\begin{tabular}{|c|c|c|c|} \hline
			Frequency (Hz) & $h^2 \Omega_{GW}$ & PLS 4 links & PLS 6 links \\ \hline
            0.00010 & 3.745195e-09 & 2.692611e-10 & 1.903964e-10 \\
            0.00100 & 2.893585e-11 & 7.873148e-13 & 5.567156e-13 \\
            0.01000 & 2.409087e-10 & 2.160153e-12 & 1.527459e-12 \\
            0.10000 & 2.851698e-06 & 7.868305e-09 & 5.563732e-09 \\
            1.00000 & 4.046701e-01 & 1.030663e-03 & 7.287887e-04 \\
			\hline
		\end{tabular}
	\end{center}
	\caption{The numerical values for some fixed frequencies of the sensitivity in $h^2 \Omega_{GW}$ and  PLS for 4.5 years and SNR=20.}
	\label{tab:PLSh2OmGWValues}
\end{table}

\newpage
\section{APPENDIX}

\appendix

\section{Noise PSD derivation}
\label{app:NoisePSDDerivation}

In this appendix, we provide a detailed  computation of the noise \ac{PSD}, 
given by formulas \eqref{eq:PSDnoiseX15}, \eqref{eq:PSDnoiseX15_EqArm},  \eqref{eq:PSDX15EqLSameLevel} and~\eqref{eq:PSDX20EqLSameLevel} above.
Introduce the following  assumptions:
\begin{enumerate}
	\item no residual laser noise;
	\item no clock noise;
	\item no optical bench noise;
	\item no backlink noise;
	\item no ranging error;
	\item no interpolation error;
	\item no effect of on board filtering of the measurements;
	\item all lasers are at the same nominal frequency, $c/\lambda$;
	\item consider only the projection of the acceleration noise on the sensitive axis, $\delta_i$ 
%	with the following convention
%	\begin{equation}
%	\delta_{ij} = 2 k_{ij} \vec{\delta}_{ij}.\vec{n}_{(i+2)'} 
%	\quad \textrm{and} \quad  
%	\delta_{i'} = - k_{i'} \vec{\delta}_{i'}.\vec{n}_{i+1} 
%	\end{equation}
%	with $k_{ij} = 2 \pi / \lambda_{ij}$ in phase and $k_{ij} = 1 / c$ in relative frequency;
	\item all the \ac{OMS} noises, optical path noises and readout noises are grouped in one term, $N_{ij}$,
	for the science interferometer; this type of noise is neglected for the other interferometers;
	\item $\theta_i^j$ account for the frequency polarity :
	\begin{equation}
	\theta_{ij}^{kl} = \left\{ 
	\begin{array}{cc}
	+1 & \rm{if} \quad \omega_{kl \rightarrow ij} - \omega_{ij} > 0 \\
	-1 & \rm{if} \quad \omega_{kl \rightarrow ij} - \omega_{ij} < 0 
	\end{array}
	\right. 
	\quad { \rm with} \quad \omega_{ij} = 2 \pi \nu_{ij}
	\label{E:defthetaij}
	\end{equation}
\end{enumerate}
with $\omega_{ij}$ the frequency of the laser on optical bench $ij$ 
and $\omega_{kl \rightarrow ij}$ the frequency of the beam received on optical bench 
$ij$ from optical bench $kl$. 

%Note that the signal of gravitational wave is not considered here, since we are
%interested to the noise only. 
%The detailed derivation is given in the Appendix and here we give the final expression for the noise  PSD.

The measurements on the two optical benches of spacecraft 1 are
\begin{equation}
\left\{
\begin{array}{lll}
s_{12}^c  & =&  \theta_{12}^{21} \ N_{12} \\
\tau_{12}  & = & 0 \\
\epsilon_{12}  & = & 2 \theta_{12}^{13} \ \delta_{12} \\
s_{12}^{sb}  & =&  \theta_{12}^{21} \ N_{12}
\end{array}
\right.
\quad \quad
\left\{
\begin{array}{lll}
s_{13}^c  & =&  \theta_{13}^{31} \ N_{13} \\
\tau_{13} & = & 0 \\
\epsilon_{13} & = & - 2 \theta_{13}^{12} \ \delta_{13}  \\
s_{13}^{sb}  & =&  \theta_{13}^{31} \ N_{13}
\end{array}
\right.
\end{equation}

%The indexing is simplifed compared to $slr$ indexing used in section~\ref{sec:RespGWX}. 
%The mapping is:
%\begin{eqnarray}
%  231 \rightarrow 12 & \quad & 32'1 \rightarrow 13 \nonumber \\
%  312 \rightarrow 23 & \quad & 13'2 \rightarrow 21 \nonumber \\
%  123 \rightarrow 31 & \quad & 21'3 \rightarrow 32
%\end{eqnarray}

We then apply the first step of \ac{TDI} to suppress the spacecraft jitter
\begin{eqnarray}
\xi_{12} & = & s_{12}^c - \theta_{12}^{21} \theta_{12}^{13}  {\epsilon_{12} - \tau_{12} \over 2}
- \theta_{12}^{21} \theta_{21}^{23} { D_{12} \epsilon_{21} - D_{12} \tau_{21} \over 2}   \\
& = & \theta_{12}^{21} N_{12} - \theta_{12}^{21} \theta_{12}^{13}   \theta_{12}^{13} \delta_{12}
- \theta_{12}^{21} \theta_{21}^{23}  D_{12} ( - \theta_{21}^{23} \delta_{21} )  \nonumber  \\
\xi_{12} & = & \theta_{12}^{21} \left( N_{12} -  \delta_{12} +  D_{12}  \delta_{21} \right)  \\
\xi_{13} & = & s_{13}^c - \theta_{13}^{31} \theta_{13}^{12}  {\epsilon_{13} - \tau_{13} \over 2}
- \theta_{13}^{31} \theta_{31}^{32} { D_{13}  \epsilon_{3} - D_{13}  \tau_{31} \over 2}  \\
& = & \theta_{13}^{31} N_{13} - \theta_{13}^{31} \theta_{13}^{12} ( - \theta_{13}^{12} \delta_{13} )
- \theta_{13}^{31} \theta_{31}^{32} D_{13}  \theta_{31}^{32} \delta_{31}  \nonumber  \\
\xi_{13}  & = & \theta_{13}^{31} \left( N_{13} + \delta_{13} - D_{13}  \delta_{31} \right)
\end{eqnarray}

Then we reduce half of the laser noise, i.e. equivalent to one laser per spacecraft (see section~4.3.3 of \cite{OttoPhD}):
\begin{eqnarray}
\eta_{12} & = & \theta_{12}^{21} \xi_{12} + { \theta_{23}^{21} D_{12} \tau_{21} - \theta_{21}^{23} D_{12} \tau_{23} \over 2} \nonumber \\
%& = & \theta_{12}^{21} \xi_{12}\\
& = &  N_{12} -  \delta_{12} +  D_{12}  \delta_{21} \\
\eta_{13} & = & \theta_{13}^{31} \xi_{13} - { \theta_{13}^{12} \tau_{13} -  \theta_{12}^{13} \tau_{12} \over 2}  \nonumber  \\
%& = & \theta_{13}^3 \xi_{13}\\
& = & N_{13} + \delta_{13} - D_{13}  \delta_{31}
\end{eqnarray}

%Then step 1.3, used to remove part of the clock noise (see section~4.3.4 of \cite{OttoPhD}) :
% \begin{eqnarray}
%   \eta_{12}  & = & Q_{12} - {\theta_{2'}^2 (b_{2'}+ b_2) \over 2} { s_{2';3}^c(t) - s_{2';3}^{sb}(t) \over a_{2'} + \theta_{2'}^1 m_{2'} - c_{2'}} \nonumber  \\
%   & = & N_{12} -  \delta_{12} +  D_{12}  \delta_{21} \\
%   \eta_{23} & = & Q_2 - \theta_{2}^{3'} a_2 { s_{2'}^c(t) - s_{2'}^{sb}(t) \over a_{2'} + \theta_{2'}^1 m_{2'} - c_{2'}}
%   - {\theta_{3'}^3 (b_{3'}+ b_3) \over 2} {  D_1 s_{3}^c(t) - D_1 s_{3}^{sb}(t) \over a_{3} + \theta_{3}^{1'} m_{3} - c_{3}} \nonumber \\
%   & = & N_2 -  \delta_{2} +  D_1  \delta_{3'} \\
%   \eta_{31} & =&  Q_3 (t) - \theta_3^{1'} a_3 { s_3^c (t)  - s_3^{sb} (t)  \over a_3 + \theta_3^{1'} m_3 -c_3} \nonumber  \\
%   & = & N_{31} -  \delta_{31} +  D_{31}  \delta_{13} \\
%   \eta_{13}  & = & Q_{13} \nonumber \\
%     & = & N_{13} + \delta_{13} - D_{13}  \delta_{31} \\
%      \eta_{32}  & = & Q{2'}(t) - \theta_{2'}^{1} a_{2'} { s_{2'}^c(t) - s_{2'}^{sb}(t) \over a_{2'} + \theta_{2'}^1 m_{2'} - c_{2'}}
%   + {\theta_{2'}^2 (b_{2'}+ b_2) \over 2} { s_{2'}^c(t) - s_{2'}^{sb}(t) \over a_{2'} + \theta_{2'}^{1} m_{2'} - c_{2'}} \nonumber \\
%   & = & N_{2'} + \delta_{21} - D_{21} \delta_{12} \\
%   \eta_{13}  & = &    Q{3'}(t) - \theta_{3'}^{2} a_{3'} { s_{3}^c(t) - s_{3}^{sb}(t) \over a_{3} + \theta_{3}^{1'} m_{3} - c_{3}}
%   + {\theta_{3'}^3 (b_{3'}+ b_3) \over 2} { s_{3}^c(t) - s_{3}^{sb}(t) \over a_{3} + \theta_{3}^{1'} m_{3} - c_{3}} \nonumber  \\
%   & = & N_{3'} + \delta_{3'} - D_{13} \delta_{2}
% \end{eqnarray}

Now he have to apply the \ac{TDI} generators. We use \ac{TDI} generator X (Michelson) as for the sensitivity
curve, and \ac{TDI}-generation 1.5 is given as  

\begin{eqnarray}
X_{1.5}
%& = & D_{12}D_{31}D_{13}  \eta_{13'2} + D_{13} D_{31} \eta_{231} + D_{13}  \eta_{123} + \eta_{32'1} \nonumber \\
%& & - \left[  D_{12}D_{21}D_{13}  \eta_{123} + D_{12}D_{21} \eta_{32'1} + D_{12} \eta_{13'2} + \eta_{231} \right]
%\\
%& = & D_{12}D_{31}D_{13}  \eta_{21} + D_{13} D_{31} \eta_{12} + D_{13}  \eta_{31} + \eta_{13} \nonumber \\
%& & - \left[  D_{12}D_{21}D_{13}  \eta_{31} + D_{12}D_{21} \eta_{13} + D_{12} \eta_{21} + \eta_{12} 
%\right]\\
& = & 
\eta_{13} + D_{13}  \eta_{31} + D_{13}  D_{31} \eta_{12} + D_{13}  D_{31} D_{12} \eta_{21}
- \eta_{12} - D_{12} \eta_{21} - D_{12} D_{21} \eta_{13} - D_{12} D_{21} D_{13}  \eta_{31} \nonumber \\
& = &  N_{13} + \delta_{13} - D_{13}  \delta_{31}  \nonumber \\
& &  + D_{13}  N_{31} - D_{13}  \delta_{31} + D_{13}  D_{31}  \delta_{13}  \nonumber \\
& &  + D_{13}  D_{31} N_{12} - D_{13}  D_{31} \delta_{12} + D_{13}  D_{31} D_{12}  \delta_{21}  \nonumber \\
& &  + D_{13}  D_{31} D_{12} N_{21} + D_{13}  D_{31} D_{12} \delta_{21} - D_{13}  D_{31} D_{12} D_{21} \delta_{12} \nonumber \\
& & - N_{12} +  \delta_{12} -  D_{12}  \delta_{21} \nonumber \\
& & - D_{12} N_{21} - D_{12} \delta_{21} + D_{12} D_{21} \delta_{12}  \nonumber  \\
& & - D_{12} D_{21} N_{13} - D_{12} D_{21} \delta_{13} + D_{12} D_{21} D_{13}  \delta_{31}  \nonumber \\
& & - D_{12} D_{21} D_{13}  N_{31} + D_{12} D_{21} D_{13}  \delta_{31} - D_{12} D_{21} D_{13}  D_{31}  
\delta_{13}
\end{eqnarray}
Factorizing, we get:
\begin{eqnarray}
X_{1.5} & = & 
  (1 - D_{12} D_{21}) N_{13} 
- (1 - D_{13}  D_{31}) N_{12} 
+ ( D_{13}  - D_{12} D_{21} D_{13}  ) N_{31}
- ( D_{12} - D_{13}  D_{31} D_{12}) N_{21} \nonumber \\
& &  (1 + D_{12} D_{21} - D_{13}  D_{31} - D_{13}  D_{31} D_{12} D_{21} ) \delta_{12} 
+ (1 - D_{12} D_{21} + D_{13}  D_{31} - D_{12} D_{21} D_{13}  D_{31} ) \delta_{13} \nonumber \\
& &  - 2 ( D_{12} - D_{13}  D_{31} D_{12}  ) \delta_{21} 
- 2 ( D_{13}   - D_{12} D_{21} D_{13}   ) \delta_{31}
\end{eqnarray}
%\cSB{please give a link between two equations above for $X_{1.5}$ and put properly eqn. numbering}

The next step is the computation of PSD.  Note that
\begin{eqnarray}
PSD \left[ (1 - D_{13}  D_{31}) N_{12} \right]
& = & \left\langle \left( 1 - e^{- i \omega ( L_{13} + L_{31} ) } \right)  \left( 1 - e^{ i \omega ( L_{13} + L_{31} ) } \right) \tilde{N}_{12} \tilde{N}_{12}^{*} \right\rangle  \nonumber \\
& = & \left\langle \left( e^{ i \omega  {L_{13} + L_{31} \over 2} }  - e^{- i \omega {L_{13} + L_{31} \over 2} } \right) \tilde{N}_{12} \tilde{N}_{12}^{*} \right\rangle  \nonumber \\
& = & 4 \sin^2 \left( \omega {L_{13} + L_{31} \over 2} \right) S_{OMS_{12}},
\end{eqnarray}
so the PSD of terms 1, 2, 3, 4, 7 and 8 can be reduced in a similar way.
The tilde denotes the Fourier transfer of the quantity.
The term 5 and 6 are given by:
\begin{eqnarray}
& & PSD [ (1 + D_{12} D_{21} - D_{13}  D_{31} - D_{13}  D_{31} D_{12} D_{21} ) \delta_{12} ] \nonumber \\
& = & \left\langle \left(  1 
+ e^{- i \omega ( L_{12} + L_{21} )} 
- e^{- i \omega ( L_{13} + L_{31} )} 
- e^{- i \omega ( L_{13} + L_{31} + L_{12} + L_{21})  } \right)  \left( ... \right)^{*} \tilde{\delta}_{12} \tilde{\delta}_{12}^{*} \right\rangle  \nonumber \\
& = & \left\langle \left( \left( 1 + e^{ - i \omega  (L_{12} + L_{21}) } \right)
\left( 1 - e^{ - i \omega  (L_{13} + L_{31}) } \right)
%e^{- i \omega ( L_{13} + L_{31} + L_{12} + L_{21})  }
\right) \left(...\right)^* \tilde{\delta}_{12} \tilde{\delta}_{12}^{*} \right\rangle  \nonumber \\
& = & \left\langle \left( e^{ i \omega  {L_{12} + L_{21} \over 2} } + e^{ - i \omega  {L_{12} + L_{21} \over 2}} \right)^2
\left( e^{ i \omega  {L_{13} + L_{31} \over 2}} - e^{ - i \omega  {L_{13} + L_{31} \over 2} } \right)^2
\delta_{12} \delta_{12}^{*} \right\rangle  \nonumber \\
& = & 16 \cos^2 \left( \omega {L_{12} + L_{21} \over 2} \right) \sin^2 \left( \omega {L_{13} + L_{31} \over 2} \right) S_{acc_{12}}.
\end{eqnarray}

Similarly we can write the  2nd generation $X$-\ac{TDI}:
\begin{eqnarray}
	X_{2.0} & = &
	\eta_{13} %
	+ D_{13}  \eta_{31} %
	+ D_{13}  D_{31} \eta_{12} %
	+ D_{13}  D_{31} D_{12} \eta_{21} %
	+ D_{13}  D_{31} D_{12} D_{21} \eta_{12} \nonumber \\ %
	& & + D_{13}  D_{31} D_{12} D_{21} D_{12} \eta_{21} %
	+ D_{13}  D_{31} D_{12} D_{21} D_{12} D_{21} \eta_{13} %
	+ D_{13}  D_{31} D_{12} D_{21} D_{12} D_{21} D_{13}  \eta_{31}  \nonumber \\
	& & - \eta_{12} 
	- D_{12} \eta_{21} 
	- D_{12} D_{21} \eta_{13}
	- D_{12} D_{21} D_{13}  \eta_{31}
	- D_{12} D_{21} D_{13}  D_{31} \eta_{13} \nonumber \\ %
	& & - D_{12} D_{21} D_{13}  D_{31} D_{13}  \eta_{31} %
	- D_{12} D_{21} D_{13}  D_{31} D_{13}  D_{31} \eta_{12} %
	- D_{12} D_{21} D_{13}  D_{31} D_{13}  D_{31} D_{12} \eta_{21} \nonumber \\ %
	& = & ( 1 - D_{12} D_{21} D_{13}  D_{31} ) 
	( ( \eta_{13} + D_{13}  \eta_{31} ) + D_{13}  D_{31}(  \eta_{12} + D_{12} \eta_{21} ) )  \nonumber \\
	& & - ( 1 - D_{13}  D_{31} D_{12} D_{21} ) 
	( ( \eta_{12} + D_{12} \eta_{21} ) + D_{12} D_{21} ( \eta_{13} +  D_{13}  \eta_{31} ) )  \nonumber
	\\
	%& = & ( 1 - D_{12} D_{21} D_{13}  D_{31} ) 
	%( (N_{13} + \delta_{13} - D_{13}  \delta_{31} 
	%+ D_{13}  N_{31} - D_{13}  \delta_{31} + D_{13}  D_{31}  \delta_{13} ) \nonumber \\
	%& & + D_{13}  D_{31}(N_{12} -  \delta_{12} + D_{12}  \delta_{21} 
	%+ D_{12} N_{21} + D_{12} \delta_{21} - D_{12} D_{21} \delta_{12} ) )  \nonumber \\
	%& & - ( 1 - D_{13}  D_{31} D_{12} D_{21} ) 
	%( (  N_{12} -  \delta_{12} + D_{12}  \delta_{21} 
	%+ D_{12} N_{21} + D_{12} \delta_{21} - D_{12} D_{21} \delta_{12} ) \nonumber \\
	%& & + D_{12} D_{21} (N_{13} + \delta_{13} - D_{13}  \delta_{31}  
	%+  D_{13}  N_{31} - D_{13}  \delta_{31} + D_{13}  D_{31}  \delta_{13} ) ) \nonumber \\
	%
	& = & ( 1 - D_{12} D_{21} D_{13}  D_{31} ) 
	( (N_{13} + D_{13}  N_{31}
	+ (1 + D_{13}  D_{31} )\delta_{13} 
	- 2 D_{13}  \delta_{31}) \nonumber \\
	& & + D_{13}  D_{31}(N_{12} + D_{12} N_{21}
	- \delta_{12} ( 1 + D_{12} D_{21} )
	+ 2 D_{12}  \delta_{21} ) )  \nonumber \\
	& & - ( 1 - D_{13}  D_{31} D_{12} D_{21} ) 
	( (  N_{12} + D_{12} N_{21}
	- ( 1 + D_{12} D_{21}) \delta_{12} 
	+ 2 D_{12}  \delta_{21} ) \nonumber \\
	& & + D_{12} D_{21} (N_{13} +  D_{13}  N_{31}
	+ ( 1 + D_{13}  D_{31} ) \delta_{13}
	- 2 D_{13}  \delta_{31} ) ). 
\end{eqnarray}
Factorizing, we get:
\begin{eqnarray}
	X_{2.0} & = &
	( 1 - D_{12} D_{21} D_{13}  D_{31} ) 
	( (1  - D_{12} D_{21} )
	(N_{13} 
	+ D_{13}  N_{31}
	+ (1 + D_{13}  D_{31} )\delta_{13} 
	- 2 D_{13}  \delta_{31}) \nonumber \\
	& & - ( 1 - D_{13}  D_{31} ) (
	N_{12} 
	+ D_{12} N_{21}
	-( 1 + D_{12} D_{21} )  \delta_{12} 
	+ 2 D_{12}  \delta_{21} ) )
	\label{eq:TimeTDIX20}
\end{eqnarray}

\textbf{Approximation}: Armlength are constant, i.e. the delay operators are 
commuting.

\begin{eqnarray}
PSD_{X_{1.5}} & = &  4 \sin^2 \left( \omega { L_{31} + L_{13} \over 2} \right) \left( S_{OMS_{12}} + S_{OMS_{21}} + 4 S_{acc_{21}} + 4 \cos^2 \left( \omega { L_{12} + L_{21} \over 2} \right) S_{acc_{12}} \right) \nonumber \\
& &+ 4 \sin^2 \left( \omega {L_{21} + L_{12} \over 2} \right) \left( S_{OMS_{13}} + S_{OMS_{3}} + 4 S_{acc_{3}} + 4 \cos^2 \left( \omega { L_{31} + L_{13} \over 2} \right) S_{acc_{13}} \right)
\end{eqnarray}

Computation of PSD for the 2nd generation \ac{TDI} requires some additional math. Starting with 
\begin{eqnarray}
 PSD_{ X_{2.0}} & = &
  \left\langle \left[  \left( 1 - e^{- i \omega ( L_{12} + L_{21} + L_{13}  + L_{31} ) } \right) \right. \right. \nonumber  \\
& & \times \left( \left(1 - e^{- i \omega ( L_{12} + L_{21} )} \right)
\left( \tilde{N}_{13} 
+ e^{- i \omega L_{13} }  \tilde{N}_{31}
+ \left( 1 + e^{- i \omega ( L_{13} + L_{31} )} \right) \tilde{\delta}_{13} 
- 2 e^{- i \omega L_{13} }  \tilde{\delta}_{31} \right) \right.   \nonumber \\
& & \left. \left. \left. - \left(1 - e^{- i \omega ( L_{13} + L_{31} )} \right) 
\left(
\tilde{N}_{12} 
+ e^{- i \omega L_{12} }  \tilde{N}_{21}
-\left( 1 + e^{- i \omega ( L_{12} + L_{21} )} \right)  \tilde{\delta}_{12} 
+ 2 e^{- i \omega L_{12} }  \tilde{\delta}_{21} \right) \right) \right] [...]^* \right\rangle \nonumber \\
\end{eqnarray}
 and using the following simplifications
\begin{eqnarray*}
  \left( 1 - e^{- i \omega ( L_{12} + L_{21} + L_{13}  + L_{31} ) } \right)(...)^* 
  & \rightarrow & 4 \sin^2 \left(  \omega \frac{L_{12} + L_{21} + L_{13}  + L_{31}}{2}\right)\\
  \left(1 - e^{- i \omega ( L_{12} + L_{21} )} \right)(...)^*
  & \rightarrow & 4 \sin^2 \left(  \omega \frac{L_{12} + L_{21} }{2}\right)\\
  \left( 1 + e^{- i \omega ( L_{12} + L_{21} )} \right)(...)^*
  & \rightarrow & 4 \cos^2 \left(  \omega \frac{L_{12} + L_{21} }{2}\right) \\
  \left(e^{- i \omega L_{12} }\right)(...)^*  & \rightarrow &  1\\
\end{eqnarray*}
we arrive at
\begin{eqnarray}
 PSD_{ X_{2.0}} & = &
  16 \sin^2 \left(  \omega \frac{L_{12} + L_{21} + L_{13}  + L_{31}}{2}\right) \nonumber \\
  & & \times \left( \sin^2 \left(  \omega \frac{L_{12} + L_{21} }{2}\right) 
  \left( S_{OMS_{13}} + S_{OMS_{31}} 
  + 4 \left( S_{acc_{31}} + \cos^2 \left(  \omega \frac{L_{13} + L_{31} }{2}\right) S_{acc_{13}} \right) \right) \right. \nonumber \\
   & & \left. + \sin^2 \left(  \omega \frac{L_{13} + L_{31} }{2}\right) 
  \left( S_{OMS_{12}} + S_{OMS_{21}} 
  + 4 \left( S_{acc_{21}} + \cos^2 \left(  \omega \frac{L_{12} + L_{21} }{2}\right) S_{acc_{12}} \right) \right) \right).
\end{eqnarray}

\textbf{Approximation 1+2}: Armlength are all equal.

\begin{eqnarray}
	S_{n,X_{1.5}} = PSD_{X_{1.5}}  =   4 \sin^2 \left( \omega L \right)
	\left( S_{OMS_{12}} + S_{OMS_{21}}  + S_{OMS_{13}} + S_{OMS_{31}} +  \right.\nonumber \\
	\left. 4 \left( S_{acc_{21}} + S_{acc_{31}}
	+ \cos^2 \left( \omega L \right)  \left( S_{acc_{12}} + S_{acc_{13}}  \right) \right) 
	\right) \nonumber 
\end{eqnarray}

\begin{eqnarray}
 PSD_{ X_{2.0}} & = &
  16 \sin^2 \left( 2 \omega L \right) \sin^2 \left(  \omega L \right) 
  \left( S_{OMS_{13}} + S_{OMS_{31}} + S_{OMS_{12}} + S_{OMS_{21}} \right.  \nonumber \\
  & & \left. + 4 \left( S_{acc_{31}} + S_{acc_{21}} 
  + \cos^2 \left(  \omega L \right) \left(  S_{acc_{13}} + S_{acc_{12}} \right) \right) \right)
\end{eqnarray}

\textbf{Approximation 1+2+3}: All noises of the same type have the same PSD ($S_{OMS_{ij}}= S_{OMS}$ and $S_{acc_{ij}} = S_{acc}$):

\begin{eqnarray}
	S_{n,X_{1.5}} = PSD_{X_{1.5}}  =   16 \sin^2 \left( \omega L \right)
	\left( S_{OMS} + \left( 3 + \cos \left( 2 \omega L \right)  \right)  S_{acc} \right) \nonumber
\end{eqnarray}

\begin{eqnarray}
 PSD_{ X_{2.0}} & = &
  64 \sin^2 \left( 2 \omega L \right) \sin^2 \left(  \omega L \right) 
  \left( S_{OMS} + 2 \left( 1 + \cos^2 \left(  \omega L \right) \right) S_{acc} \right) \nonumber \\
  & = &
  64 \sin^2 \left( 2 \omega L \right) \sin^2 \left(  \omega L \right) 
  \left( S_{OMS} + \left( 3 + \cos \left( 2 \omega L \right) \right) S_{acc} \right)
\end{eqnarray}

\section{Noise spectrum from SciRD sensitivity}
\label{app:SciRDNoise}

The official required sensitivity of \ac{LISA} is decribed in~\cite{SciRD}:
\begin{eqnarray}
  S_{h,SciRD}(f) & = & \frac{1}{2} \frac{20}{3} \left( \frac{S_{I}(f)}{(2 \pi f)^4} +S_{II}(f)\right) R(f) \nonumber \\
  S_{I}(f) & = & 5.76 \times 10^{-48} \left( 1 + \left( \frac{f_1}{f} \right)^2 \right) \textrm{s}^{-4}.\textrm{Hz}^{-1} \\
  S_{II}(f) & = & 3.6 \times 10^{-41} \textrm{Hz}^{-1} \nonumber \\
  R(f) & = & 1 + \left( \frac{f}{f_2} \right)^2 \nonumber
\end{eqnarray}
with $f_1 = 0.4 \textrm{mHz}$ and $f_2 = 25 \textrm{mHz}$. It is for a full instrument so 6 links.

The analytic approximation of the sensitivity for TDI $X_{2.0}$ (4 links) is given by equation~\eqref{eq:SensX2approx}. 
By equalizing the sensitivities, i.e. $S_{h,SciRD} = S_{h,X_{2.0}} / 2$ 
and using $R(f) \approx 1+ 0.6 (\omega L)^2$, 
we get for the noise \ac{PSD} in relative frequency (as $S_{n,X_{2.0}}$ is in relative frequency in~\eqref{eq:SensX2approx}):
\begin{eqnarray}
  S_{n,X_{2.0}} & \approx &
  (4\omega L)^2  \sin^2({\omega L}) (2 \sin{(2\omega L)})^2 \left( \frac{S_{I}(\omega)}{\omega^4} +S_{II}(\omega)\right)
\end{eqnarray}
Converting to displacement (i.e. divide by $ ( \omega / c)^2$): 
\begin{eqnarray}
    S_{n,X_{2.0},dL} & \approx &
  4 C(\omega) \left( \frac{S_{I}(\omega)}{\omega^4} L^2 +S_{II}(\omega)L^2 \right)
\end{eqnarray}
with
\begin{eqnarray}
 C(\omega) & = & 16 \sin^2({\omega L})  \sin^2{(2\omega L)})
\end{eqnarray}
We find back the values for the two noise terms (see~\ref{sec:InstrumentNoise}):
\begin{eqnarray}
  S_{oms} (\omega) & = & S_{II}(\omega)L^2 = \left( 15 \ \textrm{pm}/\sqrt{Hz} \right)^2\\
  S_{acc} (\omega) & = &  \frac{S_{I}(\omega) L^2}{4} = \left( 3 \ \textrm{fm}.\textrm{s}^{-2}/\sqrt{Hz} \right)^2 \left( 1 + \left( \frac{0.4 \textrm{mHz}}{f} \right)^2 \right)
\end{eqnarray}
entering in the noise \ac{PSD}:
\begin{eqnarray}
    S_{n,X_{2.0},dL} & \approx & 
    4 C(\omega) \left( S_{oms}(\omega) + 4 \frac{S_{acc}(\omega)}{\omega^4} \right)
\end{eqnarray}
Note the $4$ in front of the $S_{acc}$ is an approximation done in the SciRD~\cite{SciRD} of $(3+\cos(\omega L))$ since this term mainly contributes at low frequency.

\section{SNR for Phenomenological IMR models}
\label{app:SNR_SMBHB_TestCases}

We have used several phenomenological inspiral-merger-ringdown (IMR)
models: IMRPhenomA \cite{PhenomA}, IMRPhenomC \cite{PhenomC} and
IMRPhenomD \cite{PhenomD} to compute \ac{SNR} for several test 
MBHB systems.

%The IMRPhenomA model is implemented in \url{Codes/SB/NeilCompare.py}. Note that PhenomA model returns the amplitude for the optimally oriented source.
%
%The IMRPhenomC model (python module) is implemented in \url{Codes/SB/phenomC.py}. The returned amplitude is for $2,2$ mode.
%
%For the IMRPhenomD model we have used the module implemented in LDC software
%(module pyFDresponse).
%
%
%We compute the sky, polarization, inclination averaged SNR for each module
%(where appropriate) using the script \url{/Codes/SB/ComputeAverSNR_1.py}.
%$\rightarrow$ results 'PhA Sens', 'PhC Sens' and 'PhD Sens'
%
%The averaged amplitudes (SPA, IMRPhenom) are compared/checked in the
%script \url{Codes/SB/CheckHpHcAmpl.py}.

%We compute the sky, polarization, inclination averaged SNR using
%\ac{TDI} noise and long-wavelength approximation in the script
%\url{Codes/SB/ComputeAverSNR_TDI.py}.
%$\rightarrow$ results  'PhD AvTDI'.

%\subsection{SMBHB: Computation of SNR distribution using Monte-Carlo}

%We conducted Monte-Carlo study of the SNR for three IMR models using the script \url{Codes/SB/SNR_MC_1.py}.

For the Monte-Carlo simulation we randomly draw source varying inclination,
polarization and sky position with fixed intrinsic parameter (masses, spins)
as well as time to coalescence and compute SNR unig LDC tools.  We use only IMRPhenomD model.

%The first Monte-Carlo is done using the LDC data generation pipeline (git@gitlab.in2p3.fr:stas/MLDC.git:SNRs \#65b56710 ).
%
%For the sky position and orientation of the source,
%we first generate a file with random direction of sky position and orbital
%angular momentum.
%\begin{verbatim}
%Make_Rand_SkyPolInc.py --NSrc=100000
%     --seed=12345 SkyAngleL_s12345_N100000.npy
%\end{verbatim}
%
%For the other parameters, masses, redshift (distance), spin amplitudes, initial
%phase, we generate an hdf5 file modified using \verb+LISAh5_edit.py+ (example: \verb+Data/MBHB_M1e7_m1e6_z1_Chi0_chi0.hdf5+).
%
%Finally we run the code \verb+MBHB_FD_SNR_Average.py+, as example:
%\begin{verbatim}
%  /codes/LDC/software/LDCpipeline/scripts/MBHB_FD_SNR_Average.py \
%  MBHB_M1e8_m1e6_z1_Chi0_chi0.hdf5 SkyAngleL_s12345_N100000.npy \
%  0 19 MBHB_M1e8_m1e6_z1_Chi0_chi0_100000_0.txt
%\end{verbatim}
%with the arguments
%\begin{itemize}
%  \item \verb+MBHB_M1e8_m1e6_z1_Chi0_chi0.hdf5+ : hdf5 file for masses, redshift (distance), spin amplitudes, initial
%phase
%\item \verb+SkyAngleL_s12345_N100000.npy+ : file with sky position and
%orientation
%\item \verb+0 19+ : id of the run and number of runs
%\item \verb+MBHB_M1e8_m1e6_z1_Chi0_chi0_100000_0.txt+ : output file
%\end{itemize}
%
%
%

%------------------------------------------------------------------------------
%\section{Test Cases}
%\label{app:SNR_SMBHB_TestCases}
%
%This section documents different test cases for comparing the different tools.
%
%%%%%%%%%%%%%%%%%%%%%%%%%%%%%%
\subsection{Test case 1}
%
%Compute SNR for PhenomC, D, A for one TDI-X channel and the following setup:
%\begin{verbatim}
%# chi1 = chi2 = 0
%# z = 1
%# Dl = 6823
%# Source Masses 10^5, 10^6, 10^7, 10^8
%\end{verbatim}
Reference system: non-spinning $\chi_1 = \chi_2=0$, redshift $z=1$, 
luminosity distance $D_L = 6823$ Mpc, source frame individual masses
$m_i=10^5, 10^6, 10^7, 10^8$. 

\begin{figure}[htbp]
\centering
\includegraphics[width=0.48\textwidth]{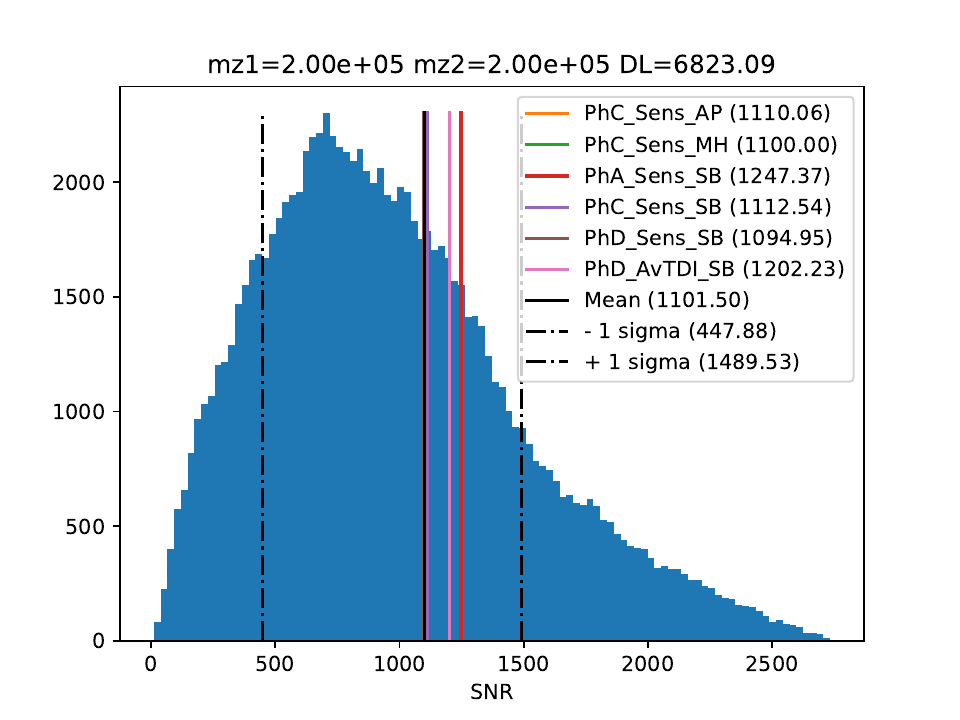}
\includegraphics[width=0.48\textwidth]{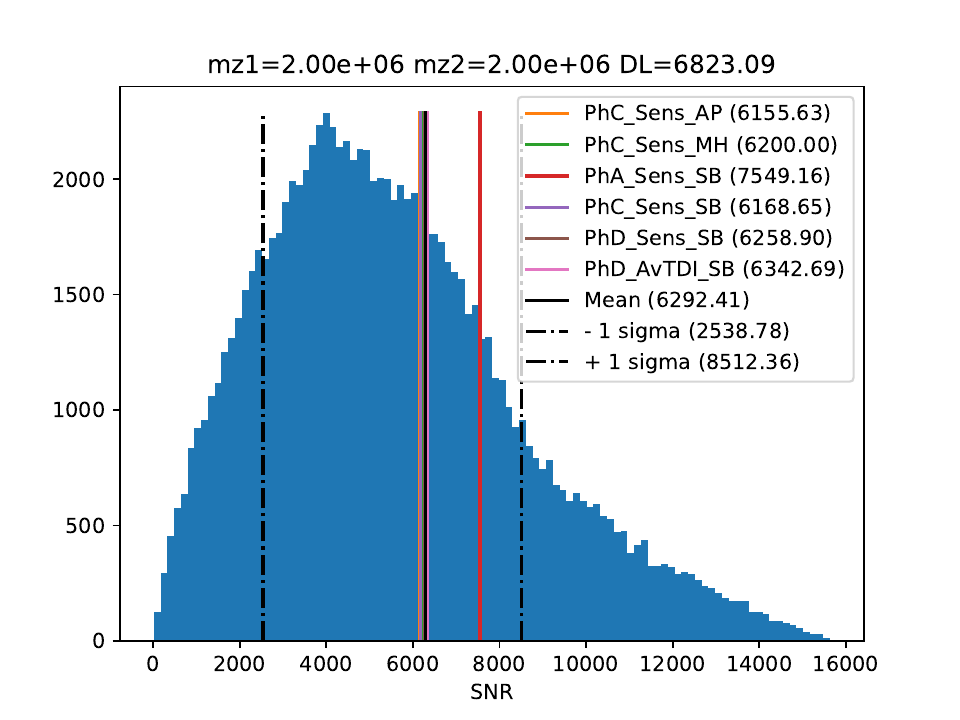}
\includegraphics[width=0.48\textwidth]{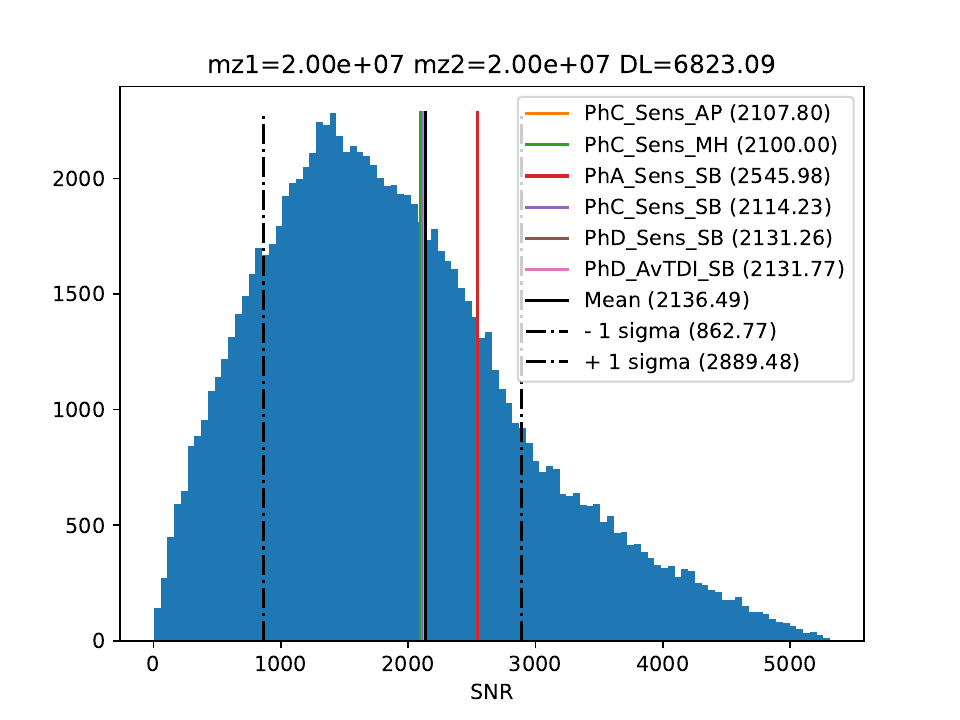}
\includegraphics[width=0.48\textwidth]{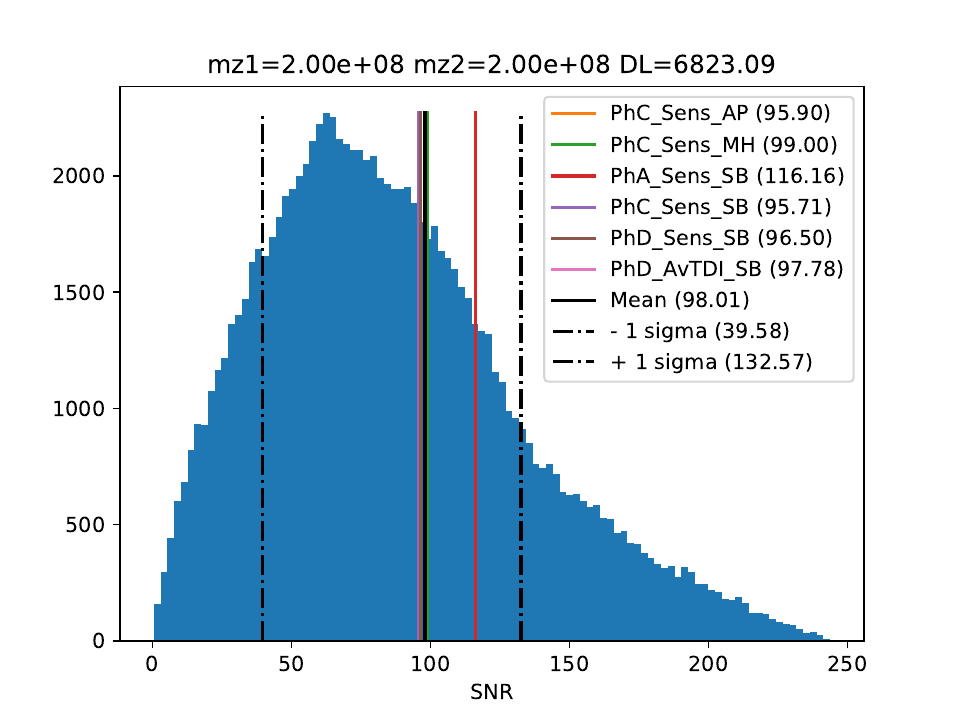}
\includegraphics[width=0.48\textwidth]{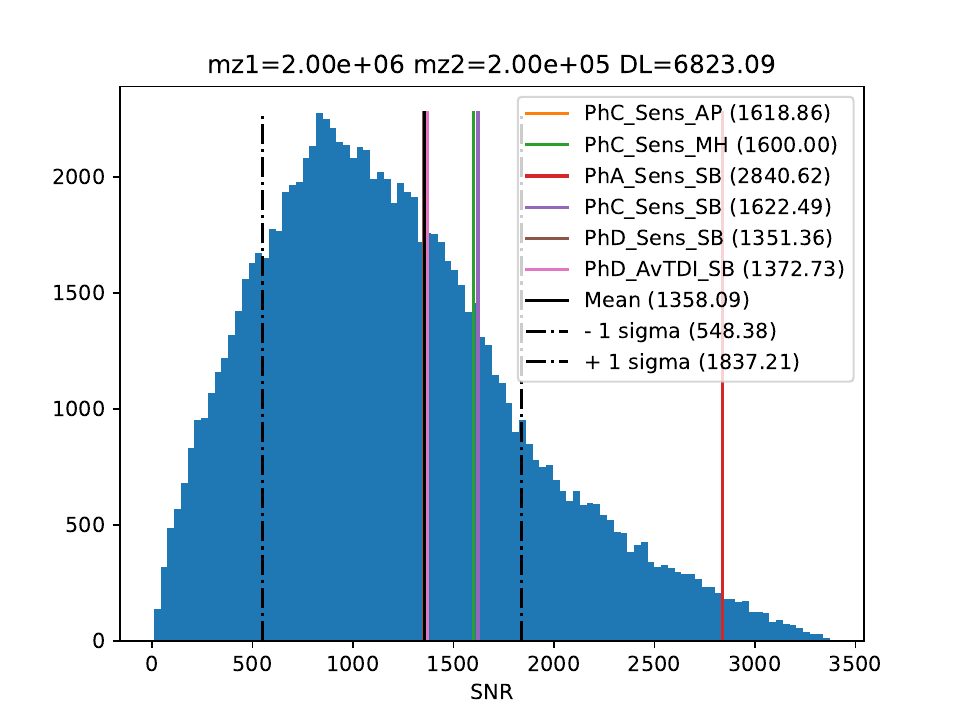}
\includegraphics[width=0.48\textwidth]{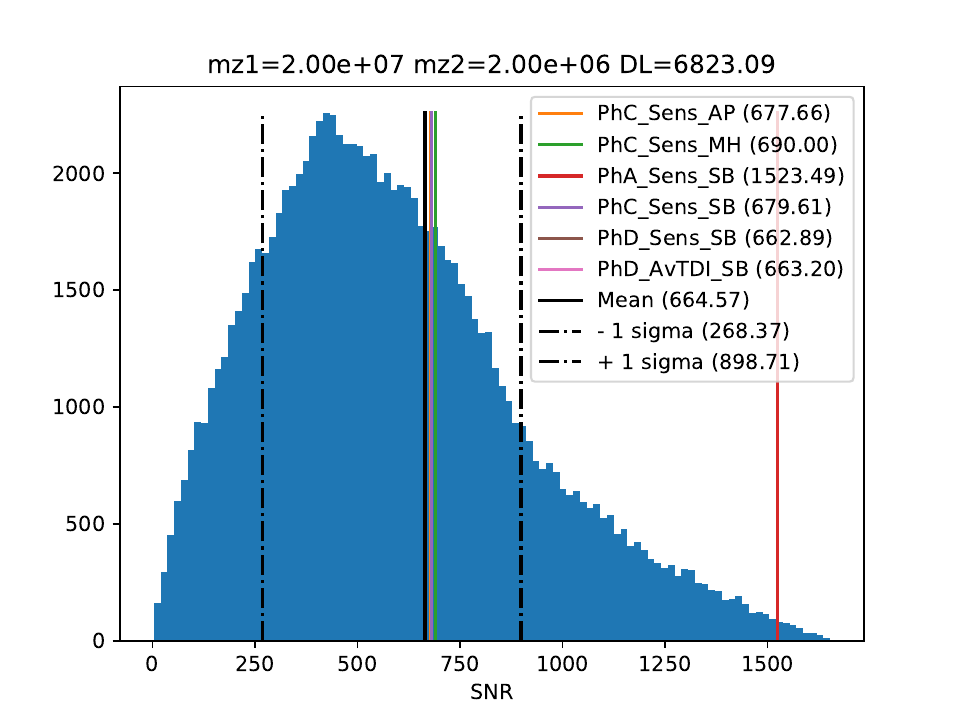}
\includegraphics[width=0.48\textwidth]{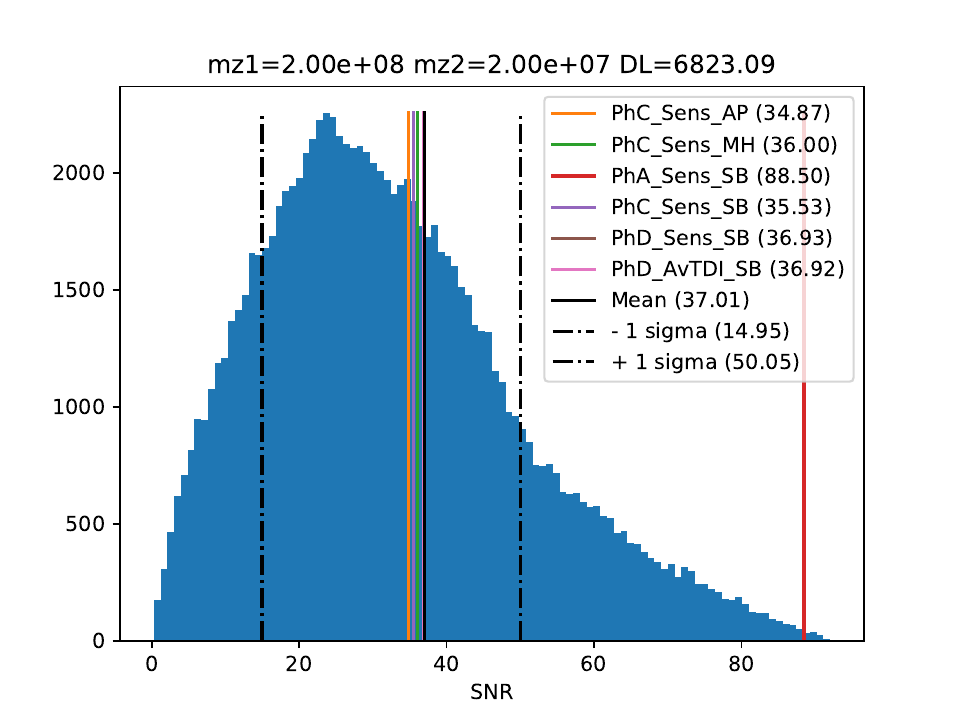}
\includegraphics[width=0.48\textwidth]{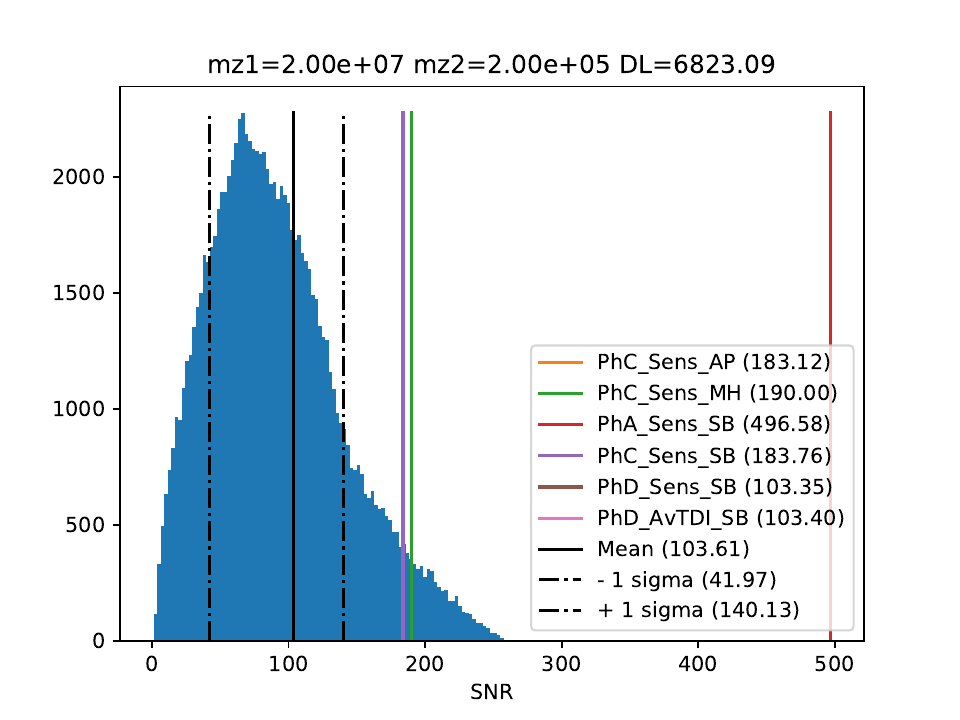}
\caption{TC1: Numerical results and analytical results for the non-spinning cases and redshift 1.
The intrinsic masses of the source are :
row 1: $10^5-10^5$ and $10^6-10^6$,
row 2: $10^7-10^7$ and $10^8-10^8$,
row 3: $10^6-10^5$ and $10^7-10^6$,
row 4: $10^8-10^7$ and $10^7-10^5$}
\label{fig:HistTC1}
\end{figure}

The table~\ref{tab:SNRsTC1} is summarizing the results with various 
methods and the figure~\ref{fig:HistTC1} is showing a comparison of the results 
including the distribution of SNRs.

\begin{table}[ht]
\begin{center}
\begin{tabular}{|c|c|c|c|c|c|}
\hline
 \multicolumn{2}{|c|}{$m_1 | m_2$} & $ 1\times10^{5} $  &  $ 1\times10^{6} $  &  $ 1\times10^{7} $  &  $ 1\times10^{8} $ \\ \hline
$ 1\times10^{5} $  & PhD Num AP & $1102_{-654}^{+388}$ & $1358_{-810}^{+479}$ & $104_{-62}^{+37}$ & -\\
 & PhC Sens AP & 1110 & 1619 & 183 & - \\
 & PhC Sens MH & 1100 & 1600 & 190 & - \\
 & PhA Sens SB & 1247 & 2841 & 497 & - \\
 & PhC Sens SB & 1113 & 1622 & 184 & - \\
 & PhD Sens SB & 1095 & 1351 & 103 & - \\
 & PhD AvTDI SB & 1202 & 1373 & 103 & - \\
\hline
$ 1\times10^{6} $  & PhD Num AP & - & $6292_{-3754}^{+2220}$ & $665_{-396}^{+234}$ & $6_{-3}^{+2}$\\
 & PhC Sens AP & -  & 6156 & 678 & 9\\
 & PhC Sens MH & -  & 6200 & 690 & 9\\
 & PhA Sens SB & -  & 7549 & 1523 & 28\\
 & PhC Sens SB & -  & 6169 & 680 & 9\\
 & PhD Sens SB & -  & 6259 & 663 & 6\\
 & PhD AvTDI SB & -  & 6343 & 663 & 6\\
\hline
$ 1\times10^{7} $  & PhD Num AP & - & - & $2136_{-1274}^{+753}$ & $37_{-22}^{+13}$\\
 & PhC Sens AP & -  & -  & 2108 & 35\\
 & PhC Sens MH & -  & -  & 2100 & 36\\
 & PhA Sens SB & -  & -  & 2546 & 88\\
 & PhC Sens SB & -  & -  & 2114 & 36\\
 & PhD Sens SB & -  & -  & 2131 & 37\\
 & PhD AvTDI SB & -  & -  & 2132 & 37\\
\hline
$ 1\times10^{8} $  & PhD Num AP & - & - & - & $98_{-58}^{+35}$\\
 & PhC Sens AP & -  & -  & -  & 96\\
 & PhC Sens MH & -  & -  & -  & 99\\
 & PhA Sens SB & -  & -  & -  & 116\\
 & PhC Sens SB & -  & -  & -  & 96\\
 & PhD Sens SB & -  & -  & -  & 97\\
 & PhD AvTDI SB & -  & -  & -  & 98\\
\hline
\end{tabular}
\end{center}
\caption{SNRs for TC1, i.e. $\chi_1 = \chi_2 = 0$ and redshift 1.}
\label{tab:SNRsTC1}
\end{table}

%%%%%%%%%%%%%%%%%%%%%%%%%%%%%
\subsection{Test case 2}

%Compute SNR for PhenomC for one TDI-X channel and the following setup:
%\begin{verbatim}
%# chi1 = chi2 = 0
%# z = 8
%# Dl = 81480.3
%# Source Masses 10^5, 10^6, 10^7, 10^8
%\end{verbatim}

Reference system: non-spinning $\chi_1 = \chi_2=0$, redshift $z=8$, 
luminosity distance $D_L = 81480.3$ Mpc, source frame individual masses
$m_i=10^5, 10^6, 10^7, 10^8$.

\begin{figure}[htbp]
\centering
\includegraphics[width=0.48\textwidth]{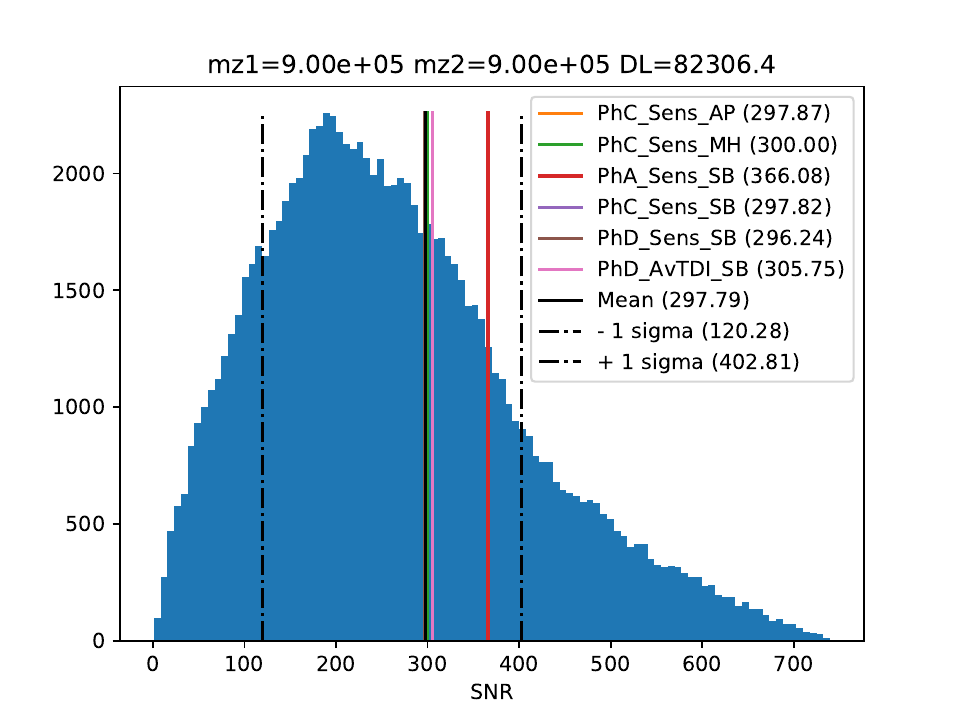}
\includegraphics[width=0.48\textwidth]{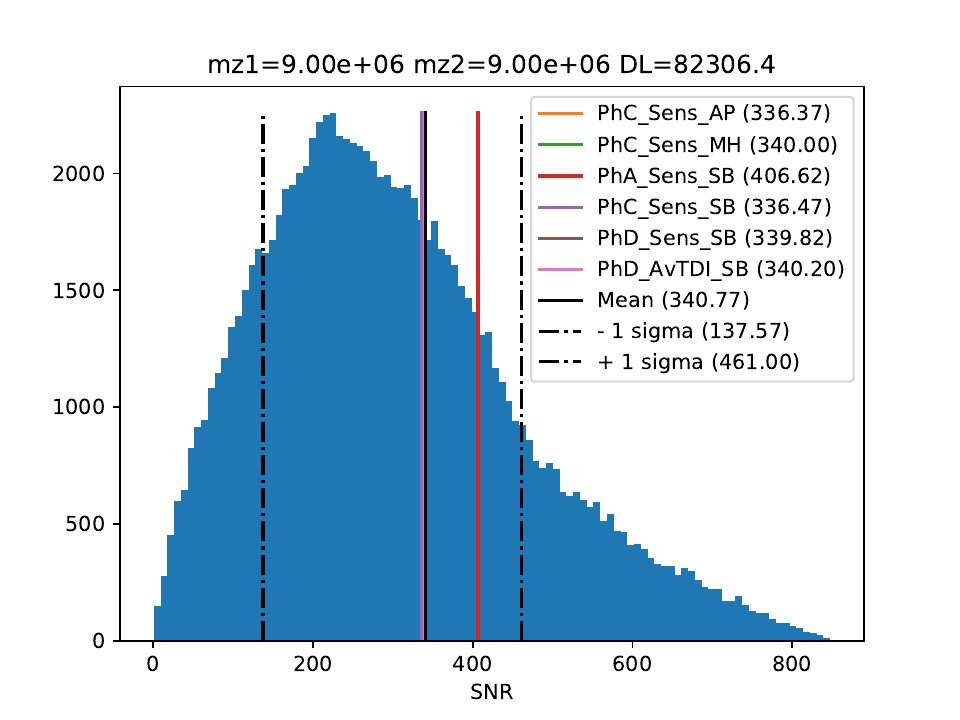}
\includegraphics[width=0.48\textwidth]{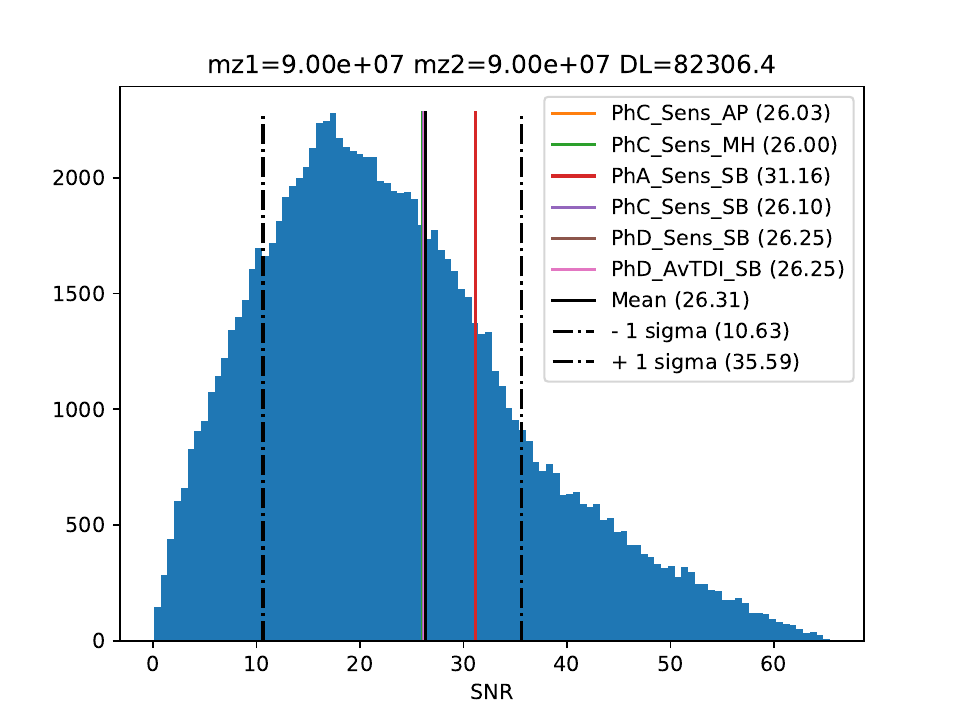}
\includegraphics[width=0.48\textwidth]{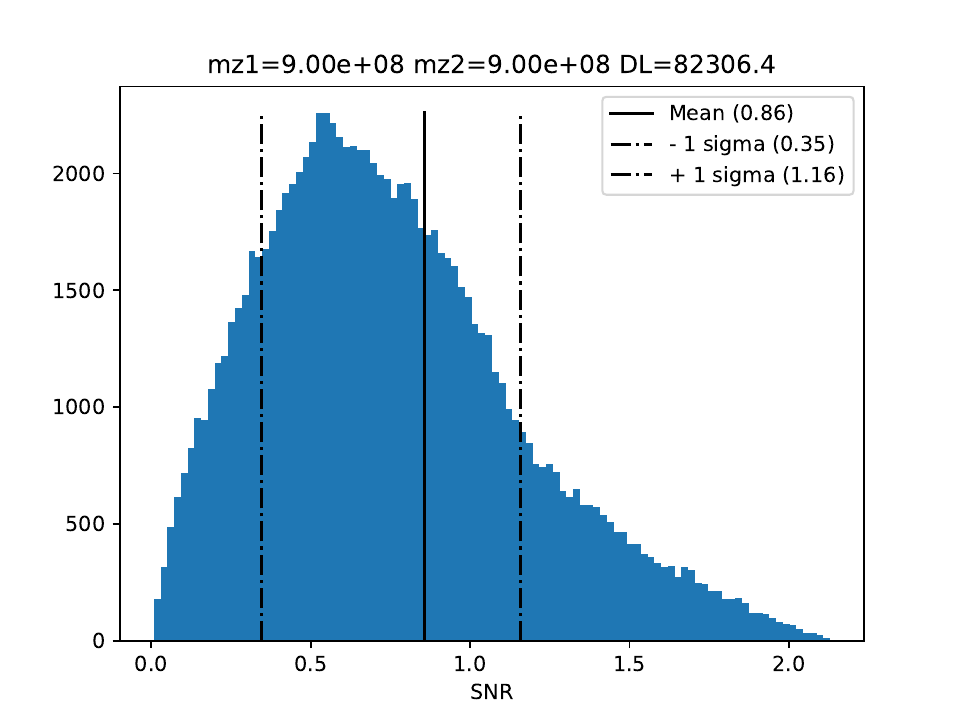}
\includegraphics[width=0.48\textwidth]{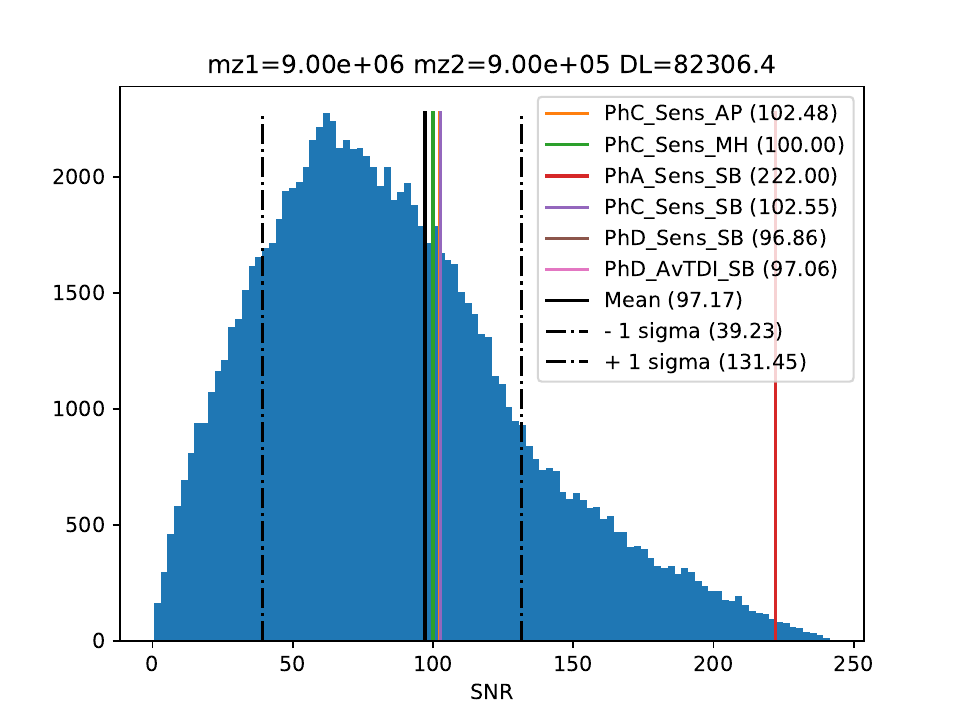}
\includegraphics[width=0.48\textwidth]{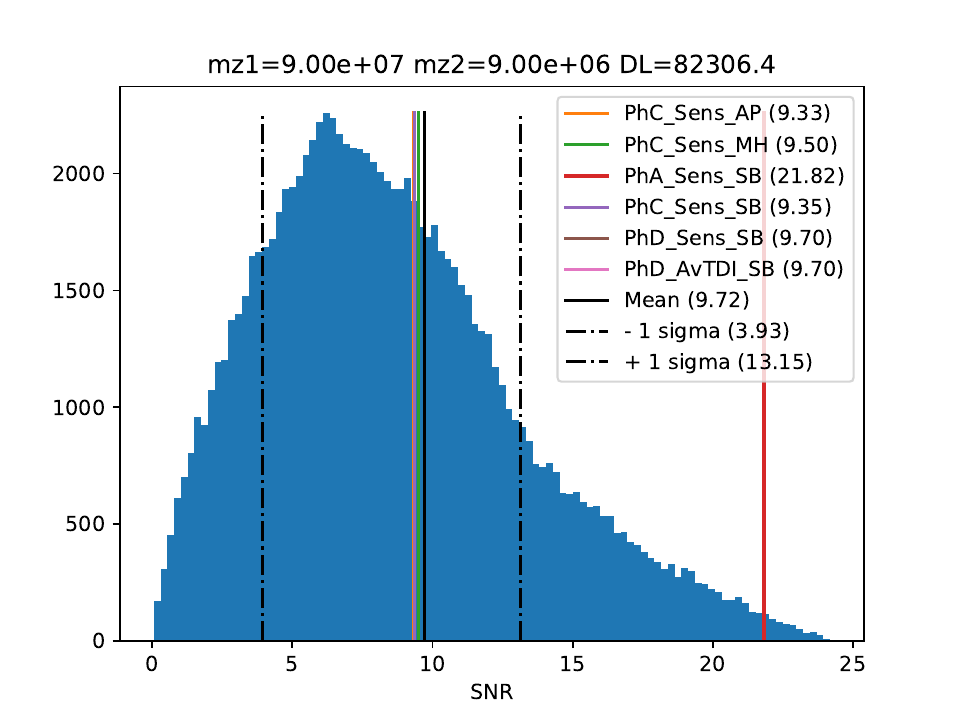}
\includegraphics[width=0.48\textwidth]{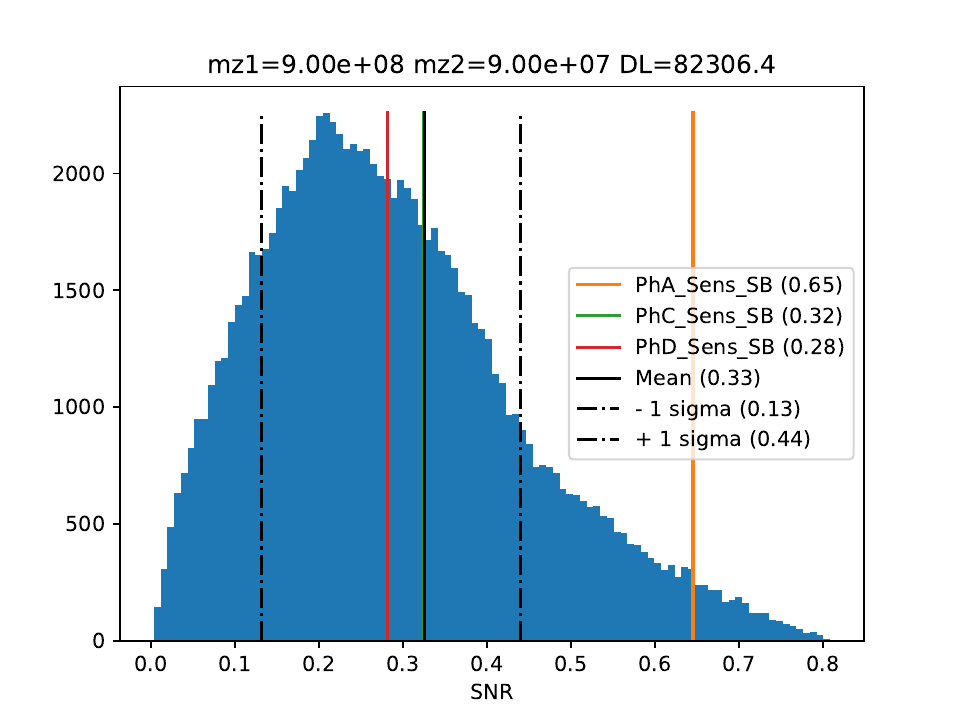}
\includegraphics[width=0.48\textwidth]{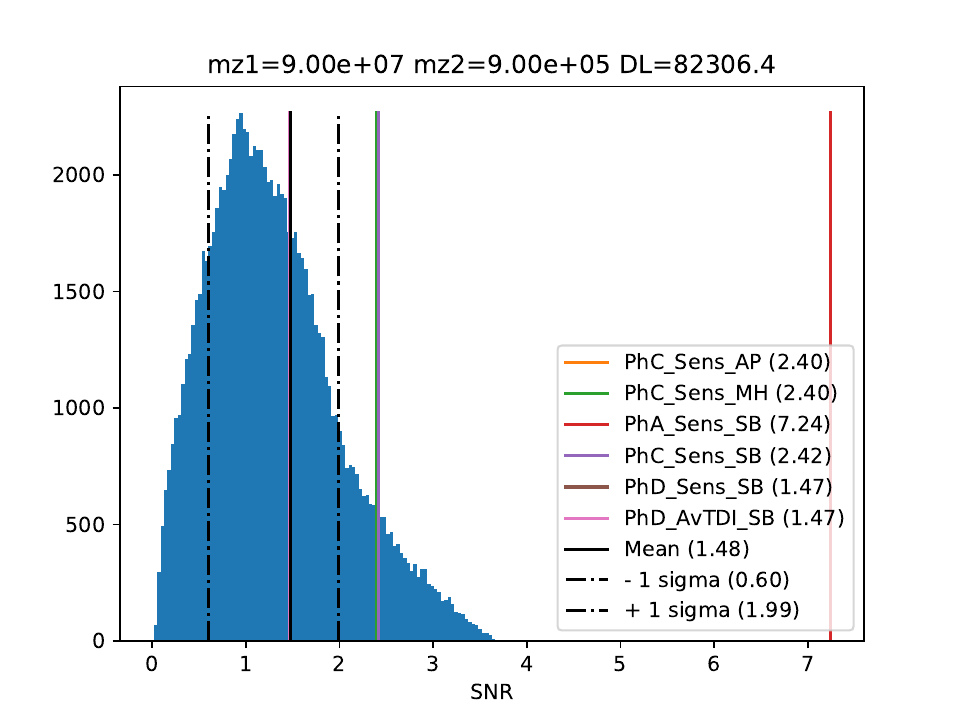}
\caption{TC2: Numerical results and analytical results for the non-spinning cases and redshift 8.
The intrinsic masses of the source are :
row 1: $10^5-10^5$ and $10^6-10^6$,
row 2: $10^7-10^7$ and $10^8-10^8$,
row 3: $10^6-10^5$ and $10^7-10^6$,
row 4: $10^8-10^7$ and $10^7-10^5$}
\label{fig:HistTC2}
\end{figure}

The table~\ref{tab:SNRsTC2} is summarizing the results with various 
methods and the figure~\ref{fig:HistTC2} is showing a comparison of the results 
including the distribution of SNRs.

%\begin{table}[htp]
%\begin{center}
%\begin{tabular}{|c|c|c|c|c|} \hline
%M1 | M2 & $ 1\times10^{5} $  &  $ 1\times10^{6} $  &  $ 1\times10^{7} $  &  $ 1\times10^{8} $ \\ \hline
%$ 1\times10^{5} $ & $ 3\times10^{2} $ & $ 1\times10^{2} $ & $ 2.4 $ & $ 0.02 $\\ \hline
%$ 1\times10^{6} $ & $ 1\times10^{2} $ & $ 3.4\times10^{2} $ & $ 9.5 $ & $ 0.063 $\\ \hline
%$ 1\times10^{7} $ & $ 2.4 $ & $ 9.5 $ & $ 26 $ & $ 0.27 $\\ \hline
%$ 1\times10^{8} $ & $ 0.02 $ & $ 0.063 $ & $ 0.27 $ & $ 0.38 $\\ \hline
%\end{tabular}
%\end{center}
%\caption{TC2 SNRs for LISAModule Code}
%\label{tab:LISAModule_SNR_TC2}
%\end{table}

\begin{table}[ht]
\begin{center}
\begin{tabular}{|c|c|c|c|c|c|}
\hline
 \multicolumn{2}{|c|}{$m_1 | m_2$} & $ 1\times10^{5} $  &  $ 1\times10^{6} $  &  $ 1\times10^{7} $  &  $ 1\times10^{8} $ \\ \hline
$ 1\times10^{5} $  & PhD Num AP & $298_{-178}^{+105}$ & $97_{-58}^{+34}$ & $1_{-1}^{+1}$ & -\\
 & PhC Sens AP & 298 & 102 & 2 & - \\
 & PhC Sens MH & 300 & 100 & 2 & - \\
 & PhA Sens SB & 366 & 222 & 7 & - \\
 & PhC Sens SB & 298 & 103 & 2 & - \\
 & PhD Sens SB & 296 & 97 & 1 & - \\
 & PhD AvTDI SB & 306 & 97 & 1 & - \\
\hline
$ 1\times10^{6} $  & PhD Num AP & - & $341_{-203}^{+120}$ & $10_{-6}^{+3}$ & $0_{-0}^{+0}$\\
 & PhC Sens AP & -  & 336 & 9 & -\\
 & PhC Sens MH & -  & 340 & 10 & -\\
 & PhA Sens SB & -  & 407 & 22 & 0\\
 & PhC Sens SB & -  & 336 & 9 & 0\\
 & PhD Sens SB & -  & 340 & 10 & 0\\
 & PhD AvTDI SB & -  & 340 & 10 & -\\
\hline
$ 1\times10^{7} $  & PhD Num AP & - & - & $26_{-16}^{+9}$ & $0_{-0}^{+0}$\\
 & PhC Sens AP & -  & -  & 26 & -\\
 & PhC Sens MH & -  & -  & 26 & -\\
 & PhA Sens SB & -  & -  & 31 & 1\\
 & PhC Sens SB & -  & -  & 26 & 0\\
 & PhD Sens SB & -  & -  & 26 & 0\\
 & PhD AvTDI SB & -  & -  & 26 & -\\
\hline
$ 1\times10^{8} $  & PhD Num AP & - & - & - & $1_{-1}^{+0}$\\
 & PhC Sens AP & -  & -  & -  & -\\
 & PhC Sens MH & -  & -  & -  & -\\
 & PhA Sens SB & -  & -  & -  & -\\
 & PhC Sens SB & -  & -  & -  & -\\
 & PhD Sens SB & -  & -  & -  & -\\
 & PhD AvTDI SB & -  & -  & -  & -\\
\hline
\end{tabular}
\end{center}

\caption{SNRs for TC2, i.e. $\chi_1 = \chi_2 = 0$ and redshift 8.}
\label{tab:SNRsTC2}
\end{table}

%%%%%%%%%%%%%%%%%%%%%%%%%%%%%
\subsection{Test case 3}

Reference system: equal spins $\chi_1 = \chi_2=0.5$, redshift $z=1$, 
luminosity distance $D_L = 6823$ Mpc, source frame individual masses
$m_i=10^5, 10^6, 10^7, 10^8$. 

%Compute SNR for one TDI-X channel and the following setup:
%\begin{verbatim}
%# chi1 = chi2 = 0.5
%# z = 1
%# Dl = 6823
%# Source Masses 10^5, 10^6, 10^7, 10^8
%\end{verbatim}

\begin{figure}[htbp]
\centering
\includegraphics[width=0.48\textwidth]{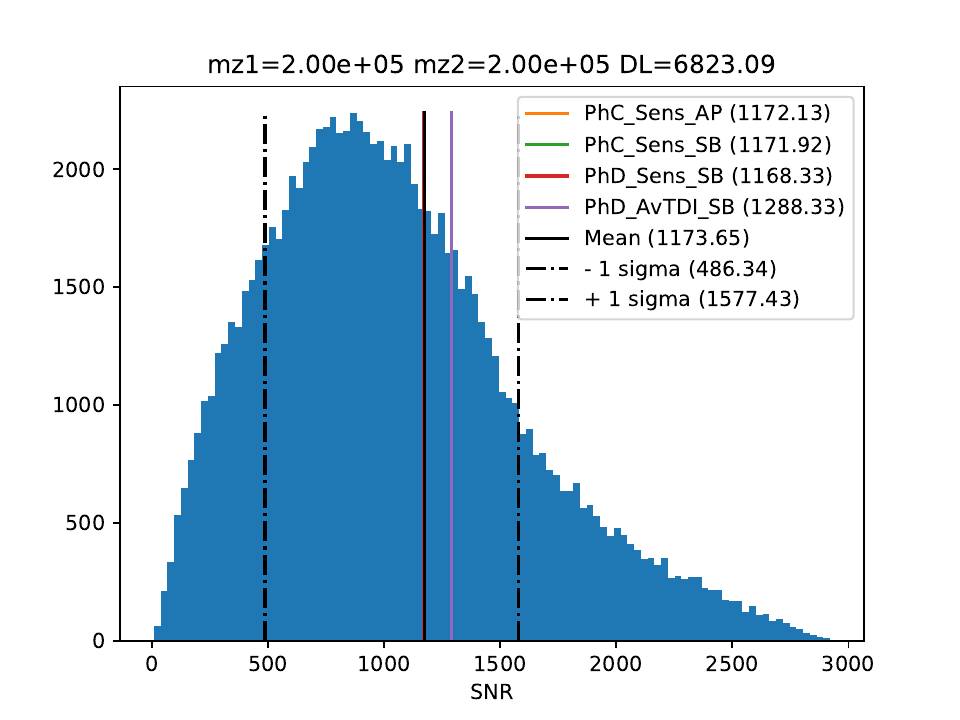}
\includegraphics[width=0.48\textwidth]{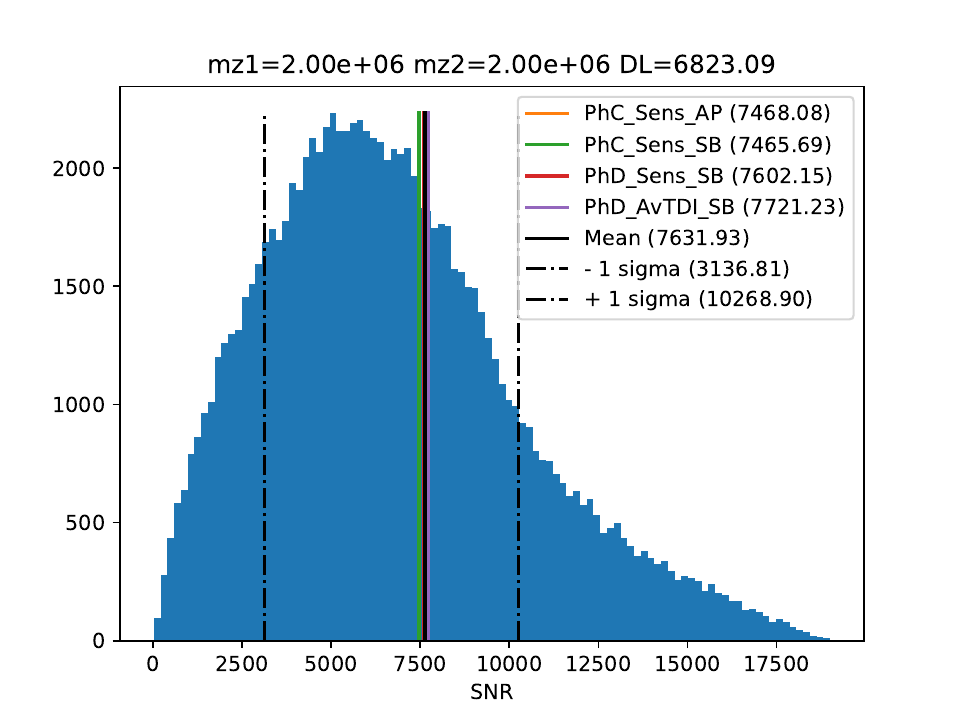}
\includegraphics[width=0.48\textwidth]{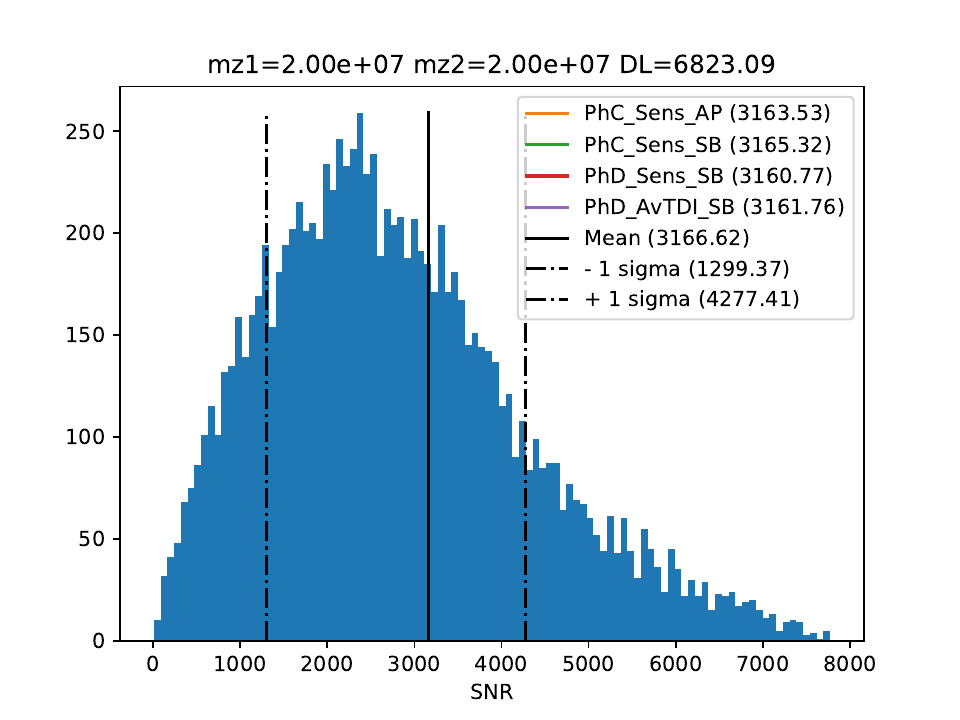}
\includegraphics[width=0.48\textwidth]{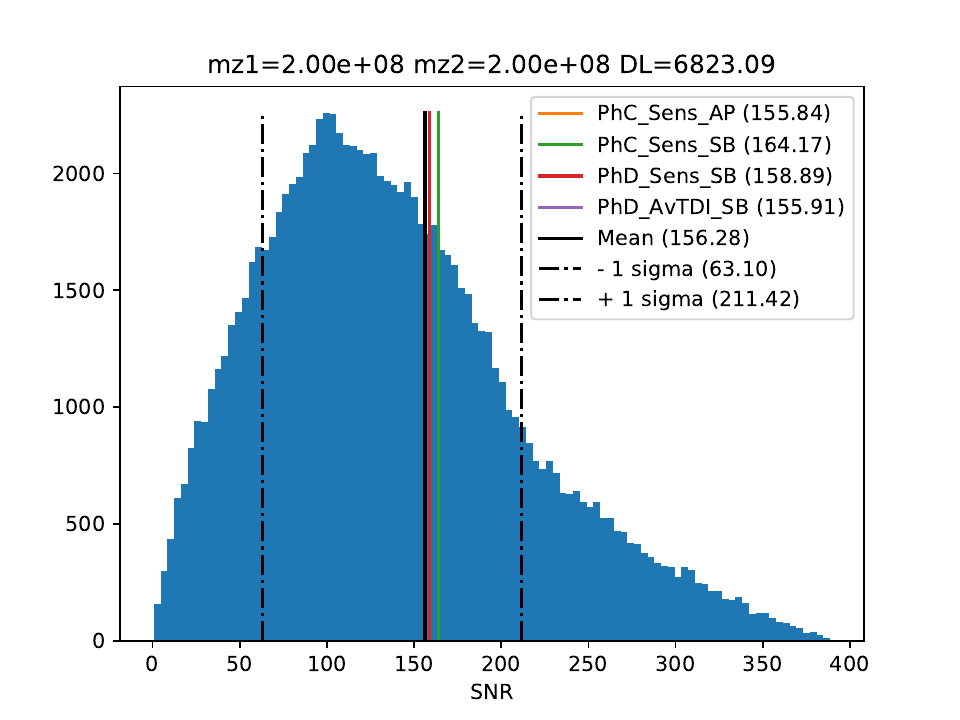}
\includegraphics[width=0.48\textwidth]{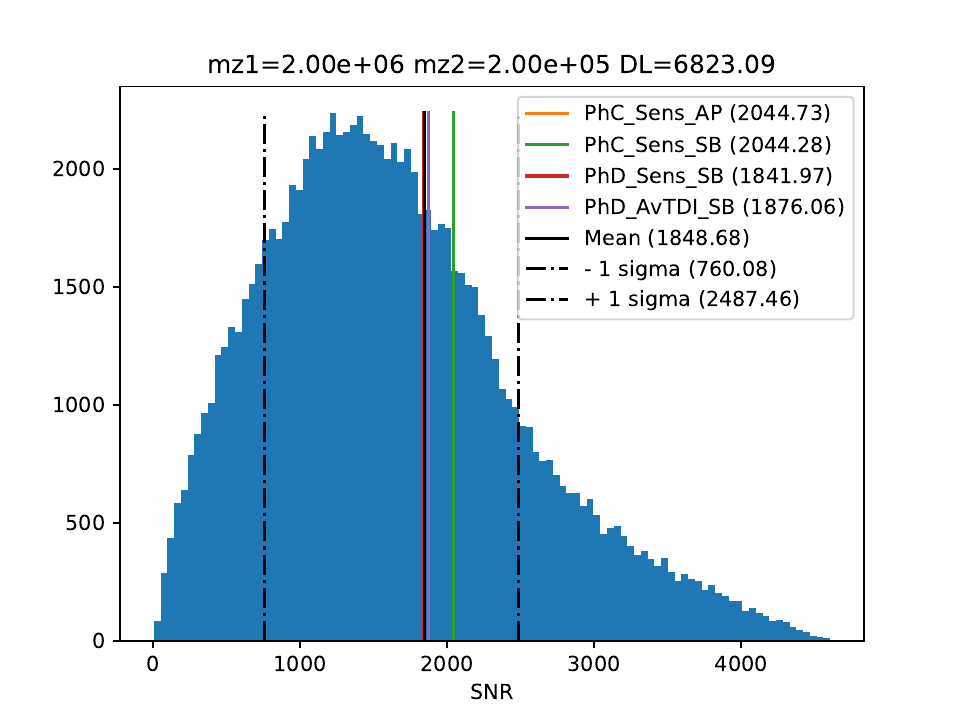}
\includegraphics[width=0.48\textwidth]{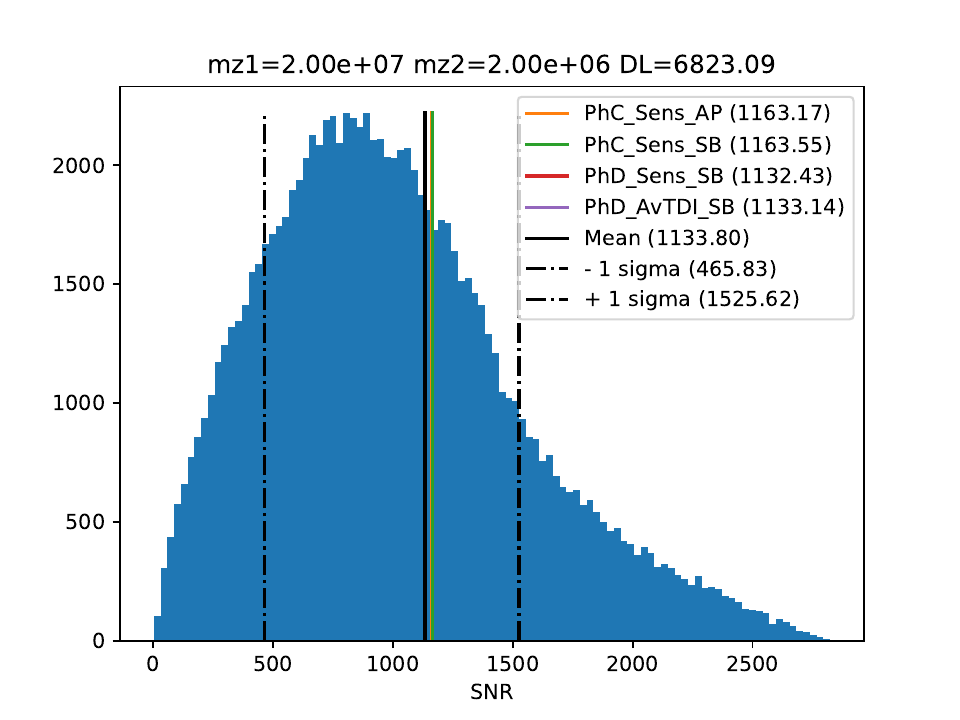}
\includegraphics[width=0.48\textwidth]{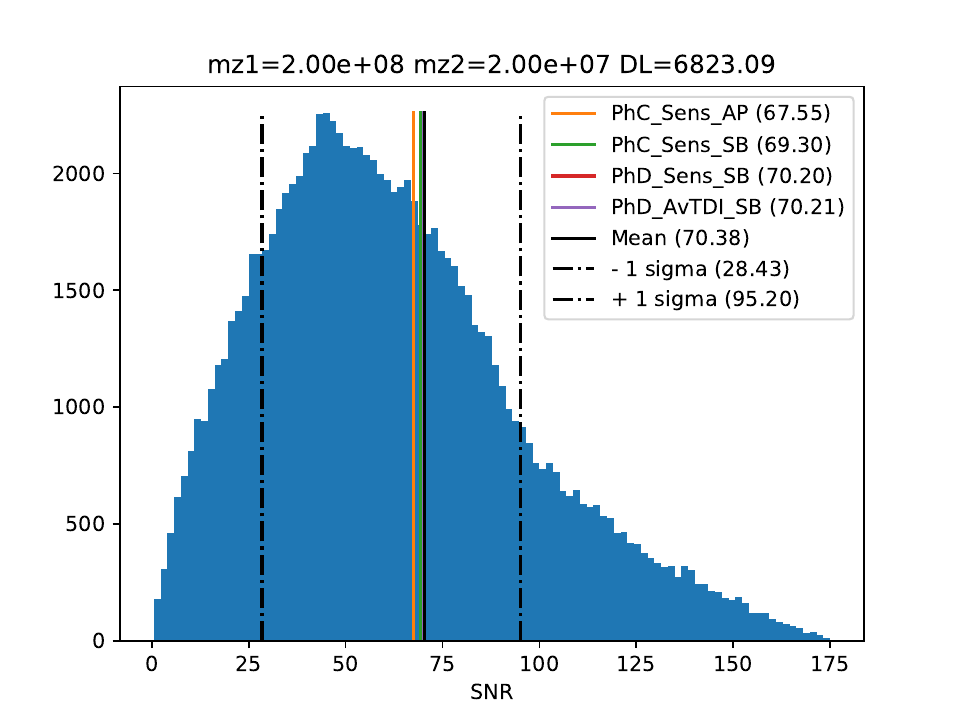}
\includegraphics[width=0.48\textwidth]{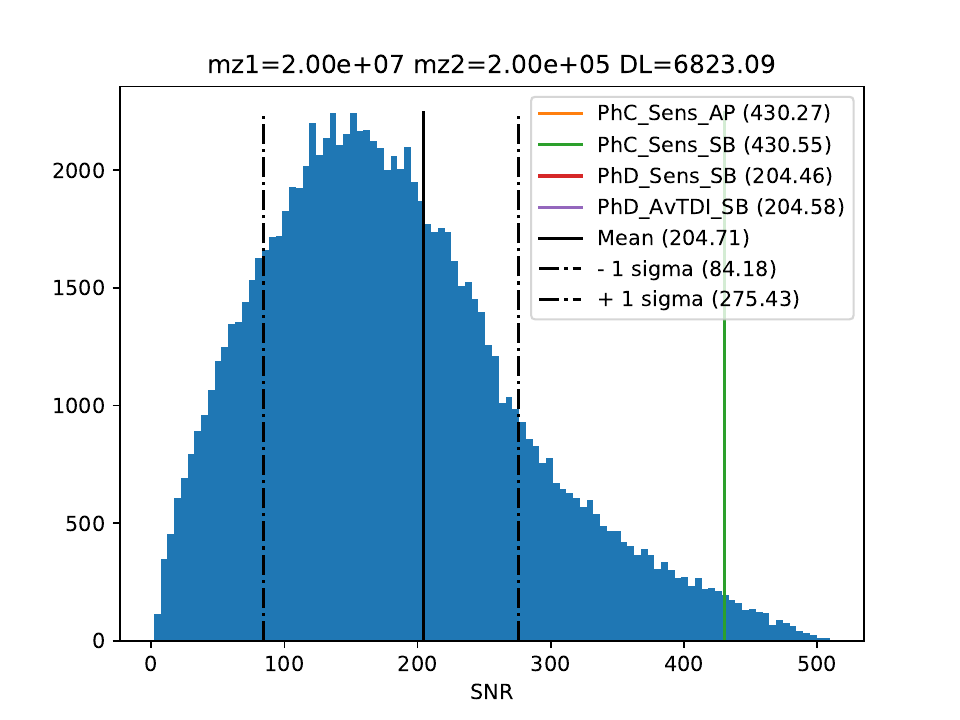}
\caption{TC3: Numerical results and analytical results for $\chi_1 = \chi_2 = 0.5$ and redshift 1.
The intrinsic masses of the source are :
row 1: $10^5-10^5$ and $10^6-10^6$,
row 2: $10^7-10^7$ and $10^8-10^8$,
row 3: $10^6-10^5$ and $10^7-10^6$,
row 4: $10^8-10^7$ and $10^7-10^5$}
\label{fig:HistTC3}
\end{figure}

The table~\ref{tab:SNRsTC3} is summarizing the results with various 
methods and the figure~\ref{fig:HistTC3} is showing a comparison of the results 
including the distribution of SNRs.

\begin{table}[ht]
\begin{center}
\begin{tabular}{|c|c|c|c|c|c|}
\hline
 \multicolumn{2}{|c|}{$m_1 | m_2$} & $ 1\times10^{5} $  &  $ 1\times10^{6} $  &  $ 1\times10^{7} $  &  $ 1\times10^{8} $ \\ \hline
$ 1\times10^{5} $  & PhD Num AP & $1174_{-687}^{+404}$ & $1851_{-1104}^{+653}$ & $205_{-122}^{+72}$ & -\\
 & PhC Sens AP & 1172 & 2045 & 430 & - \\
 & PhC Sens SB & 1172 & 2044 & 431 & - \\
 & PhD Sens SB & 1168 & 1842 & 204 & - \\
 & PhD AvTDI SB & 1288 & 1876 & 205 & - \\
\hline
$ 1\times10^{6} $  & PhD Num AP & - & $7643_{-4558}^{+2697}$ & $1135_{-677}^{+400}$ & $12_{-7}^{+4}$\\
 & PhC Sens AP & -  & 7468 & 1163 & 27\\
 & PhC Sens SB & -  & 7466 & 1164 & 29\\
 & PhD Sens SB & -  & 7602 & 1132 & 12\\
 & PhD AvTDI SB & -  & 7721 & 1133 & 12\\
\hline
$ 1\times10^{7} $  & PhD Num AP & - & - & $3169_{-1889}^{+1117}$ & $70_{-42}^{+25}$\\
 & PhC Sens AP & -  & -  & 3164 & 68\\
 & PhC Sens SB & -  & -  & 3165 & 69\\
 & PhD Sens SB & -  & -  & 3161 & 70\\
 & PhD AvTDI SB & -  & -  & 3162 & 70\\
\hline
$ 1\times10^{8} $  & PhD Num AP & - & - & - & $156_{-93}^{+55}$\\
 & PhC Sens AP & -  & -  & -  & 156\\
 & PhC Sens SB & -  & -  & -  & 164\\
 & PhD Sens SB & -  & -  & -  & 159\\
 & PhD AvTDI SB & -  & -  & -  & 156\\
\hline
\end{tabular}
\end{center}
\caption{SNRs for TC3, i.e. $\chi_1 = \chi_2 = 0.5$ and redshift 1.}
\label{tab:SNRsTC3}
\end{table}

%%%%%%%%%%%%%%%%%%%%%%%%%%%%%
\subsection{Test case 4}

Reference system: equal spins $\chi_1 = \chi_2=-0.5$, redshift $z=1$, 
luminosity distance $D_L = 6823$ Mpc, source frame individual masses
$m_i=10^5, 10^6, 10^7, 10^8$. 

%Compute SNR for one TDI-X channel and the following setup:
%\begin{verbatim}
%# chi1 = chi2 = -0.5
%# z = 1
%# Dl = 6823
%# Source Masses 10^5, 10^6, 10^7, 10^8
%\end{verbatim}

\begin{figure}[htbp]
\centering
\includegraphics[width=0.48\textwidth]{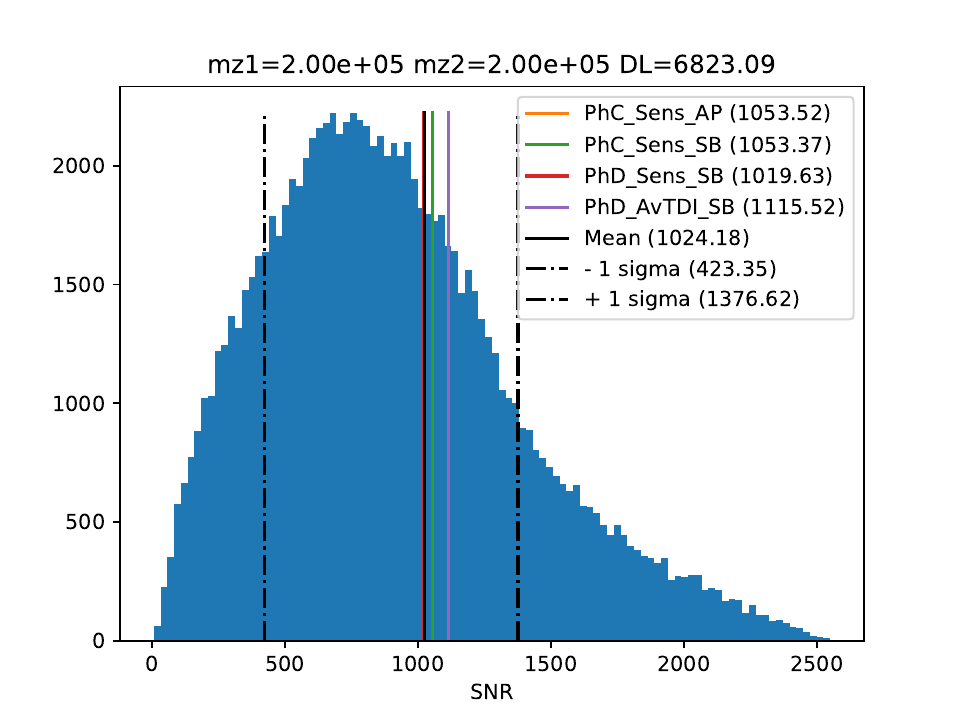}
\includegraphics[width=0.48\textwidth]{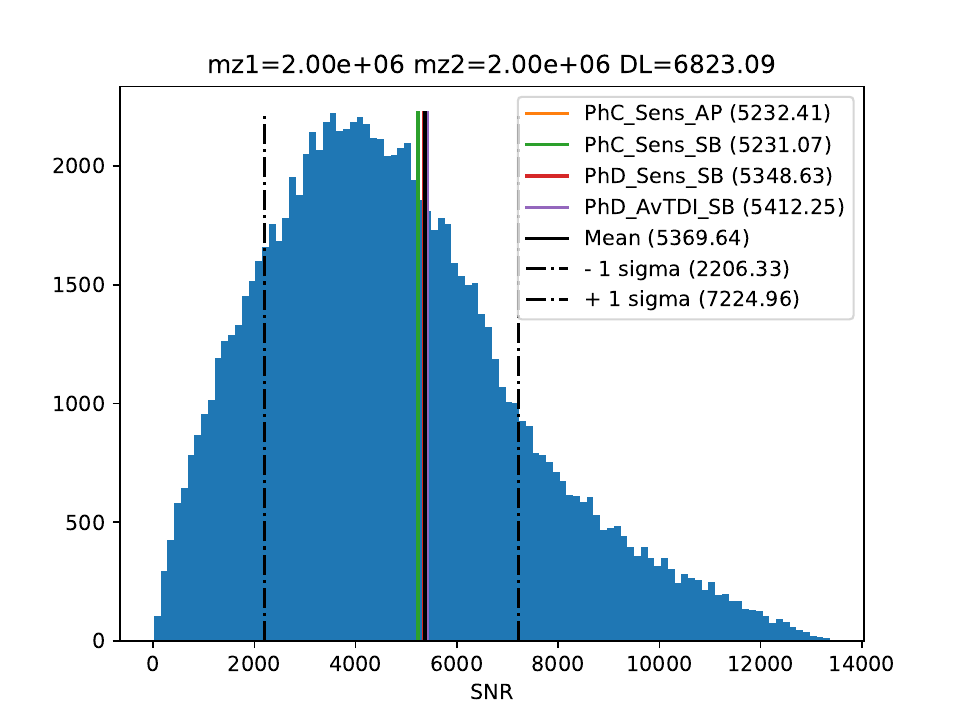}
\includegraphics[width=0.48\textwidth]{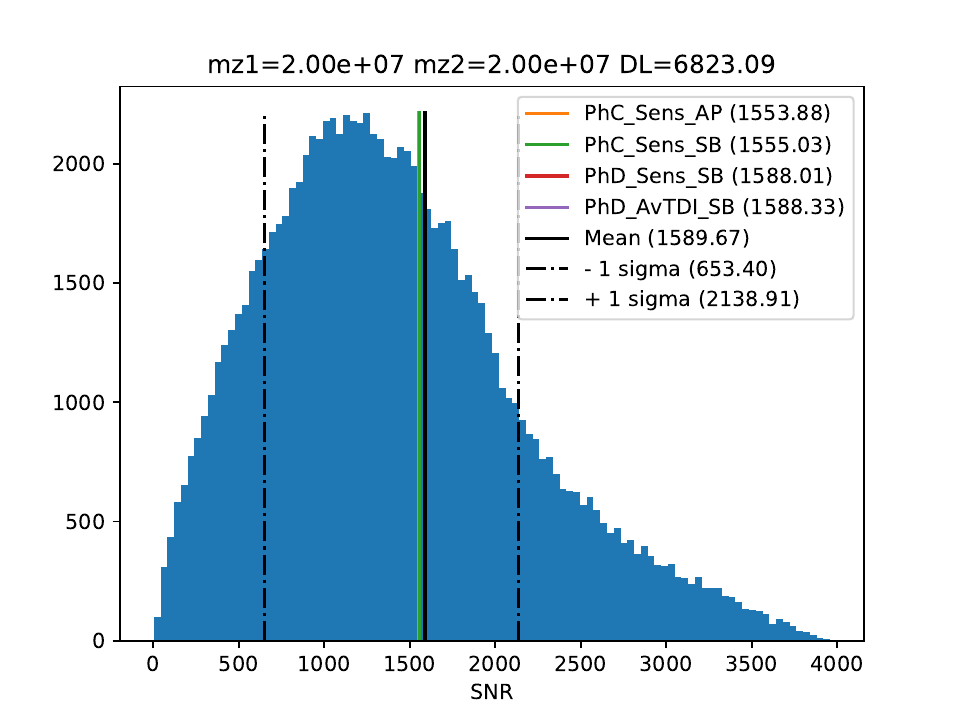}
\includegraphics[width=0.48\textwidth]{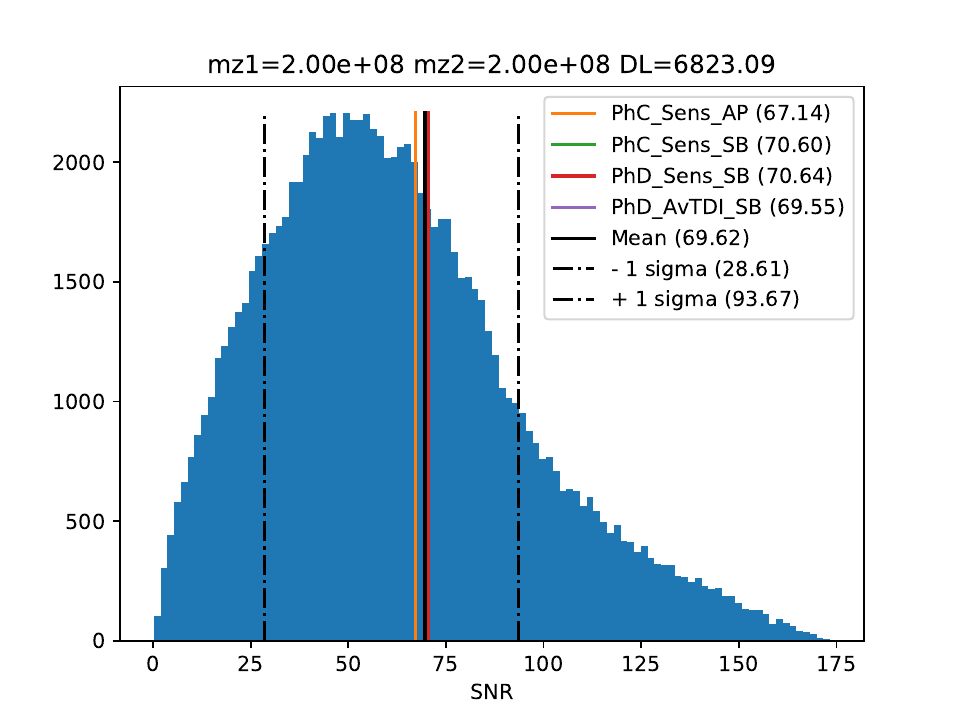}
\includegraphics[width=0.48\textwidth]{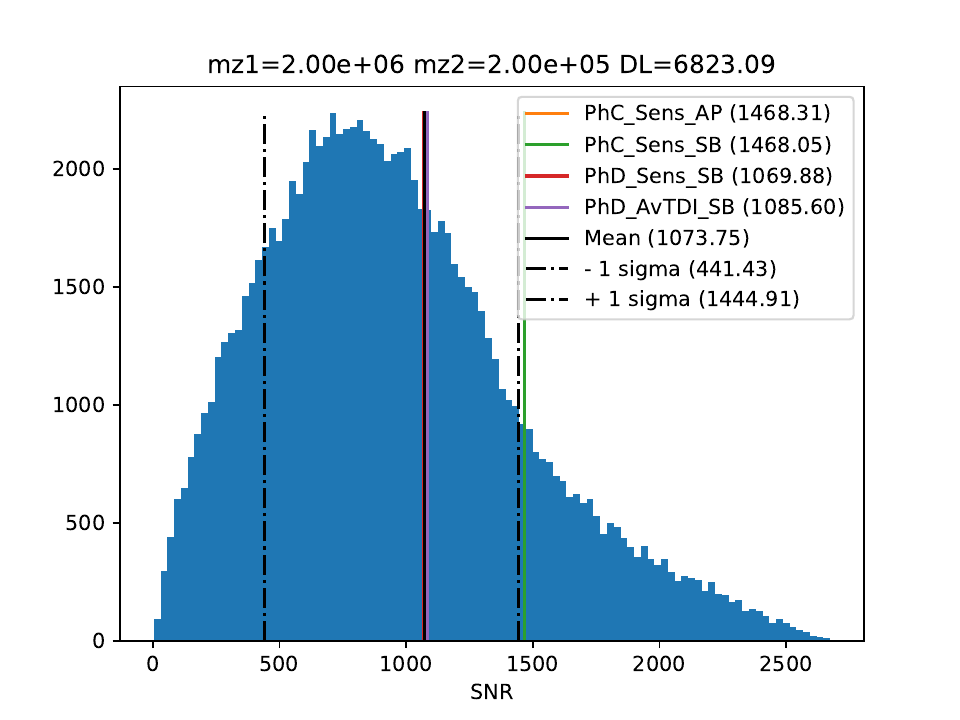}
\includegraphics[width=0.48\textwidth]{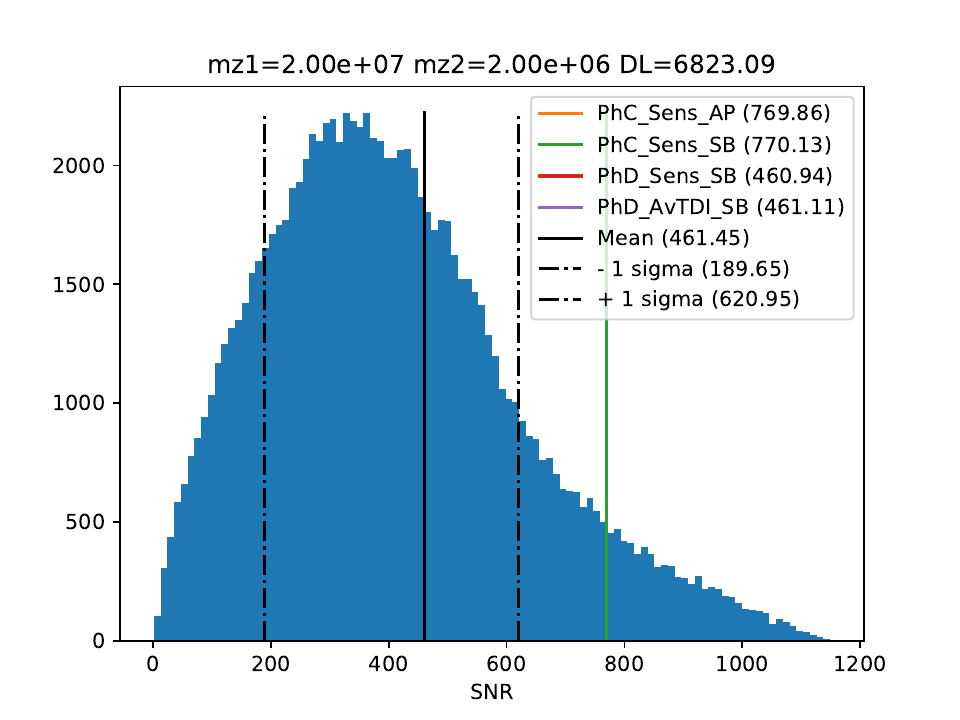}
\includegraphics[width=0.48\textwidth]{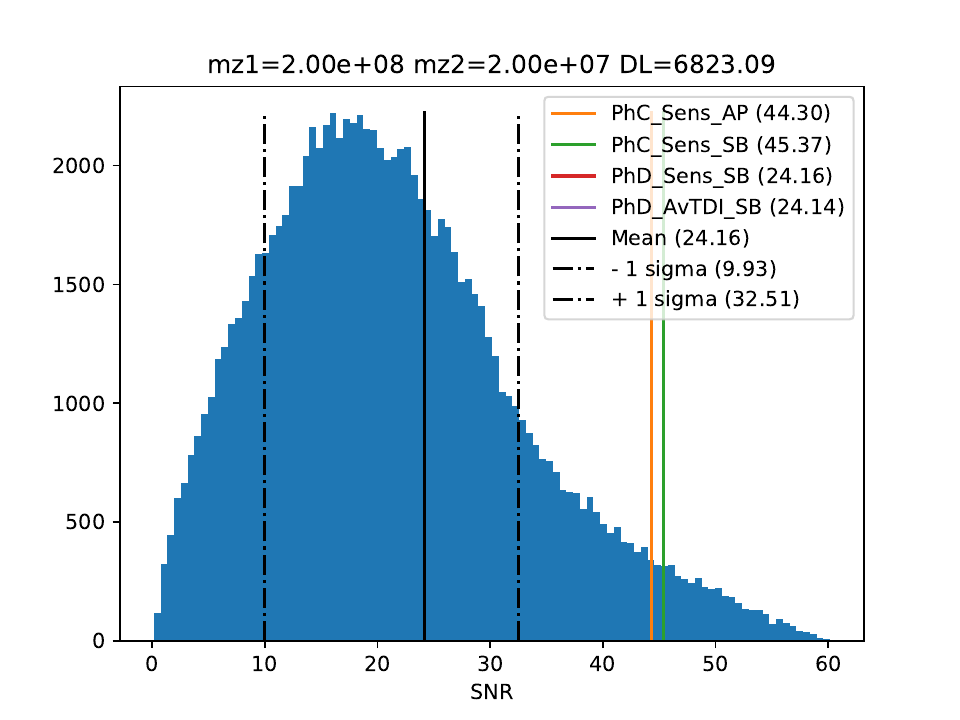}
\includegraphics[width=0.48\textwidth]{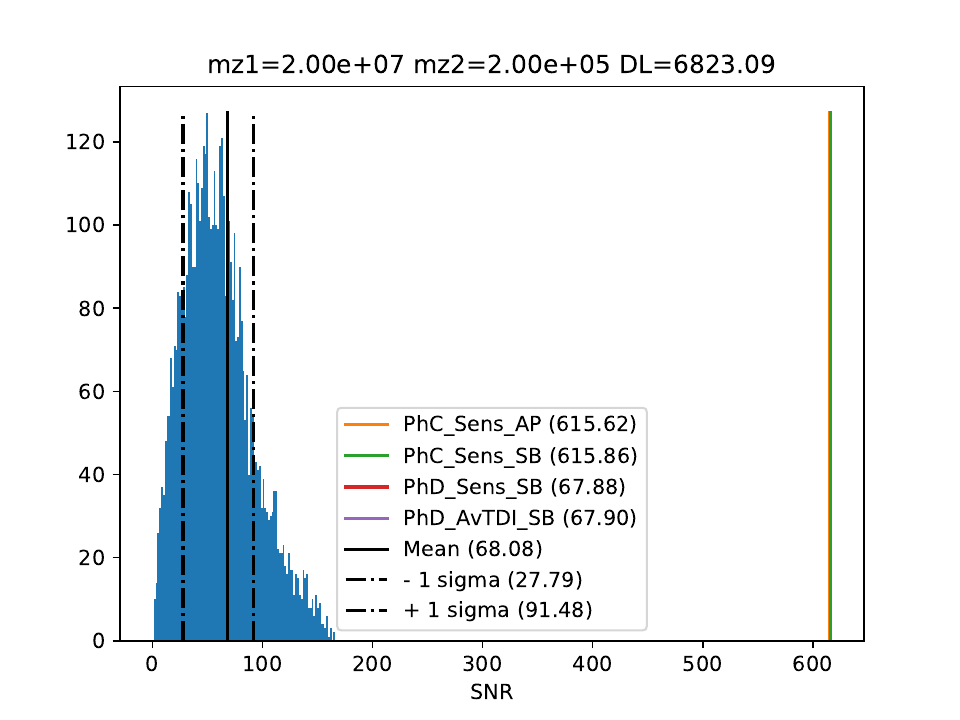}
\caption{TC4: Numerical results and analytical results for $\chi_1 = \chi_2 = 0.5$ and redshift 1.
The intrinsic masses of the source are :
row 1: $10^5-10^5$ and $10^6-10^6$,
row 2: $10^7-10^7$ and $10^8-10^8$,
row 3: $10^6-10^5$ and $10^7-10^6$,
row 4: $10^8-10^7$ and $10^7-10^5$}
\label{fig:HistTC4}
\end{figure}

The table~\ref{tab:SNRsTC4} is summarizing the results with various 
methods and the figure~\ref{fig:HistTC4} is showing a comparison of the results 
including the distribution of SNRs.

\begin{table}[ht]
\begin{center}
\begin{tabular}{|c|c|c|c|c|c|}
\hline
 \multicolumn{2}{|c|}{$m_1 | m_2$} & $ 1\times10^{5} $  &  $ 1\times10^{6} $  &  $ 1\times10^{7} $  &  $ 1\times10^{8} $ \\ \hline
$ 1\times10^{5} $  & PhD Num & $1024_{-601}^{+352}$ & $1074_{-632}^{+371}$ & $68_{-40}^{+23}$ & -\\
 & PhD Sens & 1020 & 1070 & 68 & - \\
 & PhD AvTDI & 1116 & 1086 & 68 & - \\
% & PhC Sens AP & 1054 & 1468 & 616 & - \\
 & PhC Sens & 1053 & 1468 & 616 & - \\
\hline
$ 1\times10^{6} $  & PhD Num & - & $5370_{-3163}^{+1855}$ & $461_{-272}^{+159}$ & $3_{-2}^{+1}$\\
 & PhD Sens & -  & 5349 & 461 & 3\\
 & PhD AvTDI & -  & 5412 & 461 & 3\\
% & PhC Sens AP & -  & 5232 & 770 & 45\\
 & PhC Sens & -  & 5231 & 770 & 42\\
\hline
$ 1\times10^{7} $  & PhD Num & - & - & $1590_{-936}^{+549}$ & $24_{-14}^{+8}$\\
 & PhD Sens & -  & -  & 1588 & 24\\
 & PhD AvTDI & -  & -  & 1588 & 24\\
% & PhC Sens AP & -  & -  & 1554 & 44\\
 & PhC Sens & -  & -  & 1555 & 45\\
\hline
$ 1\times10^{8} $  & PhD Num & - & - & - & $70_{-41}^{+24}$\\
 & PhD Sens & -  & -  & -  & 71\\
 & PhD AvTDI & -  & -  & -  & 70\\
% & PhC Sens AP & -  & -  & -  & 67\\
 & PhC Sens & -  & -  & -  & 71\\
\hline
\end{tabular}
\end{center}
\caption{TC4 SNRs.}
\label{tab:SNRsTC4}
\end{table}

%%%%%%%%%%%%%%%%%%%%%%%%%%%%%
\subsection{Test case 5}

Reference system: spins $\chi_1 =0.7,\;  \chi_2=0.9$, redshift $z=1$, 
luminosity distance $D_L = 6823$ Mpc, source frame individual masses
$m_i=10^5, 10^6, 10^7, 10^8$. 

%Compute SNR for one TDI-X channel and the following setup:
%\begin{verbatim}
%# chi1 = 0.7
%# chi2 = 0.9
%# z = 1
%# Dl = 6823
%# Source Masses 10^5, 10^6, 10^7, 10^8
%\end{verbatim}

\begin{figure}[htbp]
\centering
\includegraphics[width=0.48\textwidth]{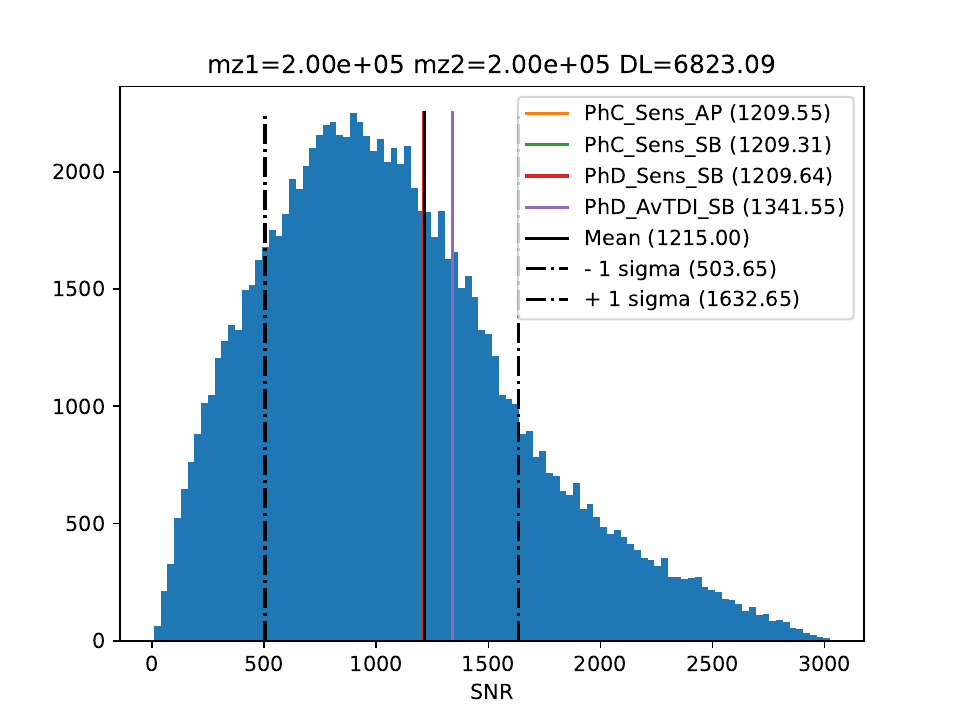}
\includegraphics[width=0.48\textwidth]{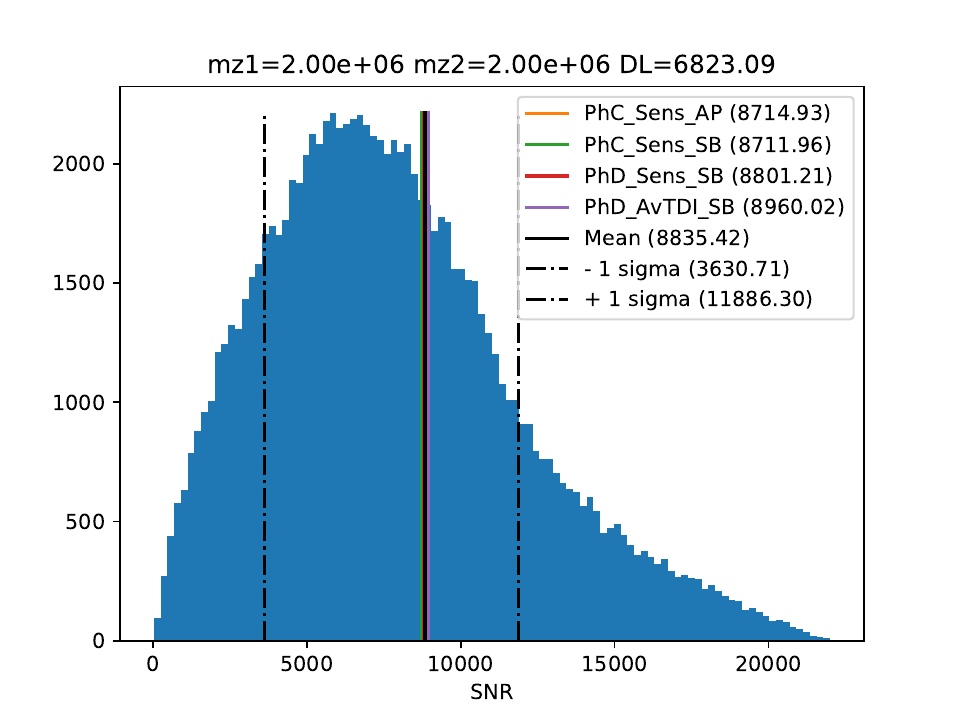}
\includegraphics[width=0.48\textwidth]{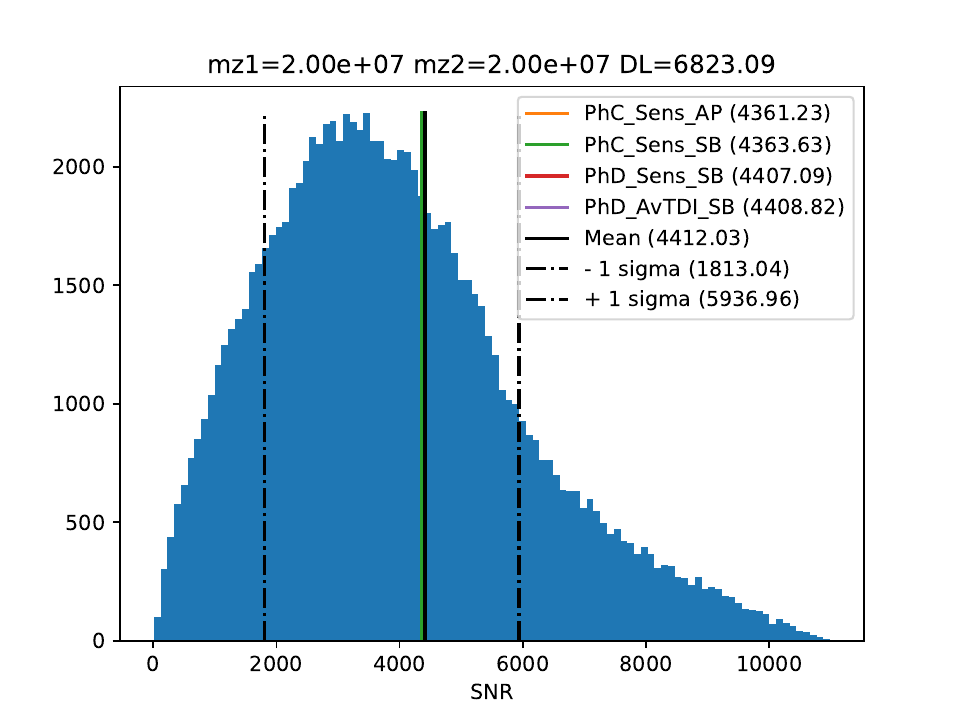}
\includegraphics[width=0.48\textwidth]{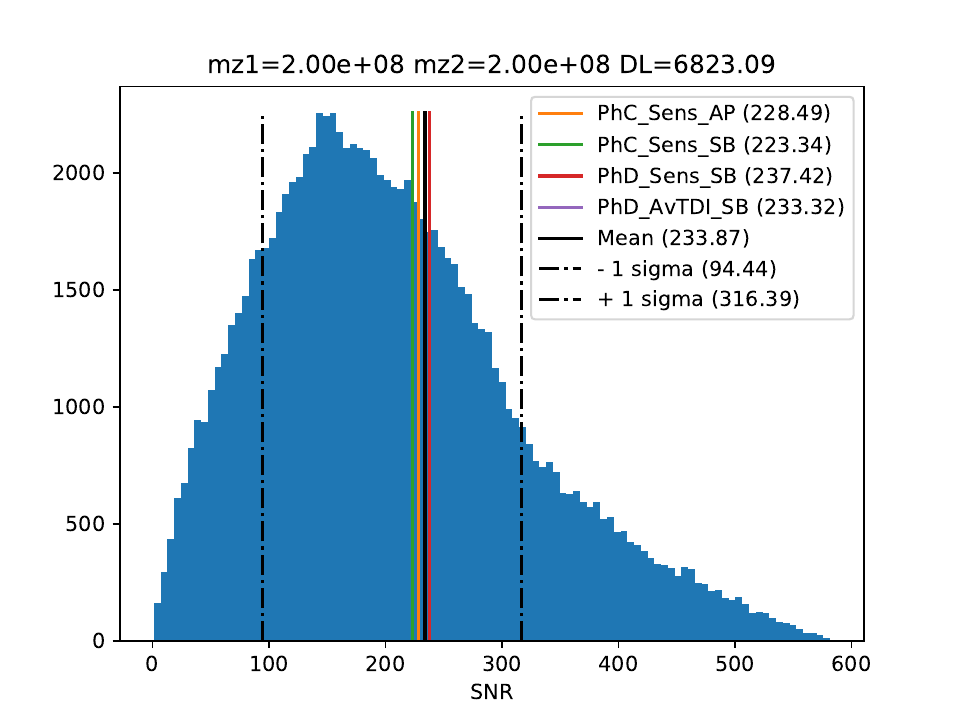}
\includegraphics[width=0.48\textwidth]{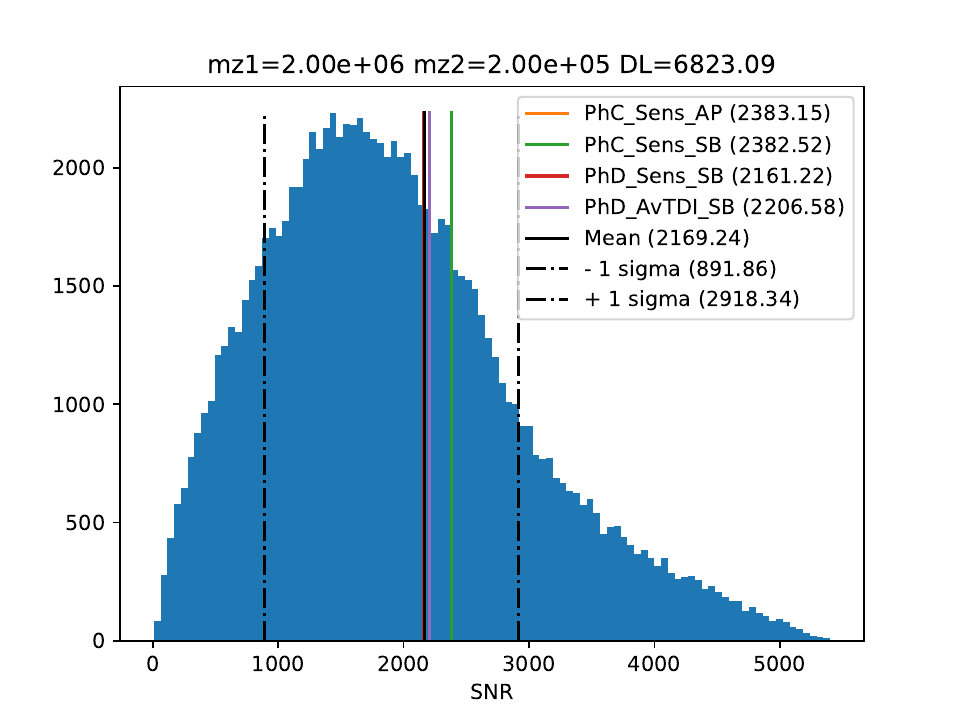}
\includegraphics[width=0.48\textwidth]{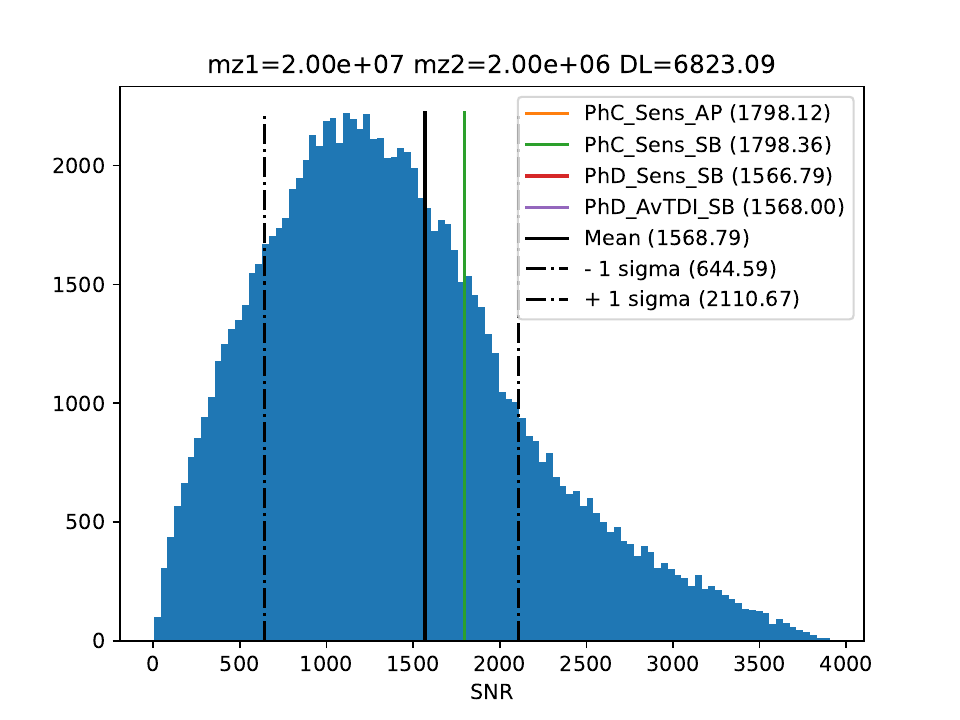}
\includegraphics[width=0.48\textwidth]{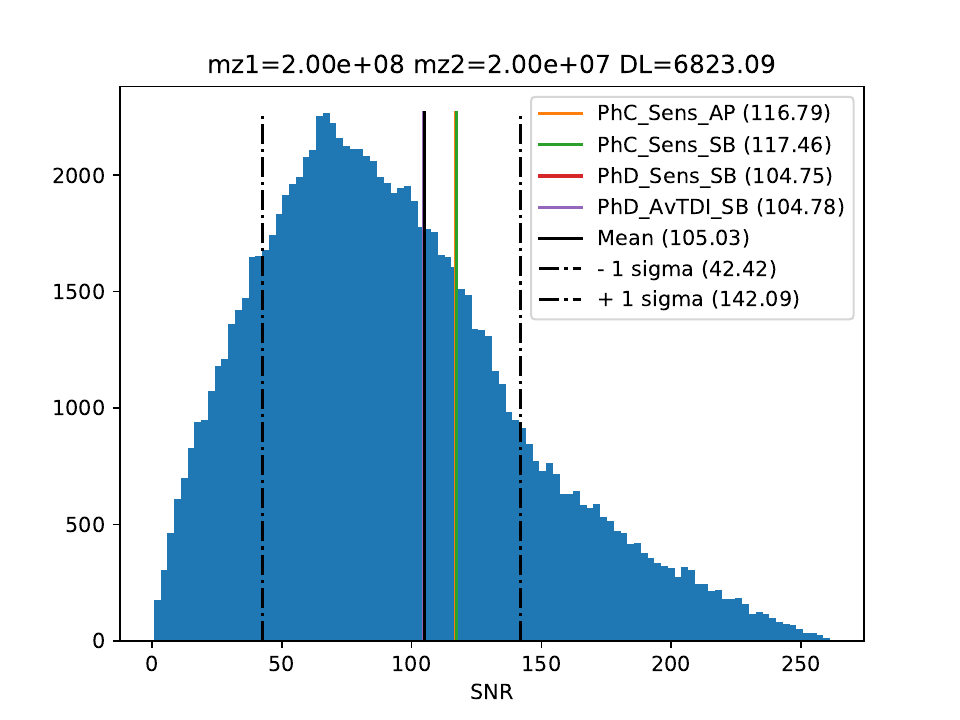}
\includegraphics[width=0.48\textwidth]{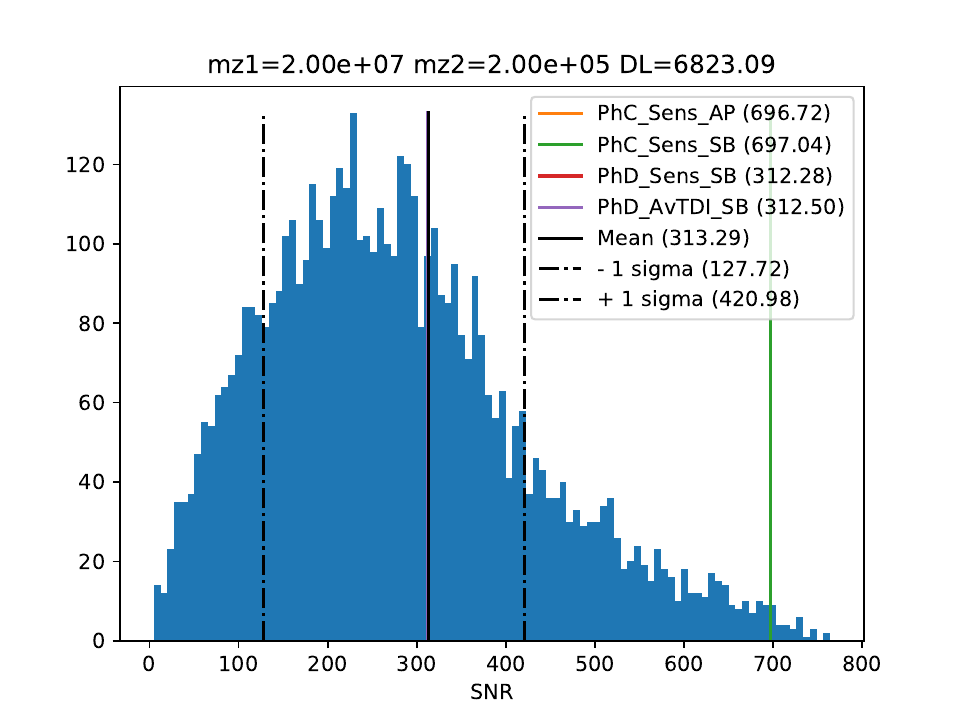}
\caption{TC4: Numerical results and analytical results for $\chi_1 = 0.7 $ and $ \chi_2 = 0.9$ and redshift 1.
The intrinsic masses of the source are :
row 1: $10^5-10^5$ and $10^6-10^6$,
row 2: $10^7-10^7$ and $10^8-10^8$,
row 3: $10^6-10^5$ and $10^7-10^6$,
row 4: $10^8-10^7$ and $10^7-10^5$}
\label{fig:HistTC5}
\end{figure}

The table~\ref{tab:SNRsTC5} is summarizing the results with various 
methods and the figure~\ref{fig:HistTC5} is showing a comparison of the results 
including the distribution of SNRs.

\begin{table}[ht]
\begin{center}
\begin{tabular}{|c|c|c|c|c|c|}
\hline
 \multicolumn{2}{|c|}{$m_1 | m_2$} & $ 1\times10^{5} $  &  $ 1\times10^{6} $  &  $ 1\times10^{7} $  &  $ 1\times10^{8} $ \\ \hline
$ 1\times10^{5} $  & PhD Num & $1215_{-711}^{+418}$ & $2169_{-1277}^{+749}$ & $313_{-186}^{+108}$ & -\\
 & PhD Sens & 1210 & 2161 & 312 & - \\
 & PhD AvTDI & 1342 & 2207 & 313 & - \\
% & PhC Sens AP & 1210 & 2383 & 697 & - \\
 & PhC Sens & 1209 & 2383 & 697 & - \\
\hline
$ 1\times10^{6} $  & PhD Num & - & $8835_{-5205}^{+3051}$ & $1569_{-924}^{+542}$ & $20_{-12}^{+7}$\\
 & PhD Sens & -  & 8801 & 1567 & 20\\
 & PhD AvTDI & -  & 8960 & 1568 & 20\\
% & PhC Sens AP & -  & 8715 & 1798 & 49\\
 & PhC Sens & -  & 8712 & 1798 & 54\\
\hline
$ 1\times10^{7} $  & PhD Num & - & - & $4412_{-2599}^{+1525}$ & $105_{-63}^{+37}$\\
 & PhD Sens & -  & -  & 4407 & 105\\
 & PhD AvTDI & -  & -  & 4409 & 105\\
% & PhC Sens AP & -  & -  & 4361 & 117\\
 & PhC Sens & -  & -  & 4364 & 117\\
\hline
$ 1\times10^{8} $  & PhD Num & - & - & - & $234_{-139}^{+83}$\\
 & PhD Sens & -  & -  & -  & 237\\
 & PhD AvTDI & -  & -  & -  & 233\\
% & PhC Sens AP & -  & -  & -  & 228\\
 & PhC Sens & -  & -  & -  & 223\\
\hline
\end{tabular}
\end{center}
\caption{TC5 SNRs}
\label{tab:SNRsTC5}
\end{table}

%%%%%%%%%%%%%%%%%%%%%%%%%%%%%%%%%%%%%%%%%%%%%%%%%%%%%%%%%%%%%
\subsection{IMR model comparison.}

We have implemented and used three IMRPhenom models in the evaluation of SNR.
We need to be aware os systematics in those models: the models have different fidelity across the parameter space. The comparable mass ratio non-spinning NR waveforms were used to fit IMRPhenomA model, moreover, only leading order  SPA amplitude is used at low frequency.  IMRPhenomC was calibrated up to mass ratio 4 and uses combination of two spins as a parameter. The most accurate model (among considered) is IMRPhenomD, it uses two spin magnitudes and was
calibrated up to mass ratio 16. Use of any of these models outside the
corresponding domain of validity might lead to erroneous results.

\begin{figure}[htbp]
\centering
\includegraphics[width=1.0\textwidth]{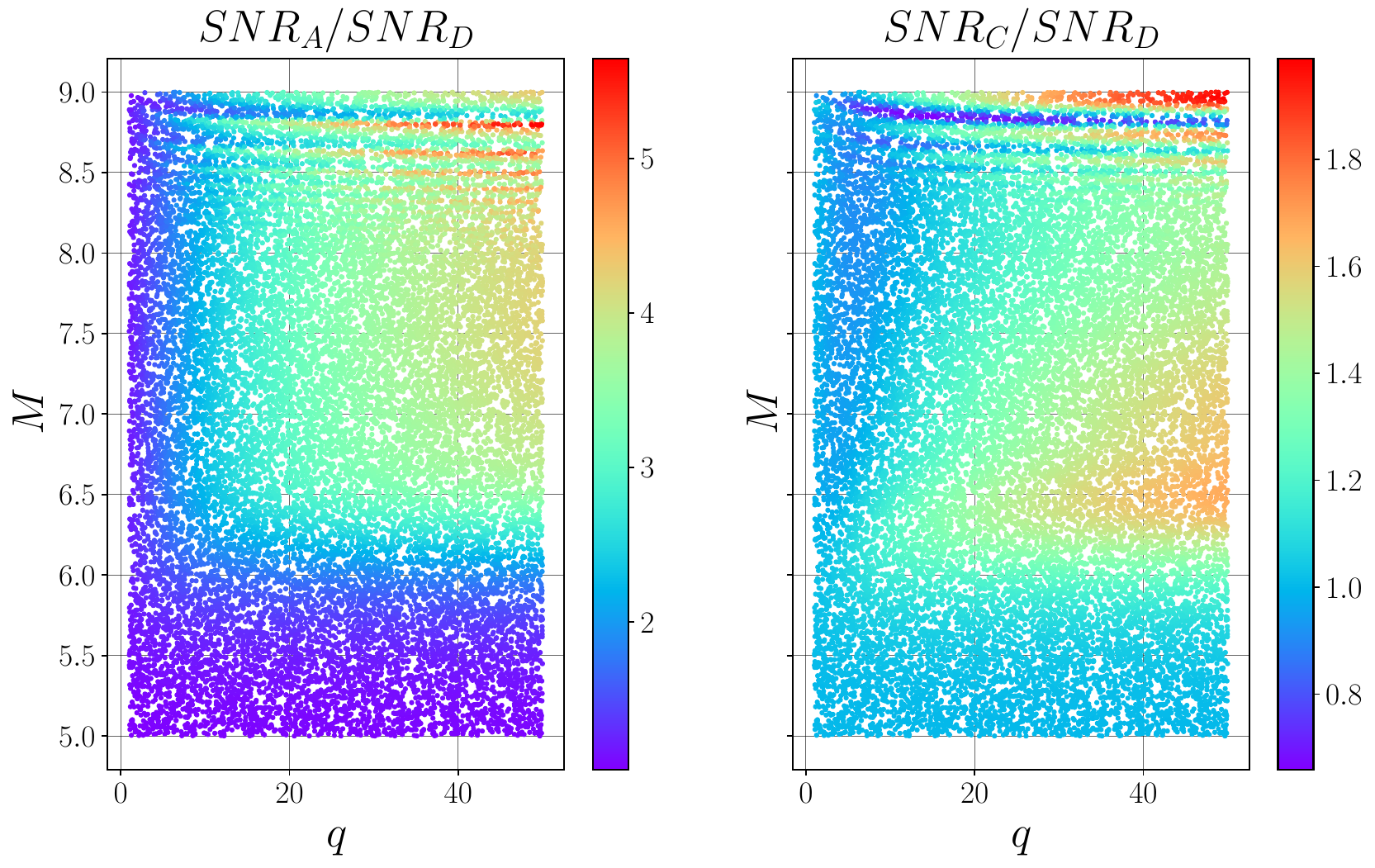}
\caption{Monte carlo simulation. Computation of SNR ratio for IMR models. Total mass $M=m_1+m_2$ vs mass ratio $q=m_1/m_2$, SNR ratio is color-coded.}
\label{fig:Mq_MC}
\end{figure}

\begin{figure}[htbp]
\centering
\includegraphics[width=1.0\textwidth]{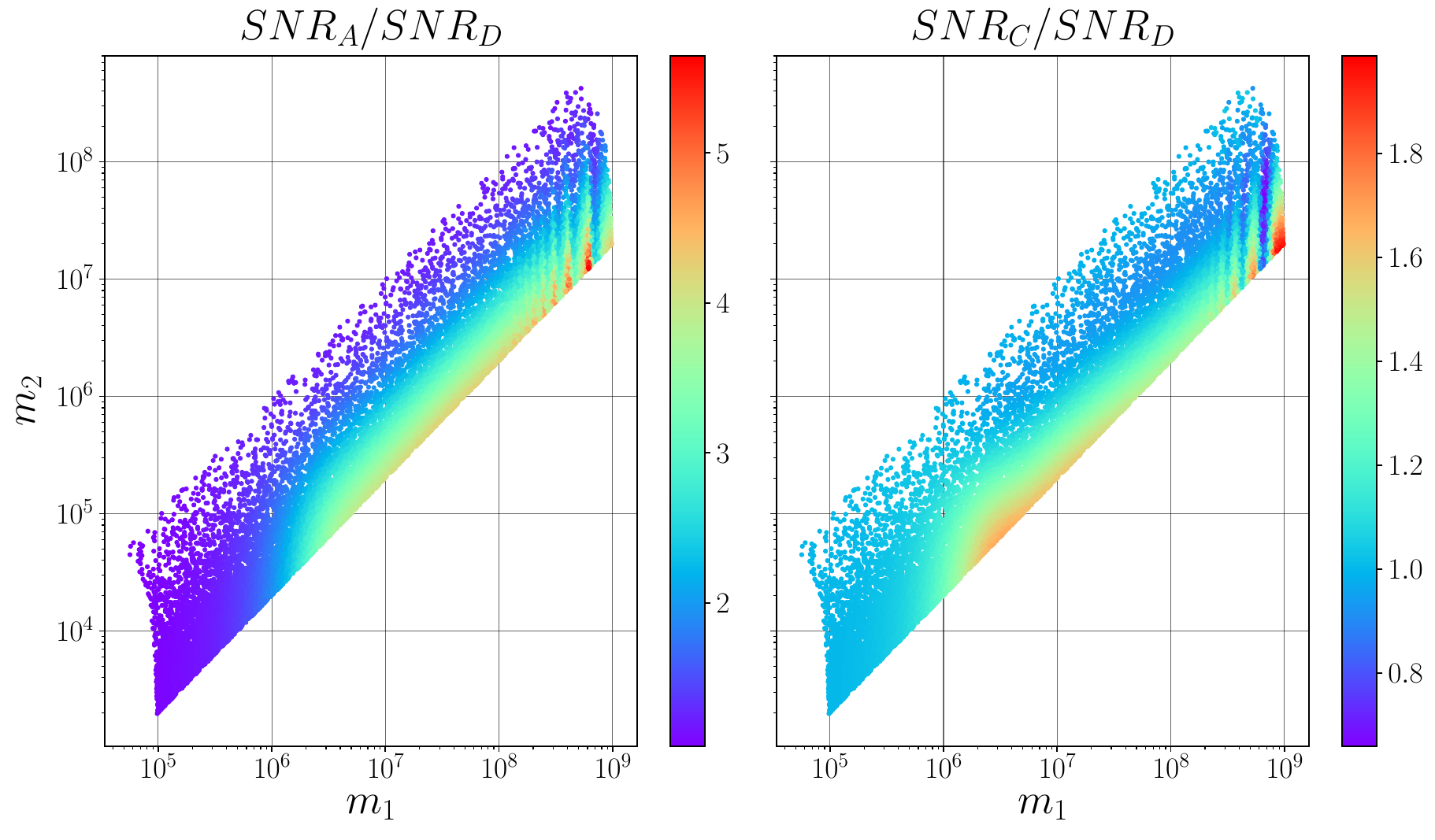}
\caption{Monte carlo simulation. Computation of SNR ratio for IMR models. Component masses (observed) are given along axes, SNR ratio is color-coded.}
\label{fig:m1m2_MC}
\end{figure}

We have performed the monte-carlo simulation in $m_1-m_2$ space
(uniform in the total mass $\log_{10}(m_1+m_2) \in U[5,9]$, uniform in the mass ratio $q=m1/m2 \in U[1,50]$) of non-spinning MBHBs.
For each model we have computed the sky, polarization, inclination
averaged SNR using three models. In two plots below [\ref{fig:Mq_MC}],
[\ref{fig:m1m2_MC}] we give (color-coded) ratios of $SNR_A/SNR_D$
and $SNR_C/SNR_D$.

%% ############################################################
\section{Combined SNR for individual source and SGWB}
\label{app:CombinedSNR}

We consider ``uncorrelated'' \ac{TDI} channels~\cite{Prince}, as for example A, E and T defined as (LDC convention):
\begin{eqnarray}
  A = \frac{Z-X}{\sqrt{2}}, \;
  E = \frac{X-2Y+Z}{\sqrt{6}}, \;
  T = \frac{X+Y+Z}{\sqrt{3}}
  \label{eq:AET}
\end{eqnarray}
Note: A, E and T are uncorrelated only theoritcially, i.e. all \ac{PSD} of X,Y and Z have to be identical as well as all the cross spectral density.

The \ac{SNR} is the quadratic sum of individual \ac{SNR} of the ``uncorrelated'' \ac{TDI} channels. 
The \ac{SNR} from the T channel is negligible so only A and E are really contributing. Therefore, the total \ac{SNR} is 
\begin{eqnarray}
  SNR_{tot}^2 = 2 \ SNR_{A,E}^{2}
\end{eqnarray}

For \textbf{individual source}, according to \eqref{eq:AvSNR2} and \eqref{eq:SensDef}, the total average \ac{SNR} is:
\begin{eqnarray}
  <SNR_{tot}^2> & = &  2 \times 4 \Re \int_{0}^{f_\textrm{max}}  df  \frac{\tilde{h}_{+}^2 +  \tilde{h}^2_{\times}}{S_h(f)} 
   =  4 \Re \int_{0}^{f_\textrm{max}}  df  \frac{\tilde{h}_{+}^2 +  \tilde{h}^2_{\times}}{S_{eq,IS}(f)}
\end{eqnarray}
The sensitivity equivalent to 6 links configuration for individual sources is $ S_{eq,IS}(f) = S_{h}(f) / 2$ as indicated in \eqref{eq:Sens6L}.

For \textbf{\ac{SGWB}}, according to \eqref{eq:SNR_SGWB}, the total average \ac{SNR} is:
\begin{eqnarray}
<SNR_{tot}^2> & = &  2 \times T_{obs} \int_{f_{min}}^{f_{max}} df { \left( h^2 \Omega_{GW} (f) \right)^2 \over \left(h^2 \Omega_{Sens} (f) \right)^2 } \\
& = &  T_{obs} \int_{f_{min}}^{f_{max}} df { \left( h^2 \Omega_{GW} (f) \right)^2 \over \left(h^2 \Omega_{Sens,eq} (f) \right)^2 }
\end{eqnarray}

with $ h^2 \Omega_{Sens,eq} = h^2 \Omega_{Sens} (f) / \sqrt{2}$.
Given the definition of $\Omega_{Sens}$ \eqref{eq:OmegaGW_Sh}, the sensitivity equivalent to 6 links configuration for \ac{SGWB} is $ S_{eq,SGWB}(f) = S_{h}(f) / \sqrt{2}$.

%% ############################################################
\section{Noise PSD for A, E and T}
\label{app:NoisePSDAET}

The \ac{TDI} A, E, T are obtained from the diagonalisation of a 3$\times$3 matrix of TDI generators~\cite{Prince}. The A,E,T defined by~\eqref{eq:AET} are eigen vectors diaogonalisation of the X,Y,Z matrix. The corresponding eigen values are:
\begin{eqnarray}
	S_{n,A_{2.0}} = S_{E_{2.0}} & = & S_{n,X_{2.0}} - S_{n,XY_{2.0}}\\
	S_{n,T_{2.0}} & = & S_{n,X_{2.0}} + 2 S_{n,XY_{2.0}}
\end{eqnarray}
with $S_{n,XY_{2.0}}$ the cross spectral density between $X_{2.0}$ and $Y_{2.0}$. 
Using the same logic as for the computation of the $PSD$ (see Appendix~\ref{app:NoisePSDDerivation}) and applying the approximations 1+2+3, we get:
\begin{eqnarray}
	S_{n,XY_{2.0}} & = & - 16 \sin \left( \omega L \right) \sin^3 \left( 2 \omega L \right) \left( S_{OMS} + 4 S_{acc}\right)
\end{eqnarray}

Then 

\begin{eqnarray}
	S_{n,A_{2.0}} = S_{n,E_{2.0}} & = & 32 \sin^2 \left( \omega L \right) \sin^2 \left( 2 \omega L \right) 
	\left( 
	\left( 2 + \cos \left( \omega L \right) \right) S_{OMS} 
	+ 2 \left( 3 + 2 \cos \left( \omega L \right) + \cos \left( 2 \omega L \right) \right) S_{acc} \right) \nonumber \\
\end{eqnarray}

$X_{2.0}$ is given by~\eqref{eq:TimeTDIX20}.
We get $Y_{2.0}$ and $Z_{2.0}$ with circular permutation of~\ref{eq:TimeTDIX20}.:
\begin{eqnarray}
	Y_{2.0} & = &
	( 1 - D_{23} D_{32} D_{21} D_{12} ) 
	( (1  - D_{23} D_{32} )
	(N_{21} 
	+ D_{21}  N_{12}
	+ (1 + D_{21}  D_{12} )\delta_{21} 
	- 2 D_{21}  \delta_{12}) \nonumber \\
	& & - ( 1 - D_{21}  D_{12} ) (
	N_{23} 
	+ D_{23} N_{32}
	-( 1 + D_{23} D_{32} )  \delta_{23} 
	+ 2 D_{23}  \delta_{32} ) )
\end{eqnarray}
\begin{eqnarray}
	Z_{2.0} & = &
	( 1 - D_{31} D_{13} D_{32} D_{23} ) 
	( (1  - D_{31} D_{13} )
	(N_{32} 
	+ D_{32}  N_{23}
	+ (1 + D_{32}  D_{13} )\delta_{32} 
	- 2 D_{32}  \delta_{23}) \nonumber \\
	& & - ( 1 - D_{32}  D_{23} ) (
	N_{31} 
	+ D_{31} N_{13}
	-( 1 + D_{31} D_{13} )  \delta_{31} 
	+ 2 D_{31}  \delta_{13} ) )
\end{eqnarray}

Then

%------------------------------------------------------------------------------

% ############################################################

%------------------------------------------------------------------------------

\FloatBarrier

%% Acronyms
\newpage
\section*{Acronyms}
\addcontentsline{toc}{section}{Acronyms}
\begin{acronym}

% A  
\acro{AC}[AC]{Alternating Current}
\acro{ADC}[ADC]{Analog to Digital Converter}
\acro{AGN}[AGN]{Active Galactic Nuclei}

\acro{AIV}[AIV]{assembly, integration, and verification}
\acro{AIVT}[AIVT]{assembly, integration, verification, and testing}
\acro{AIT}[AIT]{assembly, integration, and testing}
\acro{AK}[AK]{``Analytic Kludge''}
\acro{AKE}[AKE]{attitude absolute knowledge}
\acro{AMCVn}[AM~CVn]{class of cataclysmic variable stars}
\acro{AMR}[AMR]{Anisotropic Magnetoresistors}
\acro{AO}[AO]{Announcement of Opportunity}

\acro{AOCS}[AOCS]{attitude and orbit control system}
\acro{AOM}[AOM]{acousto-optic modulator}
\acro{ASD}[ASD]{amplitude spectral density}
\acro{AST}[AST]{autonomous star tracker}
\acro{ATA}[ATA]{\href{http://www.seti-inst.edu/ata/}{Allen Telescope Array}}

% B
\acro{BAM}[BAM]{Beam Alignment Mechanism}
\acro{BAO}[BAO]{baryonic acoustic oscillation}
\acro{BB}[BB]{Breadboard}
\acro{BBN}[BBN]{Big Bang nucleosynthesis}
\acro{BCRS}[BCRS]{Barycentric Celestial Reference System}
\acro{BH}[BH]{black hole}
\acro{BHB}[BHB]{black hole binary}

% C
\acro{CAD}[CAD]{Computer Aided Design}
\acro{CAS}[CAS]{Constellation Acquisition Sensor}
\acro{CBE}[CBE]{Current Best Estimate}
\acro{CBOD}[CBOD]{Clamp band opening device}
\acro{CCD}[CCD]{Charge-coupled Device}
\acro{CCU}[CCU]{Caging Control Unit}
\acro{CDF}[CDF]{Concurrent Design Facility}
\acro{CDM}[CDM]{Cold dark Matter}
\acro{CDR}[CDR]{Critical Design Review}
\acro{CFRP}[CFRP]{Carbon Fibre Reinforced Plastic}
\acro{CM}[CM]{Caging Mechanism}
\acro{CMD}[CMD]{Charge Management Device}
\acro{CMM}[CMM]{Coordinate Measuring Machine}
\acro{CMS}[CMS]{Charge Management System}
\acro{CMNT}[CMNT]{Colloid Micro-Newton thruster}
\acro{CMB}[CMB]{Cosmic Microwave Background}
\acro{CNES}[CNES]{Centre National d’Etudes Spatiales}
\acro{COBE}[COBE]{\href{http://lambda.gsfc.nasa.gov/product/cobe/}{COsmic Background Explorer}}
\acro{CoM}[CoM]{Centre of Mass}
\acro{COMBO}[COMBO]{\href{http://www.mpia.de/COMBO/}{Classifying Objects by Medium-Band Observations}}
\acro{COSMOS}[COSMOS]{\href{http://irsa.ipac.caltech.edu/Missions/cosmos.html}{Cosmic Evolution Survey}}
\acro{CQP}[CQP]{Calibrated Quadrant Photodiode Pair}
\acro{COTS}[COTS]{Commercial off the Shelf}
\acro{CTE}[CTE]{Coefficient of Thermal Expansion}
\acro{CTP}[CTP]{Core Technology Program}
\acro{CVM}[CVM]{Caging and Venting Mechanism}

% D
\acro{DA}[DA] {Data Analysis}
\acro{D/A}[D/A]{digital-to-analogue converter}
\acro{DCC}[DCC]{Data Computing Center}
\acro{DCCs}[DCCs]{Data Computing Centers}
\acro{DDE}[DDE]{diagnostics drive electronics}
\acro{DDPC}[DDPC]{Distributed Data Processing Centre}
\acro{DEEP2}[DEEP2]{\href{http://deep.berkeley.edu/}{Deep Extragalactic Evolutionary Probe 2}}
\acro{DF}[DF]{drag-free}
\acro{DFACS}[DFACS]{Drag-Free Attitude Control System}
\acro{DOF}[DOF]{degree of freedom}
\acro{DMU}[DMU]{Data Management Unit}
\acro{DMU}[DMS]{Document Management System}
\acro{DP}[DP]{diagnostic package}
\acro{DPC}[DPC]{Data Processing Centre}
\acro{DPLL}[DPLL]{digital phase locked loop}
\acro{DRS}[DRS]{disturbance reduction system}
\acro{DSC}[Daughter-S/C]{``Daughter'' spacecraft}
\acro{DS}[DS]{Diagnostics Subsystem}
\acro{DSN}[DSN]{Deep Space Network}
\acro{DTM}[DTM]{deterministic transfer manoeuvre}
\acro{DWS}[DWS]{differential wavefront sensing}

% E
\acro{E2E}[E2E]{End-to-End}
\acro{EBB}[EBB]{Elegant Breadboard}
\acro{ECSS}[ECSS]{European Cooperation for Space Standardization}
\acro{EDU}[EDU]{Engineering Development Unit}
\acro{EELV}[EELV]{Evolved Expendable Launch Vehicle}
\acro{EGAPS}[EGAPS]{European Galactic Plane Surveys}
\acro{EH}[EH]{Electrode Housing}
\acro{ELV}[ELV]{Expendable Launch Vehicle}
\acro{EMa}[EMa]{Electro-Magnetic}
\acro{EM}[EM]{Engineering Model}
\acro{EMRI}[EMRI]{extreme mass-ratio inspiral}
\acro{EOL}[EOL]{End-of-life}
\acro{EoM}[EoM]{Equations of Motion}
\acro{EOM}[EOM]{Electro-Optical Modulator}
\acro{EPS}[EPS]{extended Press-Schechter formalism}
\acro{EQM}[EQM]{Engineering and Qualification Model}
\acro{ESA}[ESA]{European Space Agency}
\acro{ESAC}[ESAC]{European Space Astronomy Centre in Madrid, Spain}
\acro{ESOC}[ESOC]{European Space Operations Centre}
\acro{ETU}[ETU]{Engineering Thermal Unit}
\acro{ePMS}[ePMS]{extended Phase Measurement Subsystem}

% F
\acro{FAQ}[FAQ]{Frequently Asked Questions}
\acro{FBD}[FBD]{Functional Block Diagram}
\acro{FDIR}[FDIR]{Failure Detection, Isolation, and Recovery}
\acro{FE}[FE]{finite-element (methods)}
\acro{FEE}[FEE]{front-end electronics}
\acro{FEEP}[FEEP]{field-emission electric propulsion}
\acro{FEESAU}[FEE SAU]{front-end electronics sensing and actuation unit}
\acro{FF-OGSE}[FF-OGSE]{Far-Field Optical Ground Support Equipment}  
\acro{FITS}[FITS]{Flexible Image Transport System}
\acro{FIOS}[FIOS]{Fibre Injector Optical Subassembly}
\acro{FM}[FM]{Flight Model}
\acro{FOH}[FOH]{Fibre Optic Harness}
\acro{FSU}[FSU]{fibre switching unit}
\acro{FSUA}[FSUA]{fibre switching unit assembly}
\acro{FPAG}[FPAG]{Fundamental Physics Advisory Group}
\acro{FPGA}[FPGA]{field-programmable gate array}
\acro{FR}[FR]{laser frequency  reference}
\acro{FS}[FS]{frequency  separated}
\acro{FDS}[FDS]{Frequency Distribution System}

% G
\acro{GBs}[GBs]{Galactic Binaries}
\acro{GCR}[GCR]{Galactic Cosmic Ray}
\acro{GCRS}[GCRS]{Geocentric Celestial Reference System}
\acro{GRACE-FO}[GRACE-FO]{\href{https://gracefo.jpl.nasa.gov/~}{Gravity Recovery and Climate Explorer Follow On}}
\acro{GPRM}[GPRM]{Grabbing Positioning Release Mechanism}
\acro{GR}[GR]{General Theory of Relativity}
\acro{GRS}[GRS]{Gravitational Reference Sensor}
\acro{GS}[GS]{Ground Station}
\acro{GSE}[GSE]{Ground Support Equipment}
\acro{GSFC}[GSFC]{Goddard Space Flight Center}
\acro{GTO}[GTO]{Geostationary Transfer Orbit}
\acro{GR740}[GR740]{\href{http://microelectronics.esa.int/gr740/index.html}{The ESA Next Generation Microprocessor (NGMP)}}
\acro{GW}[GW]{Gravitational Wave}

% H
\acro{HDF}[HDF]{Hierarchical Data Format}
\acro{HDRM}[HDRM]{Hold Down and Release Mechanism}
\acro{HETO}[HETO]{Heliocentric Earth Trailing Orbit}
\acro{Hg}[Hg]{mercury}
\acro{HGA}[HGA]{high-gain antenna}
\acro{HR}[HR]{High Resolution}
\acro{HST}[HST]{\href{http://hubble.nasa.gov/~}{Hubble Space Telescope}}

% I
\acro{IA}[IA]{Instrument Amplifier}
\acro{IAU}[IAU]{International Astronomical Union}
\acro{IAAS}[IAAS]{Infrastructure As A Service}
\acro{IBM}[IBM]{Internal Balance Mass}
\acro{ICC}[ICC]{Instrument Control Computer}
\acro{ICRF}[ICRF]{International Celestial Reference Frame}
\acro{ICRS}[ICRS]{International Celestial Reference System}
\acro{IDL}[IDL]{Interferometer Data Log}
\acro{I/F}[I/F]{interface}
\acro{IFO}[IFO]{Interferometer}
\acro{IFP}[IFP]{In-Field Pointing}
\acro{IGM}[IGM]{inter-galactic medium}
\acro{IMA}[IMA]{Integrated Modular Avionics}
\acro{IMBH}[IMBH]{Intermediate Mass Black Hole}
\acro{IMF}[IMF]{initial mass function}
\acro{IMR}[IMR]{Inspiral-Merger-Ringdown}
\acro{IMRI}[IMRI]{intermediate mass-ratio inspiral}
\acro{IMS}[IMS]{interferometric measurement system}
\acro{IN2P3}[IN2P3]{National Institute of Nuclear and Particle Physics}
\acro{INReP}[INReP]{Initial Noise Reduction Pipeline}
\acro{IOCR}[IOCR]{in-orbit commissioning review}
\acro{IOT}[IOT]{Instrument Operations Team}
\acro{ISH}[ISH]{Inertial Sensor Head}
\acro{ISM}[ISM]{instrument sensitivity model}
\acro{ISUK}[ISUK]{Inertial Sensor UV Kit}
\acro{IT}[IT]{Information Technology}

% J
\acro{JILA}[JILA]{\href{http://jila.colorado.edu/}{Joint Institute for Laboratory Astrophysics}}
\acro{JPL}[JPL]{\href{http://www.jpl.nasa.gov/}{Jet Propulsion Laboratory}}
\acro{JWST}[JWST]{\href{http://www.jwst.nasa.gov/}{James Webb Space Telescope}}

% K
\acro{KSC}[KSC]{\href{http://www.nasa.gov/centers/kennedy/home/index.html}{Kennedy Space Center}}

% L
\acro{LA}[LA]{Laser Assembly}
\acro{LAGOS}[LAGOS]{Laser Antenna for Gravitational-radiation Observation in Space}
\acro{LCA}[LCA]{LISA Core Assembly}
\acro{LCM}[LCM]{NGO launch composite}
\acro{LDC}[LDC]{LISA Data Challenge}
\acro{LDP}[LDP]{LISA Data Processing}
\acro{LDPG}[LDPG]{LISA Data Processing Group}
\acro{LED}[LED]{light-emitting diode}
\acro{LEM}[LEM]{Laser Electrical Module}
\acro{LEOP}[LEOP]{Launch and Early Operations Phase}
\acro{LGA}[LGA]{low-gain antenna}
\acro{LIG}[LIG]{LISA Instrument Group}
\acro{LIGO}[LIGO]{\href{http://www.ligo.caltech.edu/}{Laser Interferemeter Gravitational Wave Observatory}}
\acro{LISA}[LISA]{\href{https://www.lisamission.org/}{Laser Interferometer Space Antenna}}
\acro{LIST}[LIST]{\href{http://list.caltech.edu/}{LISA International Science Team}}
\acro{LLD}[LLD]{launch lock device}
\acro{LMC}[LMC]{Large Magellanic Cloud}
\acro{LMF}[LMF]{LISA mission formulation study}
\acro{LOA}[LoA]{Letter of Agreement}
\acro{LOM}[LOM]{Laser Optical Module}
\acro{LOS}[LOS]{line of sight}
\acro{LPF}[LPF]{\href{http://lisapathfinder.esa.int}{LISA Pathfinder}}
\acro{LPS}[LPS]{Laser Pre-stabilization System}
\acro{LTPDA}[LTPDA]{LISA Technology Package Data Analysis}
\acro{LH}[LH]{Laser Head}
\acro{LO}[LO]{Local Oscillator}
\acro{LRI}[LRI]{Laser Ranging Instrument (on GRACE-FO)}
\acro{LS}[LS]{laser system}
\acro{LSG}[LSG]{LISA Science Group}
\acro{LSO}[LSO]{last stable orbit}
\acro{LSST}[LSST]{\href{http://www.lsst.org/lsst}{Large Synoptic Survey Telescope}}
\acro{LTP}[LTP]{LISA Technology Package}
\acro{LUT}[LUT]{Look-Up Table}
\acro{LVA}[LVA]{launch vehicle adaptor}

% M
\acro{MAC}[MAC]{Mass Acceleration Curve}
\acro{MAXI}[MAXI]{\href{http://www.nasa.gov/mission_pages/station/research/experiments/MAXI.html}{Monitor of All-sky X-ray Image}}
\acro{MBH}[MBH]{Massive Black Hole}
\acro{MBHB}[MBHB]{Massive Black Hole Binary}
\acro{MCMC}[MCMC]{Markov-chain Monte Carlo}
\acro{MCU}[MCU]{Mechanism Control Unit}
\acro{MEOP}[MEOP]{maximum expected operating pressure}
\acro{MGSE}[MGSE]{Mechanical Ground Support Equipment}
\acro{MICD}[MICD]{Mechanical Interface Control Document}
\acro{MLA}[MLA]{Multi-lateral agreement}
\acro{MLB}[MLB]{motorised light band}
\acro{MLDC}[MLDC]{\href{http://astrogravs.nasa.gov/docs/mldc/}{Mock LISA Data Challenge}}
\acro{MLI}[MLI]{multi layer insulation}
\acro{MMH}[MMH]{monomethyl hydrazine}
\acro{MO}[MO]{Maser Oscillator}
\acro{MOC}[MOC]{Mission Operation Centre}
\acro{MOFPA}[MOFPA]{Master Oscillator Fibre Power Amplifier}
\acro{MOPA}[MOPA]{Master Oscillator Power Amplifier}
\acro{MON-3}[MON-3]{mixed oxides of nitrogen with 3\% nitric oxide}
\acro{MOSA}[MOSA]{Moving Optical Sub-Assembly}
\acro{MOU}[MoU]{Memorandum of Understanding}
\acro{MRD}[LISA-MRD-001]{Mission Requirement Document}
\acro{MSC}[Mother-S/C]{``Mother'' spacecraft}
\acro{MSS}[MSS]{MOSA Support Structure}

% N
\acro{NASA}[NASA]{\href{http://www.nasa.gov}{National Aeronautic and Space Administration}}
\acro{NGMP}[NGMP]{Next Generation Micro Processor}
\acro{NGO}[NGO]{New Gravitational wave Observatory}
\acro{NPMB}[NPMB]{National Program Managers Board}
\acro{NPRO}[NPRO]{non-planar ring oscillator}
\acro{NR}[NR]{numerical relativity}
\acro{NTC}[NTC]{Negative Temperature Coefficient}

% O
\acro{OAS}[OAS]{optical assembly subsystem}
\acro{OATM}[OATM]{Optical Assembly Tracking Mechanism}
\acro{OAM}[OAM]{optical assembly mechanics}
\acro{OB}[OB]{Optical Bench}
\acro{OBA}[OBA]{Optical Bench Assembly}
\acro{OBC}[OBC]{on-board computer}
\acro{OGSE}[OGSE]{Optical Ground Support Equipment}
\acro{OM}[OM]{Optical Model}
\acro{OMS}[OMS]{Optical Metrology System}
\acro{OP}[OP]{Optical Path}
\acro{ORO}[ORO]{optical read-out}
\acro{OT}[OT]{optical truss}

% P

\acro{PA}[PA]{power amplifier}
\acro{PA}[PA]{Product Assurance}
\acro{PAA}[PAA]{point-ahead angle}
\acro{PAAM}[PAAM]{point-ahead angle mechanism}
\acro{Pan-Starrs}[Pan-Starrs]{\href{http://pan-starrs.ifa.hawaii.edu/public/}{the Panoramic Survey Telescope \& Rapid Response System}}
\acro{PCU}[PCU]{power conditioning unit}
\acro{PCDU}[PCDU]{power control and distribution unit}
\acro{PCP}[PCP]{Payload Commanding and Processing}
\acro{PCP-GSE}[PCP-GSE]{Payload Commanding and Processing Ground Support Equipment}
\acro{PCS}[PCS]{payload control subsystem}
\acro{PDD}[PDD]{payload description document}
\acro{PDF}[PDF]{probability density function}
\acro{PDH}[PDH]{Pound Drever Hall}
\acro{PD}[PD]{photo diode}
\acro{PDR}[PDR]{Preliminary Design Review}
\acro{PDS}[PDS]{photo detector system}
\acro{PLS}[PLS]{Power Law Sensitivity}
\acro{PLL}[PLL]{phase-locked loop}
\acro{P/L}[P/L]{payload}
\acro{P/M}[P/M]{propulsion module}
\acro{PM}[PM]{Progress Meeting}
\acro{PMF}[PMF]{Polarization-Maintaining Fibre}
\acro{PMFDE}[PMFDE]{phase meter frequency distribution electronics}
\acro{PMFEE}[PMFEE]{phase meter front-end electronics}
\acro{PMDSP}[PMDSP]{phase meter digital signal processor}
\acro{PMS}[PMS]{Phase Measurement Subsystem}
\acro{PN}[PN]{post-Newtonian}
\acro{PRN}[PRN]{pseudo-random noise, \emph{often} pseudo-noise}
\acro{PRDS}[PRDS]{Phase Reference Distribution System}
\acro{PRDS-OGSE}[PRDS-OGSE]{Phase Reference Distribution System - Optical Ground Support Equipment}
\acro{PRT}[PRT]{Platinum Resistance Thermometers}
\acro{PSD}[PSD]{Power Spectral Density}
\acro{PSF}[PSF]{point-spread function}
\acro{PTF}[PTF]{\href{http://www.astro.caltech.edu/ptf/}{Palomar Transient Factory}}

% Q
\acro{QA}[QA]{Quality Assurance}
\acro{QM}[QM]{Qualification Model}
\acro{QNM}[QNM]{Quasi-normal mode}
\acro{QPD}[QPD]{Quadrant photodetector}
\acro{QPR}[QPR]{Quadrant Photo-Receiver }
\acro{QSO}[QSO]{Quasi-stellar object}

% R
\acro{RATS}[RATS]{Rapid Time Survey}
\acro{REF}[REF]{reference}
\acro{RF}[RF]{radio frequency}
\acro{RIN}[RIN]{Relative Intensity Noise}
\acro{RIT}[RIT]{Radio-frequency Ion Thruster}
\acro{RM}[RM]{Radiation Monitor}
\acro{RMS}[RMS]{root mean square}
\acro{RSS}[RSS]{root sum square}
\acro{RTOS}[RTOS]{Real Time Operating System}
\acro{RXTE}[RXTE]{\href{http://heasarc.gsfc.nasa.gov/docs/xte/xtegof.html}{Rossi X-Ray Timing Explorer}}
\acro{RX}[RX]{received signal}

% S
\acro{S/C}[S/C]{spacecraft}
\acro{S/S}[S/S]{subsystem}
\acro{S/C-P/M}[S/C-P/M]{spacecraft/propulsion-module}
\acro{SAVOIR}[SAVOIR]{\href{http://savoir.estec.esa.int/}{Space AVionics Open Interface aRchitecture}}
\acro{SAU}[SAU]{sensing and actuation unit}
\acro{SBCC}[SBCC]{Single Board Computer Core}
\acro{SciRD}[SciRD]{\href{https://www.cosmos.esa.int/documents/678316/1700384/SciRD.pdf/25831f6b-3c01-e215-5916-4ac6e4b306fb?t=1526479841000}{Science Requirement Document}}
\acro{SCOE}[SCOE]{Special Check Out Equipment}
\acro{SEP}[SEP]{Solar Energetic Particle}
\acro{SEPD}[SEPD]{single-element photo diode}
\acro{SIM}[SIM]{Space Interferometry Mission}
\acro{SIRD}[SIRD]{Science Implementation Requirement Document}
\acro{SIP}[SIP]{Science Implementation Plan}
\acro{SIPs}[SIPs]{Science Implementation Plans}
\acro{SISO}[SISO]{single input/single output}
\acro{SDP}[SDP]{system data pool}
\acro{SDSS}[SDSS]{\href{http://www.sdss.org/}{Sloan Digital Sky Survey}}
\acro{SGS}[SGS]{Science Ground Segment}
\acro{SGWB}[SGWB]{Stochastic Gravitational Wave Background}
\acro{SL}[SL]{Scatter Light}
\acro{SMF}[SMF]{Single Mode Fiber}
\acro{SQUID}[SQUID]{Superconducting Quantum Interference Device}
\acro{SMBH}[SMBH]{super-massive black hole}
\acro{SMC}[SMC]{Small Magellanic Cloud}
\acro{SMP}[SMP]{Science Management Plan}
\acro{SNR}[SNR]{Signal-to-Noise Ratio}
\acro{SOBH}[SOBHB]{Stellar Origin Black Hole Binary}
\acro{SOCD}[SOCD]{Science Operations Concept Document}
\acro{SOAD}[SOAD]{Science Operations Assumptions Document}
\acro{SOC}[SOC]{Science Operation Centre}
\acro{SOVT}[SOVT]{Science Operations Verifications Tests}
\acro{SPA}[SPA]{Stationary Phase Approximation}
\acro{SPC}[SPC]{Science Programme Committee}
\acro{SRP}[SRP]{solar radiation pressure}
\acro{SRS}[SRS]{spacecraft reference system}
\acro{SSB}[SSB]{Solar System barycenter}
\acro{SST}[SST]{Science Study Team}
\acro{STM}[STM]{Structural and Thermal Model}
\acro{STR}[STR]{Coarse Star Tracker}
\acro{SWT}[SWT]{Science Working Team}

% T
\acro{TBC}[TBC]{to be confirmed}
\acro{TBD}[TBD]{to be determined}
\acro{TC}[TC]{Telecommand}
\acro{TCB}[TCB]{Barycentric Coordinate Time}
\acro{TCG}[TCG]{Geocentric Coordinate Time}
\acro{TCM}[TCM]{trajectory correction manoeuvre}
\acro{TC/TM}[TC/TM]{telecommand/telemetry}
\acro{TCLS}[TCLS]{\href{http://www.tcls-arm-for-space.eu/} {Triple Core LockStep}}
\acro{TDI}[TDI]{Time Delay Interferometry}
\acro{TECC}[TECC]{Transient Event Coordination Committee}
\acro{THE}[THE]{On-Board Clock Time}
\acro{TM}[TM]{test mass, \emph{often proof mass}}
\acro{TM-OGSE}[TM-OGSE]{Test-Mass Optical Ground Equipment}
\acro{TNO}[TNO]{Nederlandse Organisatie voor Toegepast Natuurwetenschappelijk Onderzoek}
\acro{TOBA}[TOBA]{Telescope and Optical Bench Assembly}
\acro{TOGA}[TOGA]{Telescope, Optical bench and Gravitational reference sensor Assembly}
\acro{TSP}[TSP]{Temporal and Spatial Partitioning}
\acro{TP}[TP]{Telescope Pointing}
\acro{TPS}[TPS]{Spacecraft Proper Time}
\acro{TRL}[TRL]{Technology Readiness Level}
\acro{TRP}[TRP]{temperature reference points}
\acro{TS}[TS]{Telescope}
\acro{TTC}[TT\&C]{telemetry, tracking, and command}
\acro{TTL}[TTL]{Tilt-To-Length}
\acro{TWTA}[TWTA]{traveling-wave tube amplifier}
\acro{TX}[TX]{transmit signal}

% U
\acro{USO}[USO]{ultra-stable oscillator}
\acro{UTC}[UTC]{Coordinated Universal Time}
\acro{ULU}[ULU]{UV light unit}
\acro{UV}[UV]{ultra-violet}

% V
\acro{VAST}[VAST]{\href{http://www.physics.usyd.edu.au/sifa/vast/index.php}{Variables and Slow Transients, An ASKAP Survey for Variables and Slow Transients is a Survey Science Project for the Australian SKA Pathfinder}}
\acro{VB}[VB]{Verification Binary}
\acro{VC}[VC]{Vacuum Chamber}
\acro{VMS}[VMS]{very massive star}

% W
\acro{WD}[WD]{White Dwarf}
\acro{WG}[WG]{Working Group}
\acro{WMAP}[WMAP]{\href{http://lambda.gsfc.nasa.gov/product/map/current/}{Wilkison Microwave Anisotropy Probe}}
\acro{WP}[WP]{Work Package}
\acro{WPs}[WPs]{Work Packages}
\acro{WR}[WR]{Wide Range}

% X
\acro{XML}[XML]{Extensible Markup Language}

% Y

% Z

\end{acronym}

\bibliographystyle{plain}
\addcontentsline{toc}{section}{References}
\bibliography{refs}

\end{document}